\documentclass[aps,pra,twocolumn,superscriptaddress]{revtex4-2}  

\usepackage{amsmath,amssymb,amsfonts,bm,physics,latexsym,comment} 
\usepackage[pdftex]{graphicx,color}
\usepackage{tikz}

\newcommand{\Jperp}{J_\perp}

\begin{document}
%%%%%%%%%%%%%%%%%%%%%%%%%%%%%%%%%%%%%%%%%%%%%%%%%
% Paper Information
%%%%%%%%%%%%%%%%%%%%%%%%%%%%%%%%%%%%%%%%%%%%%%%%%
\title{
Symmetry-protected topological phases and competing orders\\ in a spin-$\frac12$ XXZ ladder with a four-spin interaction
}
\author{Takuhiro Ogino}%\email{}
\affiliation{Institute for Solid State Physics, University of Tokyo, Kashiwa, Chiba 277-8581, Japan}
\author{Shunsuke Furukawa}
\affiliation{Department of Physics, Keio University, Kohoku-ku, Yokohama, Kanagawa 223-8522, Japan}
\author{Ryui Kaneko}
\affiliation{Department of Physics, Kindai University, Higashi-Osaka, Osaka 577-8502, Japan}
% \affiliation{Institute for Solid State Physics, University of Tokyo, Kashiwa, Chiba 277-8581, Japan}
\author{Satoshi Morita}
\affiliation{Institute for Solid State Physics, University of Tokyo, Kashiwa, Chiba 277-8581, Japan}
\author{Naoki Kawashima}
\affiliation{Institute for Solid State Physics, University of Tokyo, Kashiwa, Chiba 277-8581, Japan}

\date{\today}

%%%%%%%%%%%%%%%%%%%%%%%%%%%%%%%%%%%%%%%%%%%%%%%%
% Abstract
%%%%%%%%%%%%%%%%%%%%%%%%%%%%%%%%%%%%%%%%%%%%%%%%
\begin{abstract}
We study a spin-$\frac12$ XXZ model with a four-spin interaction on a two-leg ladder. 
By means of effective field theory and matrix product state calculations, 
we obtain rich ground-state phase diagrams that consist of eight distinct gapped phases. 
Four of them exhibit spontaneous symmetry breaking with either a magnetic or valence-bond-solid (VBS) long-range order. 
The other four are featureless, i.e., the bulk ground state is unique and does not break any symmetry. 
The featureless phases include the rung singlet (RS) and Haldane phases 
as well as their variants, the RS* and Haldane* phases, 
in which twisted singlet pairs $\left( \ket{\uparrow\downarrow}+\ket{\downarrow\uparrow} \right)/\sqrt{2}$ are formed between the two legs. 
We argue and demonstrate that 
Gaussian transitions with the central charge $c=1$ occur between the featureless phases and between the ordered phases 
while Ising transitions with $c=1/2$ occur between the featureless and ordered phases. 
The two types of transition lines cross 
% in the SU$(2)$-symmetric case, 
at the SU$(2)$-symmetric point,
where the criticality is described by the SU$(2)_2$ Wess-Zumino-Witten theory with $c=3/2$. 
The RS-Haldane* and RS*-Haldane transitions give examples of topological phase transitions. 
Interestingly, the RS* and Haldane* phases, which have highly anisotropic nature, appear even in the vicinity of the isotropic case. 
We demonstrate that all the four featureless phases are distinguished by topological indices in the presence of 
%the spin rotational dihedral symmetry $D_2$, the inter-leg exchange symmetry $\sigma$, and the translational symmetry. 
certain symmetries. 
\end{abstract}
\maketitle

%%%%%%%%%%%%%%%%%%%%%%%%%%%%%%%%%%%%%%%%%%%%%%%%
% Main text
%%%%%%%%%%%%%%%%%%%%%%%%%%%%%%%%%%%%%%%%%%%%%%%%

%%%%%%%%%%%%%%%%%%%%%%%%%%%%%%%%%%%%%%%%%%%%%%%%
\section{Introduction}
The concept of symmetry-protected topological (SPT) phases has been proposed and fruitfully developed over the last decade 
\cite{PhysRevB.80.155131,PhysRevB.82.155138,Weneaal3099, Senthil15,Zeng2019book,Tasaki2020book}. 
A SPT phase is a {\it featureless} gapped phase that does not exhibit spontaneous symmetry breaking
but is distinguished from a trivial phase (i.e. a phase that includes a site-factorized product state) as long as certain symmetries are imposed. 
For example, Haldane phases \cite{HALDANE1983464,PhysRevLett.50.1153,Affleck_1989} of odd-integer-spin chains are distinguished from trivial phases 
in the presence of the discrete spin rotation symmetry $D_2(=\mathbb{Z}_2\times \mathbb{Z}_2)$, time-reversal symmetry, or bond-centered inversion symmetry 
\cite{PhysRevB.80.155131,PhysRevB.81.064439,PhysRevB.85.075125,Ogata2020,ogata2019mathbb}. 
The Haldane phases with odd integer spins have been characterized 
by a string order parameter \cite{PhysRevB.40.4709}, hidden symmetry breaking \cite{PhysRevB.45.304,Oshikawa_1992}, 
a twist operator \cite{PhysRevLett.89.077204,PhysRevLett.121.140604}, 
degeneracy in the edge states \cite{Kennedy_1990},
the entanglement spectrum \cite{PhysRevB.81.064439}, 
and topological indices \cite{PhysRevB.81.064439, PhysRevB.86.125441, PhysRevB.83.035107, PhysRevB.84.235128}.
Classification of SPT phases of bosons has been achieved 
by using the projective representations of the symmetry group in one dimension \cite{PhysRevB.83.035107,PhysRevB.84.235128,PhysRevB.84.165139} 
and group cohomology theory in higher dimensions \cite{ChenX12,ChenX13}. 

%############################
\begin{figure}[b]
\begin{center}
\vspace{0cm}
\begin{tikzpicture}[scale=0.8]
\draw (-0.8,0) -- (-0.2,0);\draw (-0.8,2) -- (-0.2,2);
\node at (0,-0.5) {$\bm{S}_{2,j-2}$}; \node at (0,2.5) {$\bm{S}_{1,j-2}$};
\draw  (0,0) circle (0.2cm);\draw (0,2) circle (0.2cm); \draw (0,0.2) -- (0,1.8);
\draw [<->] (0.2,0) -- (1.8,0);\draw (0.2,2) -- (1.8,2);
\node at (2,-0.5) {$\bm{S}_{2,j-1}$}; \node at (2,2.5) {$\bm{S}_{1,j-1}$};
\draw (2,0) circle (0.2cm);\draw (2,2) circle (0.2cm);\draw [<->](2,0.2) -- (2,1.8);
\draw (2.2,0) -- (3.8,0);\draw (2.2,2) -- (3.8,2);
\node at (4,-0.5) {$\bm{S}_{2,j}$}; \node at (4,2.5) {$\bm{S}_{1,j}$};
\draw (4,0) circle (0.2cm);\draw (4,2) circle (0.2cm);\draw (4,0.2) -- (4,1.8);
\draw (4.2,0) -- (5.8,0);\draw (4.2,2) -- (5.8,2);
\node at (6,-0.5) {$\bm{S}_{2,j+1}$}; \node at (6,2.5) {$\bm{S}_{1,j+1}$};
\draw (6,0) circle (0.2cm);\draw (6,2) circle (0.2cm);\draw (6,0.2) -- (6,1.8);
\draw (6.2,0) -- (7.8,0);\draw (6.2,2) -- (7.8,2);
\node at (8,-0.5) {$\bm{S}_{2,j+2}$}; \node at (8,2.5) {$\bm{S}_{1,j+2}$};
\draw (8,0) circle (0.2cm);\draw  (8,2) circle (0.2cm);\draw (8,0.2) -- (8,1.8);
\draw (8.2,0) -- (8.8,0);\draw (8.2,2) -- (8.8,2);
\node at (1,0.4) {$J$}; \node at (2.4,1) {$J_\perp$}; \node at (5,1) {$J_4$};\draw (5,1) circle (0.8cm);
\end{tikzpicture}
\caption{\label{fig:system}
XXZ ladder with a four-spin interaction, 
which is described by the Hamiltonian \eqref{eq:modelmain}. 
The spin-$\frac12$ operator at the $j$-th site on the $\alpha$-th leg is denoted by $\bm{S}_{\alpha,j}$. 
We set $J=1$ throughout the paper. 
}
\end{center}
\end{figure}
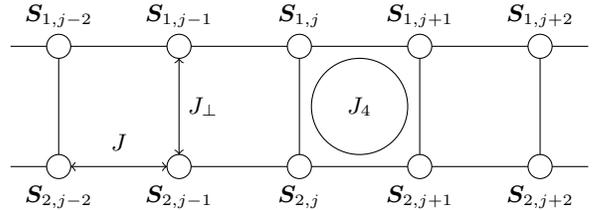
%############################

%############################
\begin{figure}
\includegraphics[width=78mm]{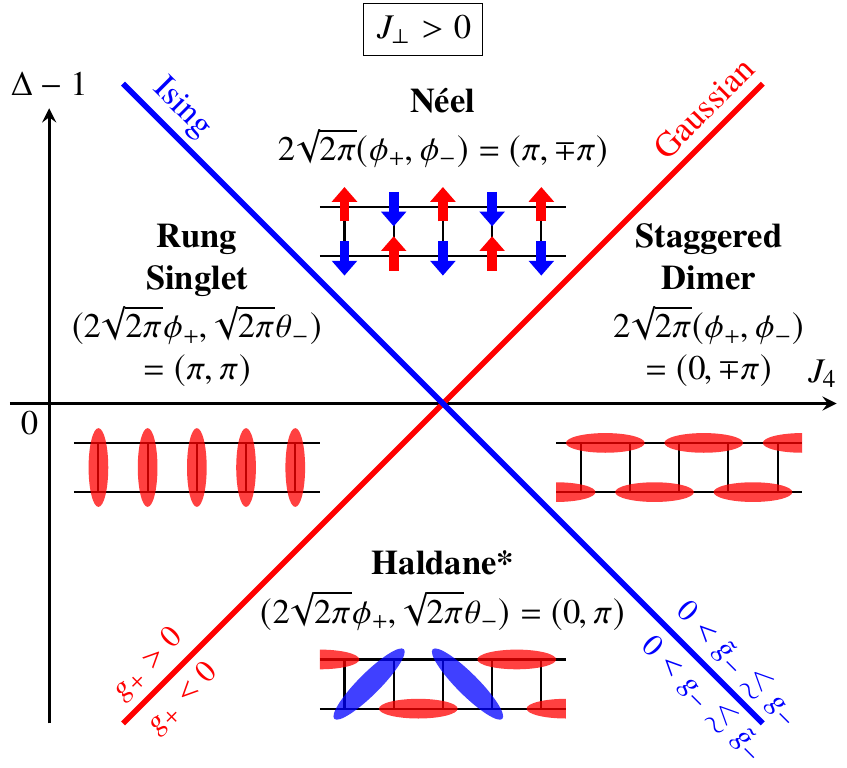} 
\caption{\label{fig:PhaseDiagrambos1}
Schematic phase diagram of the model \eqref{eq:modelmain} 
%on the $\Delta$-$J_4$ plane in the weak couplings region where $0<J_\perp, J_4 \ll 1$.
for $0<J_\perp, J_4 \ll 1$, obtained with the effective Hamiltonian \eqref{eq:Heff}. 
The value of $\Jperp$ is fixed. 
There are four phases and two phase transition lines. 
Each phase is characterized by the locking positions of the bosonic fields. 
%The red solid line is the gaussian transition with $c=1$ and the blue solid line is the $1+1$ D Ising transition with $c=1/2$. 
The red solid line shows the Gaussian transition with the central charge $c=1$ 
while the blue solid line shows the Ising transition with $c=1/2$.
%Two spins enclosed by a red dashed line is a singlet, $(\ket{\uparrow\downarrow} - \ket{\downarrow\uparrow})/\sqrt{2}$
%and two spins enclosed by a blue dashed line is a twisted singlet, $(\ket{\uparrow\downarrow} + \ket{\downarrow\uparrow})/\sqrt{2}$.
Two spins paired by a red oval form a singlet $(\ket{\uparrow\downarrow} - \ket{\downarrow\uparrow})/\sqrt{2}$ 
while those paired by a blue oval form a twisted singlet $(\ket{\uparrow\downarrow} + \ket{\downarrow\uparrow})/\sqrt{2}$. 
The Haldane* state is related with the conventional Haldane state on a ladder (see Fig.\ \ref{fig:PhaseDiagrambos2}) 
through the $\pi$ rotation of the spins on one of the chains about the $z$-axis; 
a singlet pair of spins on different chains turns into a twisted one by this transformation. 
}
\end{figure}
%############################

%############################
\begin{figure}
\includegraphics[width=78mm]{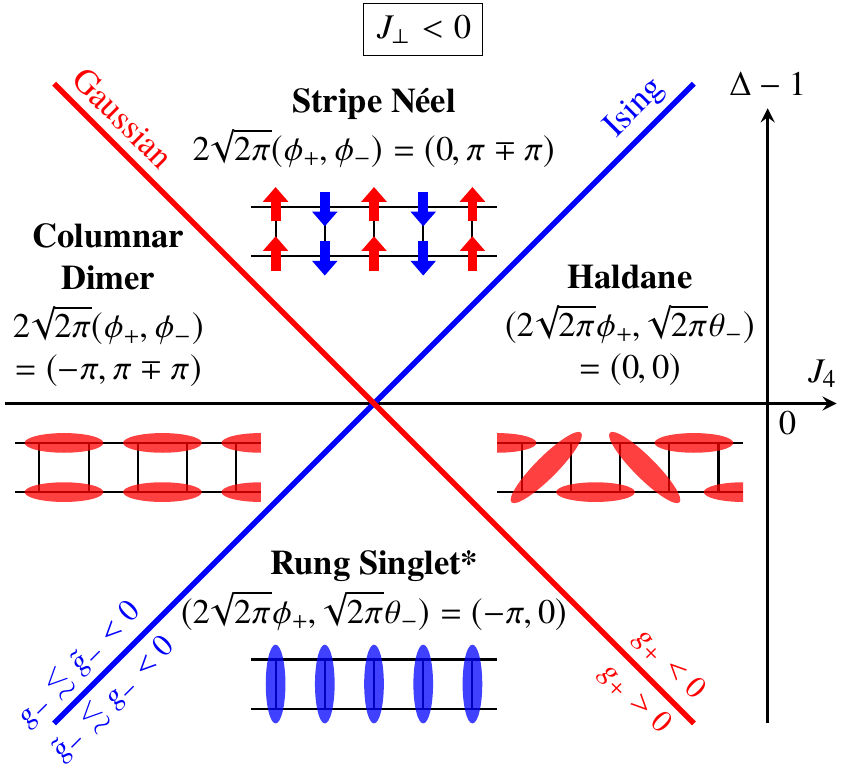} 
\caption{\label{fig:PhaseDiagrambos2}
Schematic phase diagram of the model \eqref{eq:modelmain} 
%on the $\Delta$-$J_4$ plane in the weak couplings region where $ J_\perp, J_4 < 0 $ and $|J_\perp|, |J_4| \ll 1$. 
for $J_\perp, J_4 < 0 $ with $|J_\perp|, |J_4| \ll 1$, obtained with the effective Hamiltonian \eqref{eq:Heff}.  
The value of $\Jperp$ is fixed. 
%There are four phases and two phase transition lines. 
There are four phases, each characterized by the field locking positions. 
%The red solid line is the gaussian transition with $c=1$ and the blue solid line is the $1+1$ D Ising transition with $c=1/2$. 
As in Fig.\ \ref{fig:PhaseDiagrambos1}, the red and blue solid lines show Gaussian and Ising transitions, respectively.  
%Two spins enclosed by a red dashed line is a singlet, $(\ket{\uparrow\downarrow} - \ket{\downarrow\uparrow})/\sqrt{2}$
%and two spins enclosed by a blue dashed line is a twisted singlet, $(\ket{\uparrow\downarrow} + \ket{\downarrow\uparrow})/\sqrt{2}$.
Red and blue ovals show a singlet and a twisted singlet, respectively. 
The Haldane state is a superposition of various singlet-covering states, and the inset for the Haldane phase is an example of such singlet covering.  
}
\end{figure}
%############################

% [ SPT phases on a ladder ]--------------------------------------
A simple extension of the spin-$1$ Haldane phase can be found in a spin-$\frac12$ two-leg ladder. 
Consider a spin-$\frac12$ ladder with Heisenberg interactions $J$ and $\Jperp$ along the legs and rungs, respectively, 
where $J$ is assumed to be antiferromagnetic; see Fig.\ \ref{fig:system}. 
Field-theoretical analyses indicate that the presence of the rung interaction $\Jperp\ne 0$ immediately leads to gapped phases 
whose properties depend on the sign of $\Jperp$ \cite{PhysRevLett.69.2419,PhysRevB.50.9911,PhysRevB.53.8521}. 
In the gapped phase for ferromagnetic $\Jperp<0$, effective spin-$1$ degrees of freedom emerge on the rungs and collectively form the Haldane state. 
This state can equivalently be viewed as a superposition of various singlet-covering states on a ladder. 
In contrast, the gapped phase for antiferromagnetic $\Jperp>0$ can be understood from the limit $\Jperp\to\infty$, 
where the ground state is a product state of singlet pairs on the rungs, known as the rung singlet (RS) state. 
The gapped phases for $\Jperp<0$ and $\Jperp>0$ are thus called the Haldane and RS phases, respectively. 
These phases are distinguished in terms of two types of string order parameters 
\cite{doi:10.1143/JPSJ.62.2845,doi:10.1143/JPSJ.64.1967,PhysRevB.62.14965,PhysRevB.77.205121} [shown in Eq.\ \eqref{eq:stringOP} later]. 

Liu {\it et al.} \cite{PhysRevB.86.195122} have conducted a detailed classification of SPT phases 
in a spin-$\frac12$ two-leg ladder for the symmetry group $D_2\times\sigma$, 
where $\sigma$ is the symmetry with respect to the interchange of the two legs. 
They have found three new SPT phases termed $t_\mu~(\mu=x,y,z)$, 
all of which have symmetry-protected two-fold degenerate edge states on each end of an open ladder, 
with unique responses to magnetic fields. 
For example, the edge states in the $t_z$ phase are split by the magnetic field along the $z$ direction, 
but not by the fields in the $x$ and $y$ directions. 
The $t_\mu$ phase is related with the Haldane phase (termed $t_0$ in Ref.\ \cite{PhysRevB.86.195122}) 
under the $\pi$ rotation of spins about the $\mu$-axis on the first leg, $U_1^\mu(\pi):=\exp\qty(i\pi\sum_j S_{1,j}^\mu)$, which provides some intuition on the ground state. 
Specifically, a singlet pair between spins on different legs turns into a twisted singlet 
$\ket{1,z}:=\left( \ket{\uparrow\downarrow}+\ket{\downarrow\uparrow} \right)/\sqrt{2}$ under $U_1^z(\pi)$. 
It has also been argued that the trivial phase (i.e., the phase that includes rung-factorized states) is divided into four phases, 
termed rung-$\ket{0,0}$ and rung-$\ket{1,\mu}~(\mu=x,y,z)$, if the translational symmetry is further imposed. 
The rung-$\ket{0,0}$ phase corresponds to the conventional RS phase, and the rung-$\ket{1,\mu}$ phase is obtained 
by operating the unitary transformation $U_1^\mu(\pi)$ on the rung-$\ket{0,0}$ phase. 
A highly anisotropic XYZ model on a ladder has been studied, which exhibits competition 
among the $t_0$, $t_z$, rung-$\ket{0,0}$, and  rung-$\ket{1,z}$ phases and a variety of magnetic phases (see also Ref.\ \cite{PhysRevB.87.081106} and Appendix \ref{App:XYZ}). 
Fuji has given a field-theoretical description of a variety of SPT and symmetry-broken phases of a spin-$\frac12$ XXZ ladder \cite{Fuji15,PhysRevB.93.104425}. 
Following him, we henceforth use the terms, the Haldane* and RS* phases, to refer to the $t_z$ and rung-$\ket{1,z}$ phases, respectively, 
where a star indicates the operation of $U_1^z(\pi)$ or equivalently the twist of singlet pairs between the legs. 
As the study of these phases has so far been limited in literature, 
it is worthwhile to further investigate when these phases emerge and how they compete with other phases in concrete spin models.   

%A ferromagnetic Heisenberg coupling $\Jperp<0$ on each rung of the ladder leads to effective spin-$1$ degrees of freedom, 
%which can collectively form the Haldane state \cite{PhysRevB.34.6372,doi:10.1143/JPSJ.60.1347,PhysRevLett.69.2419,PhysRevB.50.9911,PhysRevB.53.8521}. 
%Meanwhile, an antiferromagnetic coupling $\Jperp>0$ results in a product state of singlet pairs on the rungs, 
%which is known as the rung singlet (RS) state 

% [ This paper ]--------------------------------------
In this paper, we study a simple extension of the spin-$\frac12$ Heisenberg ladder 
that has XXZ anisotropy $\Delta$ and a four-spin interaction $J_4$ (Fig.\ \ref{fig:system}). 
The Hamiltonian of our model is given by
\begin{align}\label{eq:modelmain}
H  =& J \sum_{\alpha = 1,2} \sum_{j}  (\bm{S}_{\alpha,j}\cdot\bm{S}_{\alpha,j+1})_\Delta
	 + \Jperp \sum_j (\bm{S}_{1,j}\cdot\bm{S}_{2,j})_\Delta \notag\\
   &+ J_4\sum_{j}(\bm{S}_{1,j}\cdot\bm{S}_{1,j+1})(\bm{S}_{2,j}\cdot\bm{S}_{2,j+1}),
\end{align}
where
\begin{equation}\label{eq:model}
 (\bm{S}_{\alpha,i}\cdot\bm{S}_{\beta,j})_\Delta := S^x_{\alpha,i} S^x_{\beta,j} + S^y_{\alpha,i} S^y_{\beta,j} + \Delta S^z_{\alpha,i} S^z_{\beta,j} .
\end{equation}
The four-spin interaction with $J_4>0$ ($J_4<0$) introduces effective repulsion (attraction) between singlet pairs on the upper and lower edges of each plaquette. 
Throughout this paper, we take the leg interaction $J=1$ as the unit of energy. 
In contrast to the highly anisotropic models studied in Refs.\ \cite{PhysRevB.86.195122,PhysRevB.87.081106,Fuji15}, 
we investigate a regime around the isotropic case $\Delta=1$. 
By means of effective field theory for $|\Jperp|,|J_4|\ll 1$, 
we obtain rich ground-state phase diagrams that consist of eight distinct gapped phases, 
as schematically shown in Figs.\ \ref{fig:PhaseDiagrambos1} and \ref{fig:PhaseDiagrambos2}. 
We also perform numerical calculations based on the variational uniform matrix product state (VUMPS) algorithm 
\cite{PhysRevB.97.045145,10.21468/SciPostPhysLectNotes.7} for $\Jperp=\pm 1$ 
to demonstrate that essentially the same phase structures continue for $|\Jperp|,|J_4|=O(1)$, 
as shown in Figs.\ \ref{fig:PhaseDiagram1} and \ref{fig:PhaseDiagram2} later. 
A notable feature as compared to Refs.\ \cite{PhysRevB.86.195122,PhysRevB.87.081106} is 
that the four featureless phases, i.e., the RS, RS*, Haldane, and Haldane* phases 
compete not only with magnetic phases but also with staggered dimer (SD) and columnar dimer (CD) phases, 
which have valence bond solid (VBS) long-range orders breaking the translational symmetry. 
It is also remarkable that the RS* and Haldane* phases, which have highly anisotropic wave functions involving twisted singlet pairs, 
appear even in the vicinity of the isotropic case $\Delta=1$. 
Below we briefly review previous studies on the model \eqref{eq:modelmain} and highlight our contributions. 

% [ Isotropic case]--------------------------------------
The model \eqref{eq:modelmain} has been studied intensively in the isotropic case $\Delta=1$
\cite{PhysRevLett.78.3939, PhysRevLett.81.5406,PhysRevB.61.6747, 
PhysRevB.80.014426,PhysRevB.88.104403, PhysRevB.82.214420,PhysRevLett.122.027201}, 
which corresponds to the horizontal axes of Figs.\ \ref{fig:PhaseDiagrambos1} and \ref{fig:PhaseDiagrambos2}. 
The SD and CD phases can be interpreted as a consequence of effective repulsion or attraction between singlet pairs due to $J_4$. 
These phases have two-fold degenerate ground states below an excitation gap and are characterized by the order parameters
\begin{subequations}\label{eq:O_SD_CD}
\begin{align}
\expval{\mathcal{O}_{\text{SD}}(j)} &= \frac14 
\langle \bm{S}_{1,j-1}\cdot\bm{S}_{1,j} - \bm{S}_{2,j-1}\cdot\bm{S}_{2,j} \notag\\
&~~~~~~-\bm{S}_{1,j}\cdot\bm{S}_{1,j+1} + \bm{S}_{2,j}\cdot\bm{S}_{2,j+1} \rangle, \label{eq:O_SD}\\
\expval{\mathcal{O}_{\text{CD}}(j)} &= \frac14 
\langle \bm{S}_{1,j-1}\cdot\bm{S}_{1,j} + \bm{S}_{2,j-1}\cdot\bm{S}_{2,j} \notag\\
&~~~~~~-\bm{S}_{1,j}\cdot\bm{S}_{1,j+1} - \bm{S}_{2,j}\cdot\bm{S}_{2,j+1} \rangle. \label{eq:O_CD}
\end{align}
\end{subequations}
The translational symmetry as well as a $Z_2$ symmetry 
with respect to the rung-centered reflection (${\bm S}_{\alpha,j} \mapsto {\bm S}_{\alpha,-j}$) 
are spontaneously broken in these phases. 
Field-theoretical analyses for weak inter-chain couplings ($|\Jperp|,|J_4| \ll 1$) 
\cite{PhysRevLett.78.3939,PhysRevB.66.134423,PhysRevB.82.214420,PhysRevLett.122.027201} suggest 
that the RS-SD and Haldane-CD transitions are continuous and described by the SU$(2)_2$ Wess-Zumino-Witten (WZW) theory with the central charge $c=3/2$, 
although a possibility of a first-order transition is not excluded. 
In this approach, the transition points are estimated to be $J_4\approx 2.05\Jperp$ for both signs of $\Jperp$ 
for $|J_\perp|,|J_4| \ll 1$ \cite{PhysRevB.82.214420}. 
More detailed phase diagrams on the $\Jperp$-$J_4$ plane have been obtained numerically \cite{PhysRevB.80.014426, PhysRevB.88.104403, PhysRevLett.122.027201}. 
In particular, the exact diagonalization result of Ref.\ \cite{PhysRevB.80.014426} for the RS-SD transition 
is consistent with the $c=3/2$ criticality for $0.5\lesssim \Jperp \lesssim 1.5$. 

% [ Anisotropic case]--------------------------------------
We have started our analysis of the anisotropic case $\Delta\ne 1$ in Ref.\ \cite{ogino2020continuous}, 
and we extend it substantially in the present work. 
The obtained phase diagrams in Figs.\ \ref{fig:PhaseDiagrambos1} and \ref{fig:PhaseDiagrambos2} look like mirror images of each other
\footnote{
A useful way to relate the phases in Figs.\ \ref{fig:PhaseDiagrambos1} and \ref{fig:PhaseDiagrambos2} is to shift the lower leg of the ladder by one lattice spacing. 
In the bosonization formalism, this can be expressed as the shifts of $2\sqrt{\pi}\phi_2$ and $\sqrt{\pi}\theta_2$ by $-\pi$ and $\pi$, respectively, 
as seen in Eq.\ \eqref{eq:Spin_dim_bos}. 
This corresponds to the shifts of $2\sqrt{2\pi}\phi_\pm$ and $\sqrt{2\pi}\theta_-$ by $\mp \pi$ and $-\pi$, respectively. 
The reason for this qualitative relation between Figs.\ \ref{fig:PhaseDiagrambos1} and \ref{fig:PhaseDiagrambos2} is as follows. 
In the bosonized expressions of spin and dimer operators in Eq.\ \eqref{eq:Spin_dim_bos}, the staggered components play a leading role. 
Therefore, an antiferromagnetic interaction on a rung, $\Jperp (\bm{S}_{1,j}\cdot\bm{S}_{2,j})_\Delta$ with $\Jperp>0$, 
is qualitatively similar to a ferromagnetic interaction in a diagonal direction, $-\Jperp (\bm{S}_{1,j}\cdot\bm{S}_{2,j-1})_\Delta$, 
which is then transformed to a ferromagnetic interaction on a rung after the shift of the lower leg. 
A similar transformation can also be performed for $J_4$. 
}. 
The Haldane* and RS* phases, which are both featureless, appear for easy-plane anisotropy $\Delta<1$. 
The N\'eel and stripe N\'eel (SN) phases, which have magnetic long-range orders along the $z$-axis, appear for easy-axis anisotropy $\Delta>1$. 
The N\'eel and SN phases have two-fold degenerate ground states below an excitation gap, and are characterized by the order parameters
\begin{subequations}\label{eq:O_Neel_SN}
\begin{align}
\expval{\mathcal{O}_{\text{N\'eel}}(j)} &= \frac14 \expval{S^z_{1,j} - S^z_{2,j} - S^z_{1,j+1} + S^z_{2,j+1}}, \label{eq:O_Neel}\\
\expval{\mathcal{O}_{\text{SN}}(j)} &= \frac14 \expval{S^z_{1,j} + S^z_{2,j} - S^z_{1,j+1} - S^z_{2,j+1}}. \label{eq:O_SN}
\end{align}
\end{subequations}
A $Z_2$ symmetry with respect to the global $\pi$ rotation of spins about the $x$-axis
($S_{\alpha,j}^{y} \mapsto -S_{\alpha,j}^{y}$ and $S_{\alpha,j}^{z} \mapsto -S_{\alpha,j}^{z}$)  
is spontaneously broken in these phases. 

Our previous work \cite{ogino2020continuous} has focused on the nature of the phase transitions 
in the case of $\Jperp,J_4>0$ and easy-axis anisotropy $\Delta>1$, 
which corresponds to the upper half of Fig.\ \ref{fig:PhaseDiagrambos1}. 
The N\'eel-SD transition is especially intriguing as it is between two ordered phases breaking different symmetries 
and may be viewed as a one-dimensional (1D) variant of deconfined quantum critical point \cite{Senthil1490,PhysRevB.99.075103,PhysRevB.99.165143,PhysRevB.99.205153}. 
We have presented evidence that this transition belongs to the Gaussian universality class with $c=1$. 
We have also demonstrated that the RS-N\'eel transition belongs to the Ising universality class with $c=1/2$. 

Our extended analysis in the present work demonstrates 
that the Gaussian and Ising transition lines found in Ref.\ \cite{ogino2020continuous} cross in the isotropic case $\Delta=1$ 
and further continue to the easy-plane regime $\Delta<1$, as shown in Fig.\ \ref{fig:PhaseDiagrambos1}. 
We also analyze the case of $\Jperp,J_4<0$, and find similar crossing of the Gaussian and Ising transition lines, as shown in Fig.\ \ref{fig:PhaseDiagrambos2}. 
In both the cases, the Gaussian transitions occur between featureless phases in the easy-plane regime $\Delta<1$, 
giving examples of topological phase transitions.  

It is interesting to ask in what way the four featureless phases are distinguished. 
The following two types of string order parameters have been used to characterize the Haldane and RS phases 
\cite{doi:10.1143/JPSJ.62.2845,doi:10.1143/JPSJ.64.1967,PhysRevB.62.14965,PhysRevB.77.205121}: 
\begin{subequations}\label{eq:stringOP}
\begin{align}
 O_\mathrm{odd}^z (r) := - \bigg\langle &\qty(S_{1,0}^z+S_{2,0}^z) \exp\qty[ i\pi \sum_{j=1}^{r-1} \qty(S_{1,j}^z+S_{2,j}^z) ] \notag\\
 &\times\qty(S_{1,r}^z+S_{2,r}^z) \bigg\rangle ,\\
 O_\mathrm{even}^z (r) := - \bigg\langle &\qty(S_{1,0}^z+S_{2,1}^z) \exp\qty[ i\pi \sum_{j=1}^{r-1} \qty(S_{1,j}^z+S_{2,j+1}^z)  ] \notag\\
 &\times\qty(S_{1,r}^z+S_{2,r+1}^z) \bigg\rangle . 
\end{align}
\end{subequations}
Physically, these string order parameters detect the parity of the number $Q$ of valence bonds cut by a vertical line between two adjacent rungs \cite{PhysRevB.77.205121}. 
We obtain $ \lim_{r\rightarrow\infty} O_{\text{odd/even}}^z(r) \neq 0$ if $Q$ is odd/even. 
Therefore, the Haldane and RS phases are characterized by non-vanishing of $O_\mathrm{odd}^z (r)$ and $O_\mathrm{even}^z (r)$ for $r\to\infty$, respectively. 
As these order parameters remain unchanged under the operation of $U_1^z(\pi)$, 
the Haldane* and RS* phases are characterized similarly. 
However, these string order parameters do not distinguish between the Haldane and Haldane* phases or between the RS and RS* phases. 
We therefore calculate topological indices \cite{PhysRevB.81.064439, PhysRevB.86.125441, PhysRevB.83.035107, PhysRevB.84.235128} associated with $D_2\times \sigma$ and the translational symmetry, 
and demonstrate that these indices can distinguish all the four featureless phases. 

The rest of this paper is organized as follows. 
In Sec.~\ref{sec:EFT}, we present a field-theoretical analysis for weak inter-chain couplings, 
and derive the phase diagrams in Figs.\  \ref{fig:PhaseDiagrambos1} and \ref{fig:PhaseDiagrambos2}. 
In Secs.~\ref{sec:ResultJp} and \ref{sec:ResultJpinv}, 
we present numerical results for $\Jperp=1$ and $\Jperp=-1$, respectively. 
In Sec.~\ref{sec:SPT}, 
we analyze topological indices that distinguish the four featureless phases. 
In Sec.~\ref{sec:conclusion}, %we draw our conclusion. 
we present a summary and an outlook for future studies. 
In Appendix \ref{App:XYZ}, we apply the field-theoretical formulation in Sec.~\ref{sec:EFT} 
to an XYZ ladder studied by Liu {\it et al.} \cite{PhysRevB.86.195122}, 
and provide a qualitative description of their numerical results. 
In Appendices \ref{App:ES} and \ref{App:isotropic}, we provide some supplemental numerical results. 

%%%%%%%%%%%%%%%%%%%%%%%%%%%%%%%%%%%%%%%%%%%%%%%%
\section{Effective field theory}\label{sec:EFT}
%%%%%%%%%%%%%%%%%%%%%%%%%%%%%%%%%%%%%%%%%%%%%%%%

% [ Section introduction ]--------------------------------------
For weak inter-chain couplings with $|\Jperp|,|J_4| \ll 1$, the ground-state phase diagram of the model \eqref{eq:modelmain} can be studied 
by means of effective field theory based on bosonization \cite{giamarchi2003quantum,gogolin2004bosonization}. 
In our recent work \cite{ogino2020continuous}, we have applied this formalism 
to study the case of $\Jperp>0$ and $\Delta\ge 1$, i.e., the upper half of Fig. \ref{fig:PhaseDiagrambos1}. 
Here we present a more detailed analysis that reveals the rich phase diagrams in Figs.\ \ref{fig:PhaseDiagrambos1} and \ref{fig:PhaseDiagrambos2}. 
Our formulation is an extension of those in Refs.\ \cite{PhysRevLett.69.2419,PhysRevB.50.9911,PhysRevB.53.8521,PhysRevLett.78.3939,PhysRevB.82.214420,PhysRevLett.122.027201,PhysRevB.66.134423}, 
and we take similar notations as those in Refs.\ \cite{PhysRevB.82.214420,ogino2020continuous}.

%************************************************
\subsection{Bosonization }
%************************************************

% [ Subsection introduction ]--------------------------------------
Here we briefly describe the bosonization formulation to obtain the low-energy effective Hamiltonian 
of the model \eqref{eq:modelmain} for $|\Jperp|,|J_4| \ll 1$. 
While this formulation has also been described in our previous paper \cite{ogino2020continuous}, 
we summarize it for the purposes of selfcontainedness and fixing our notations. 

% [ Decoupled chains ]--------------------------------------
We start from the two decoupled XXZ chains obtained for $\Jperp=J_4=0$. 
Each chain labeled by $\alpha=1,2$ is described effectively by the quantum sine-Gordon Hamiltonian
\begin{equation}\label{eq:sineGordon}
\begin{split}
 H_\alpha^\mathrm{eff} = &\int \mathrm{d}x \frac{v}{2} \left[K^{-1} \left(\partial_x\phi_\alpha\right)^2+ K \left(\partial_x\theta_\alpha\right)^2 \right] \\
 &- \frac{v\lambda}{2\pi} \cos(4\sqrt{\pi}\phi_\alpha), 
\end{split}
\end{equation}
where $\phi_\alpha (x)$ and $\theta_\alpha(x)$ are a dual pair of bosonic fields. 
The Gaussian part of Eq.\ \eqref{eq:sineGordon} is known as the Tomonaga-Luttinger liquid (TLL) theory 
and characterized by the spin velocity $v$ and the TLL parameter $K$. 
The TLL parameter $K$ monotonically decreases as a function of $\Delta$, reaching $K=1/2$ at $\Delta=1$. 
When $\Delta$ exceeds unity, the $\lambda$ term with the scaling dimension $4K$ becomes relevant (i.e., $4K<2$) in the renormalization group (RG) sense, 
and induces the N\'eel order in the $z$ direction in each isolated chain. 
However, once the inter-chain couplings $\Jperp$ and $J_4$ are introduced, 
these couplings have much more significant impact on the low-energy physics than the $\lambda$ term 
as we discuss later. 

% [ Spin and dimer operators ]--------------------------------------
The spin and dimer operators on each chain are related to the bosonic fields as 
\begin{subequations}\label{eq:Spin_dim_bos}
\begin{align}
&S_{\alpha, j}^z = \frac{a}{\sqrt{\pi}} \partial_x \phi_\alpha + (-1)^j a_1 \cos(2\sqrt{\pi} \phi_\alpha) + \cdots, \label{eq:Sz_bos}\\
&S_{\alpha,j}^+ = e^{i\sqrt{\pi}\theta_\alpha} \qty[b_0 (-1)^j + b_1 \cos(2\sqrt{\pi}\phi_\alpha) + \cdots],\label{eq:Sp_bos}\\
&  S_{\alpha,j}^z S_{\alpha,j+1}^z  = (-1)^j d_{z} \sin(2\sqrt{\pi}\phi_\alpha) + \cdots,\\
& S_{\alpha,j}^x S_{\alpha, j+1}^x + S_{\alpha,j}^y S_{\alpha,j+1}^y = (-1)^j d_{xy} \sin(2\sqrt{\pi}\phi_\alpha) + \cdots, 
\end{align}
\end{subequations}
where $S_{\alpha,j}^\pm:=S_{\alpha,j}^x\pm iS_{\alpha,j}^y$, $a$ is the lattice constant, and $a_1$, $b_0$, $b_1$, $d_z$, and $d_{xy}$ are non-universal coefficients 
\cite{LUKYANOV1997571,PhysRevB.58.R583,PhysRevB.82.214420,PhysRevB.96.134429}. 
At $\Delta=1$, these coefficients satisfy $a_1 = b_0 =: \bar{a}$ and $d_z=d_{xy}/2 =: \bar{d}$ because of the SU$(2)$ symmetry. 

% [ Effective H for coupled chains ]--------------------------------------
Treating $J_\perp$ and $J_4$ perturbatively and expressing them using Eq.\ \eqref{eq:Spin_dim_bos}, 
we obtain the low-energy effective Hamiltonian of the model (\ref{eq:modelmain}) as 
\begin{equation}\label{eq:Heff}
\begin{split}
H^\mathrm{eff}=
& \int \dd x \sum_{q=\pm} \frac{v_q}{2} \qty[K_q^{-1}\qty(\partial_x\phi_q) ^2+ K_q \qty(\partial_x\theta_q)^2] \\
&+ g_+ \cos(2\sqrt{2\pi}\phi_+) + g_- \cos(2\sqrt{2\pi}\phi_-) \\
&+ \tilde{g}_- \cos(\sqrt{2\pi}\theta_-) + \cdots, 
\end{split}
\end{equation}
where $\phi_{\pm} := (\phi_1 \pm \phi_2)/\sqrt{2}$ and $\theta_{\pm} := (\theta_1 \pm \theta_2)/\sqrt{2}$ are symmetric and antisymmetric combinations of the bosonic fields and 
\begin{equation}\label{eq:coeff_cos}
 g_\pm = \frac1a \qty(J_{\perp} \Delta \frac{a_1^2}{2} \mp J_4 \frac{(3d)^2}{2}),~~
 \tilde{g}_-=\frac{1}{a} J_\perp b_0^2,
\end{equation}
with $3d:=d_{xy}+d_z$. 
The velocities $v_\pm$ and the TLL parameters $K_\pm$ in the symmetric and antisymmetric channels 
are in general modified from $v$ and $K$ in the decoupled XXZ chains by the effects of the inter-chain couplings. 
In Eq.\ \eqref{eq:Heff}, we focused on the most important terms around the isotropic case $\Delta=1$. 
Indeed, the $g_\pm$ and $\tilde{g}_-$ terms have the scaling dimensions $2K_\pm$ and $1/(2K_-)$, respectively, 
which are all equal to unity in the limit of the decoupled Heisenberg chains. 
These terms are much more relevant than the $\lambda$ term with the scaling dimension $2K_++2K_-$ in Eq.\ \eqref{eq:sineGordon}. 
Therefore, the $\lambda$ term can be ignored unless the anisotropy $\Delta-1$ is significantly larger than $J_\perp$ and $J_4$. 

%************************************************
\subsection{Expected phase diagrams}\label{sec:bos_phases}
%************************************************

% [ Separation into the symmetric and antisymmetric channels ]--------------------------------------
The symmetric and antisymmetric channels are separated in the effective Hamiltonian \eqref{eq:Heff}. 
The symmetric channel is described by the sine-Gordon model, 
in which the strongly relevant $g_+$ term locks $\phi_+$ at distinct positions depending on the sign of $g_+$. 
A Gaussian transition with the central charge $c=1$ is expected at $g_+=0$. 
The antisymmetric channel is described by the dual-field double sine-Gordon model, in which the strongly relevant $g_- $ and $\tilde{g}_-$ terms compete. 
When $K_-=1/2$, in particular, both the terms have the same scaling dimensions of unity, 
and the long-distance physics can be determined by examining which of $|g_-|$ and $|\tilde{g}_-|$ is larger 
(in this case, the model is known as the self-dual sine-Gordon model \cite{LECHEMINANT2002502}). 
Namely, $|g_-|>|\tilde{g}_-|$ ($|g_-|<|\tilde{g}_-|$) leads to the locking of $\phi_-$ ($\theta_-$). 
Based on a mapping to Majorana fields, an Ising transition with the central charge $c=1/2$ has been shown to occur at $|g_-|=|\tilde{g}_-|$ \cite{PhysRevB.53.8521,LECHEMINANT2002502}.
When $K_-$ deviates slightly from $1/2$, a similar picture is still expected to hold as the change in $K_-$ is a marginal perturbation. 

% [ Obtained phase diagrams ]--------------------------------------
Based on this argument and the coupling constants in Eq.\ \eqref{eq:coeff_cos}, 
we obtain the schematic phase diagrams around the isotropic case $\Delta=1$ as shown in Figs.\ \ref{fig:PhaseDiagrambos1} and \ref{fig:PhaseDiagrambos2}. 
Each phase is characterized by the locking positions of the bosonic fields. 
Remarkably, all the eight possible types of field locking in the effective Hamiltonian \eqref{eq:Heff} occur in the two diagrams.
In both the diagrams, a Gaussian transition in the symmetric channel occurs at $J_4 = \Delta \qty(a_1/3d)^2J_\perp$ (red line) 
while an Ising transition in the antisymmetric channel occurs at $J_4 \approx \qty[ (2b_0^2-\Delta a_1^2)/(3d)^2 ]J_\perp$ (blue line). 
The two transition lines cross at $J_4=  \qty(\bar{a}/3\bar{d})^2J_\perp$ in the isotropic case $\Delta=1$, 
where the central charge is expected to be $c=3/2$ 
\cite{PhysRevB.80.014426,PhysRevB.82.214420,PhysRevB.88.104403, PhysRevLett.78.3939,PhysRevLett.122.027201,PhysRevB.66.245106,PhysRevB.66.134423}; 
this point was estimated to be $J_{4,c}=2.05J_\perp$ \cite{PhysRevB.82.214420}. 
Below we discuss the phases appearing in these diagrams. 

% [ Ordered phases ]--------------------------------------
There are four ordered phases, the N\'eel, SN, SD, and CD phases. 
Using Eq.\ \eqref{eq:Spin_dim_bos}, 
we can express the order operators in Eqs.\ \eqref{eq:O_SD_CD} and \eqref{eq:O_Neel_SN} in terms of the bosonic fields as 
\begin{subequations}\label{eq:O_N_SN_SD_CD_bos}
\begin{align}
 \mathcal{O}_{\text{N\'eel}}(j) 
 &= -(-1)^j a_1 \sin \qty(\sqrt{2\pi}\phi_+) \sin \qty(\sqrt{2\pi}\phi_-), \\
 \mathcal{O}_{\text{SN}}(j) 
 &= (-1)^j a_1 \cos \qty(\sqrt{2\pi}\phi_+) \cos \qty(\sqrt{2\pi}\phi_-) ,\\
 \mathcal{O}_{\text{SD}}(j) 
 &=-(-1)^j (3d) \cos \qty(\sqrt{2\pi}\phi_+) \sin \qty(\sqrt{2\pi}\phi_-), \\
 \mathcal{O}_{\text{CD}}(j) 
 &=-(-1)^j (3d) \sin \qty(\sqrt{2\pi}\phi_+) \cos \qty(\sqrt{2\pi}\phi_-).
\end{align}
\end{subequations}
We can easily see that these operators acquire finite expectation values in the corresponding phases. 
For example, in the CD phase with $2\sqrt{2\pi}(\phi_+,\phi_-)=(-\pi,\pi\mp\pi)$ in Fig.\ \ref{fig:PhaseDiagrambos2}, 
we have $\expval{ \mathcal{O}_{\text{CD}}(j) } =\pm (-1)^j c_{\text{CD}}$, 
where $c_{\text{CD}}$ is a constant independent of $j$. 

% [ Featureless phases ]--------------------------------------
There are also four featureless phases, the RS, RS*, Haldane, and Haldane* phases. 
In these phases, the expectation values of all the operators in Eq.\ \eqref{eq:O_N_SN_SD_CD_bos} vanish as $\phi_-$ fluctuates entirely owing to the locking of $\theta_-$. 
These phases are instead characterized by the two types of string correlations in Eq.\ \eqref{eq:stringOP}, 
which have the following bosonized expressions \cite{NAKAMURA20031000}: 
\begin{subequations}\label{eq:stringOP_bos}
\begin{align}
 O_\mathrm{odd}^z (r) &\sim \expval{\cos\qty(\sqrt{2\pi}\phi_+(0)) \cos\qty(\sqrt{2\pi}\phi_+(ra))}, \\ 
 O_\mathrm{even}^z (r) &\sim \expval{\sin\qty(\sqrt{2\pi}\phi_+\qty(\frac{a}{2})) \sin\qty(\sqrt{2\pi}\phi_+\qty(ra+\frac{a}{2}))}.
\end{align}
\end{subequations}
We note that only the field in the symmetric channel is involved in these correlations. 
The Haldane and Haldane* phases with the field locking around $2\sqrt{2\pi}\phi_+=0$ exhibit non-vanishing values of $O_\mathrm{odd}^z (r)$ for $r\to\infty$. 
Similarly, the RS and RS* phases with the field locking around $2\sqrt{2\pi}\phi_+=\pi$ 
\footnote{While the RS and RS* phases have the field locking around $2\sqrt{2\pi}\phi_+=\pi$ and $-\pi$, respectively, 
in Figs.\ \ref{fig:PhaseDiagrambos1} and \ref{fig:PhaseDiagrambos2}, 
these locking positions are equivalent. 
This is because the field $\phi_\alpha$ on each leg $\alpha(=1,2)$ is compactified as $2\sqrt{\pi}\phi_\alpha\equiv 2\sqrt{\pi}\phi_\alpha+2\pi$. 
}
exhibit non-vanishing $O_\mathrm{even}^z (r)$ for $r\to\infty$. 

% [ Haldane-Haldane* and RS-RS* relations ]--------------------------------------
The difference between the Haldane and Haldane* phases or between the RS and RS* phases 
resides in the locking positions of $\sqrt{2\pi}\theta_-$, 
which cannot be detected by the string correlations \eqref{eq:stringOP_bos}. 
In fact, this difference can be used to show that the two phases are related by the unitary transformation 
$U_1^z(\pi)=\exp\qty( i\pi\sum_j S_{1,j}^z )$ \cite{Fuji15}. 
Indeed, under this transformation, the field $\sqrt{\pi}\theta_1$ is shifted by $\pi$ as seen in Eq.\ \eqref{eq:Sp_bos}, 
and then the field $\sqrt{2\pi}\theta_- = \sqrt{\pi}(\theta_1-\theta_2)$ is also shifted by $\pi$.  
Identification of the Haldane* and RS* phases in the bosonization approach is based on this observation. 

% [ Possible perturbations ]--------------------------------------
We have so far neglected possible perturbations to the effective Hamiltonian \eqref{eq:Heff} 
which have larger scaling dimensions than the $g_\pm$ and $\tilde{g}$ terms. 
If such perturbations also become relevant, they can potentially change the nature of the phase transitions. 
In Ref.\ \cite{ogino2020continuous}, we have addressed this issue for the case of the N\'eel-SD transition. 
Since the antisymmetric channel remains gapped at this transition, we can focus on the symmetric channel. 
As a possible perturbation, we can consider, for example, 
a higher-frequency cosine potential $\cos\left( 4\sqrt{2\pi}\phi_+ \right)$ with the scaling dimension $8K_+$, 
which may lead to a first-order phase transition \cite{PhysRevB.81.094430}. 
We have numerically demonstrated that $K_+$ stays around $0.6$ along the N\'eel-SD transition line in the parameter range investigated, 
and thus the above higher-frequency cosine term is likely to remain sufficiently irrelevant. 
As we have $K_+=1/2$ in the limit of decoupled Heisenberg chains, 
we can expect that our picture based on the leading terms in the effective Hamiltonian \eqref{eq:Heff} 
would hold at least for weak inter-chain couplings. 

%************************************************
\subsection{Correlation functions around the Gaussian transitions}\label{sec:bos_critical}
%************************************************

% [ Subsection introduction ]--------------------------------------
We have argued that Gaussian transitions occur between the ordered phases and between the featureless phases. 
We discuss the behavior of correlation functions around the presumed Gaussian transitions. 
As the case of the N\'eel-SD transition has been discussed previously \cite{ogino2020continuous}, we focus on the other cases. 

% [ SN-CD transition ]--------------------------------------
The SN-CD transition can be studied in a similar manner as the N\'eel-SD transition. 
We consider the following correlation functions 
\begin{subequations}\label{eq:corr_SN_CD}
\begin{align}
 C_{\text{SN}}(r)
 &:= (-1)^r \expval{ \mathcal{O}_{\text{SN}}(r) \mathcal{O}_{\text{SN}}(0) }, \label{eq:corr_SN} \\
 C_{\text{CD}}(r)
 &:= (-1)^r \expval{ \mathcal{O}_{\text{CD}}(r) \mathcal{O}_{\text{CD}}(0) }. \label{eq:corr_CD} 
\end{align}
\end{subequations}
Each of these correlation functions would show a long-range order (i.e., convergence to a non-vanishing value for $r\to\infty$) in the concerned phase 
and an exponential decay in the other phase. 
At the Gaussian transition, both of these functions would show a critical power-law decay. 
This can be shown in the following way. 
As seen in the bosonized expressions in Eq.\ \eqref{eq:O_N_SN_SD_CD_bos}, the operators 
$\mathcal{O}_{\text{SN}}(j)$ and $\mathcal{O}_{\text{CD}}(j)$ involve both the fields $\phi_\pm$. 
As the symmetric channel is described by the Gaussian theory at the transition point, 
the symmetric component $\cos\qty(\sqrt{2\pi}\phi_+)$ or $\sin\qty(\sqrt{2\pi}\phi_+)$ 
shows a critical correlation with the decay exponent $K_+$. 
In contrast, as $\sqrt{2\pi}\phi_-$ remains locked around $(\pi \mp \pi)/2$, the antisymmetric component $\cos\qty(\sqrt{2\pi}\phi_-)$ shows a correlation 
that converges to a non-vanishing constant above the length scale proportional to the inverse of the excitation gap. 
Therefore, above this scale, the correlation functions in total exhibit the power-law behavior $C_{\text{SN/CD}}(r)\sim r^{-K_+}$. 
In numerical calculations, the transition can be identified by plotting the two correlation functions in logarithmic scales 
and finding the point at which they both become linear and parallel to each other. 
We conduct this analysis in Sec.\ \ref{sec:CD_SN}. 

% [ Topological transitions ]--------------------------------------
Next we consider the RS-Haldane* and RS*-Haldane transitions. 
We consider the two types of string correlation functions \eqref{eq:stringOP}. 
Based on the bosonized expressions \eqref{eq:stringOP_bos} and field locking positions in Figs.\ \ref{fig:PhaseDiagrambos1} and \ref{fig:PhaseDiagrambos2}, 
we find that $O_\mathrm{odd}^z (r)$ changes from an exponential decay to a long-range order across these transitions 
while $O_\mathrm{even}^z (r)$ changes in the opposite way. 
At the transition point, where the symmetric channel is described by the Gaussian theory, 
we have $O_\mathrm{odd/even}^z (r)\sim r^{-K_+}$. 
Again, in numerical calculations, the transition point can be identified by linear and parallel behavior of the two correlation functions plotted in logarithmic scales. 
We conduct this analysis in Secs.\ \ref{sec:RSHS} and \ref{eq:H_RSS}. 

\section{Numerical results for $J_\perp = 1$}\label{sec:ResultJp}
%%%%%%%%%%%%%%%%%%%%%%%%%%%%%%%%%%%%%%%%%%%%%%%%

%%%%%%%%%%%%%%%%%%%%%%%%%%%%%%%%%%%%%%%%%%%%%%%%%%%%%%%%%%%%%%%%%%%%%%%%%%%%%%%%%%%%%%%%%%%%
\begin{figure}
\includegraphics[width=90mm]{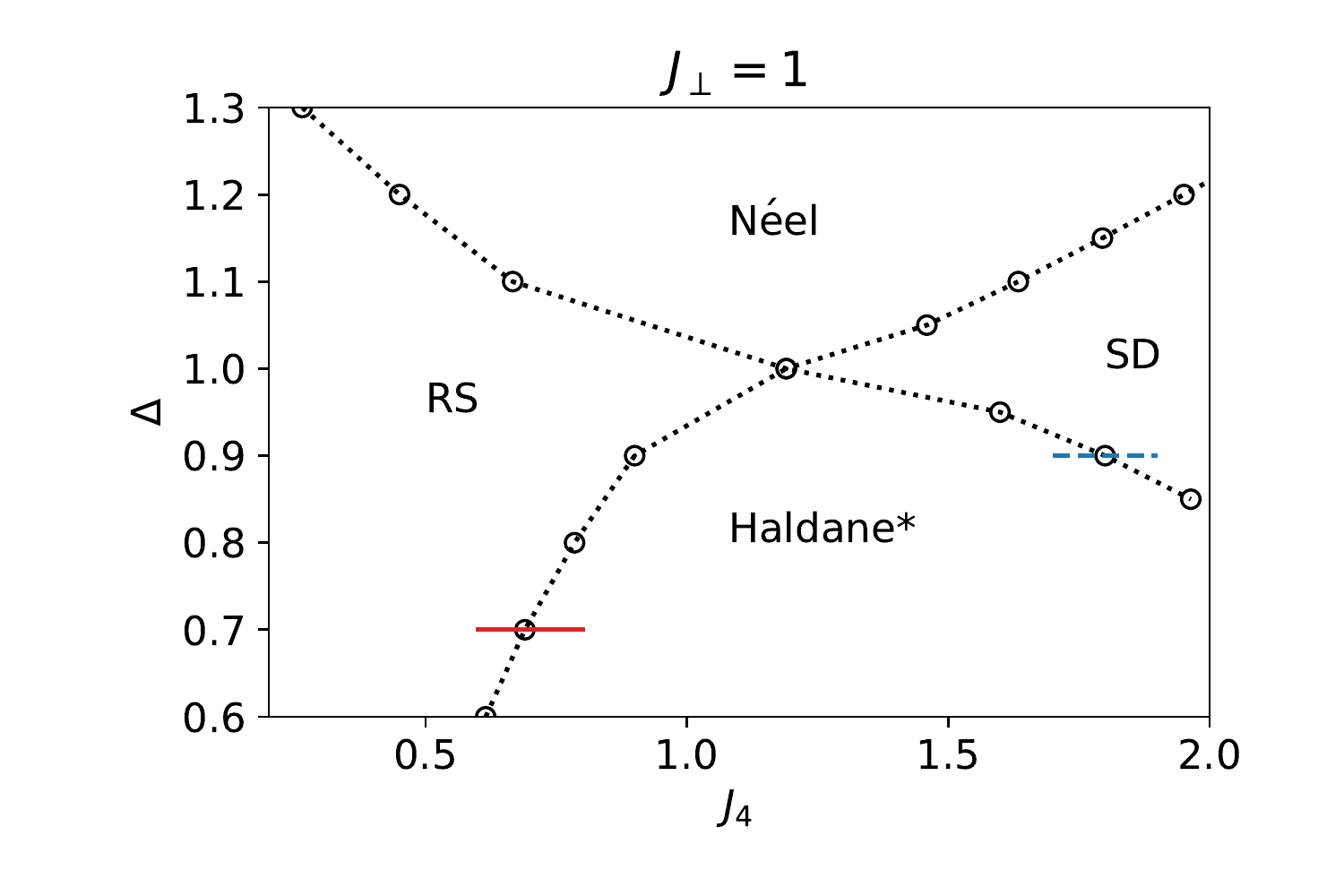}
\caption{
Phase diagram of the model (\ref{eq:modelmain}) on the $J_4$-$\Delta$ plane with $J=J_\perp=1$. 
Circles indicate transition points obtained numerically, 
and dotted lines connecting them are our assumption of the phase boundaries. 
The transition points in the easy-plane regime $\Delta<1$ are obtained in the present work 
while those in the easy-axis regime $\Delta>1$ are taken from our previous work \cite{ogino2020continuous}. 
%The transition points (circles) in the easy-plane regime $\Delta<1$ are obtained in the present work 
%while those for $\Delta\ge 1$ are taken from the previous works \cite{PhysRevB.80.014426,ogino2020continuous}. 
%The model with $\Delta \geq 1$ is calculated in Refs.~
%and we calculated the region where $\Delta < 1$.
%The circles ($\circ$) indicate the transition points obtained 
%through the analysis of the string correlation functions and the order parameters.
%The dotted lines are our assumption of the phase boundaries.
We will focus on the RS-Haldane* transition along the red solid line with $\Delta=0.7$ (Sec.~\ref{sec:RSHS}),
and the Haldane*-SD transition along the blue dashed line with $\Delta=0.9$ (Sec.~\ref{sec:HSSD}). 
The transition point $J_4\simeq 1.19$ in the isotropic case $\Delta=1$ is estimated in Appendix \ref{App:isotropic}, 
and is consistent with the previous works \cite{PhysRevB.80.014426,ogino2020continuous}. 
}
\label{fig:PhaseDiagram1}
\end{figure}
%%%%%%%%%%%%%%%%%%%%%%%%%%%%%%%%%%%%%%%%%%%%%%%%%%%%%%%%%%%%%%%%%%%%%%%%%%%%%%%%%%%%%%%%%%%%

% [ VUMPS ]--------------------------------------
We have performed numerical calculations directly for the infinite system 
by applying the VUMPS algorithm \cite{PhysRevB.97.045145,10.21468/SciPostPhysLectNotes.7}. 
In this algorithm, a variational state is prepared in the form of a uniform matrix product state (MPS), 
and the ground state is obtained by iteratively optimizing constituent tensors to lower the variational energy. 
In applying it to the present ladder system, we regard two sites on each rung as a single effective site with the local Hilbert space dimension of four. 
In this section and the next, we adopt the two-site unit cell implementation in Ref.\ \cite{PhysRevB.97.045145} 
%This is suitable for our purpose 
as all the phases discussed in Sec.\ \ref{sec:EFT} have the unit cell consisting of at most two effective sites (i.e., two rungs). 
To ensure the sufficient convergence, we only used the data points that have the gradient norm $\|B\| < 10^{-10}$,
where $B$ is the gradient of energy per site with respect to the elementary tensor in VUMPS.
The same method was employed in our previous work \cite{ogino2020continuous}. 

% [ Phase diagram ]--------------------------------------
In this section, we fix $ J = J_\perp = 1 $ and study the ground-state phase diagram in the $J_4$-$\Delta$ plane. 
The obtained phase diagram is shown in Fig.~\ref{fig:PhaseDiagram1}, 
which is qualitatively consistent with the field-theoretical prediction in Fig.\ \ref{fig:PhaseDiagrambos1}. 
As the case of $\Delta\ge 1$ has been analyzed previously \cite{PhysRevB.80.014426,ogino2020continuous}, 
we focus on the easy-plane regime $\Delta<1$ throughout this section. 
Below we explain how the RS-Haldane*-SD phase boundaries are obtained 
and how the critical properties at these transitions are characterized in our numerical analysis. 

%The N\'eel and SD phase break different $\mathbb{Z}_2$ symmetries and 
%the RS and Haldane* phase do not break any symmetry.

%************************************************
\subsection{RS-Haldane* transition} \label{sec:RSHS}
%************************************************

%%%%%%%%%%%%%%%%%%%%%%%%%%%%%%%%%%%%%%%%%%%%%%%%%%%%%%%%%%%%%%%%%%%%%%%%%%%%%%%%%%%%%%%%%%%%
\begin{figure}
\includegraphics[width=90mm]{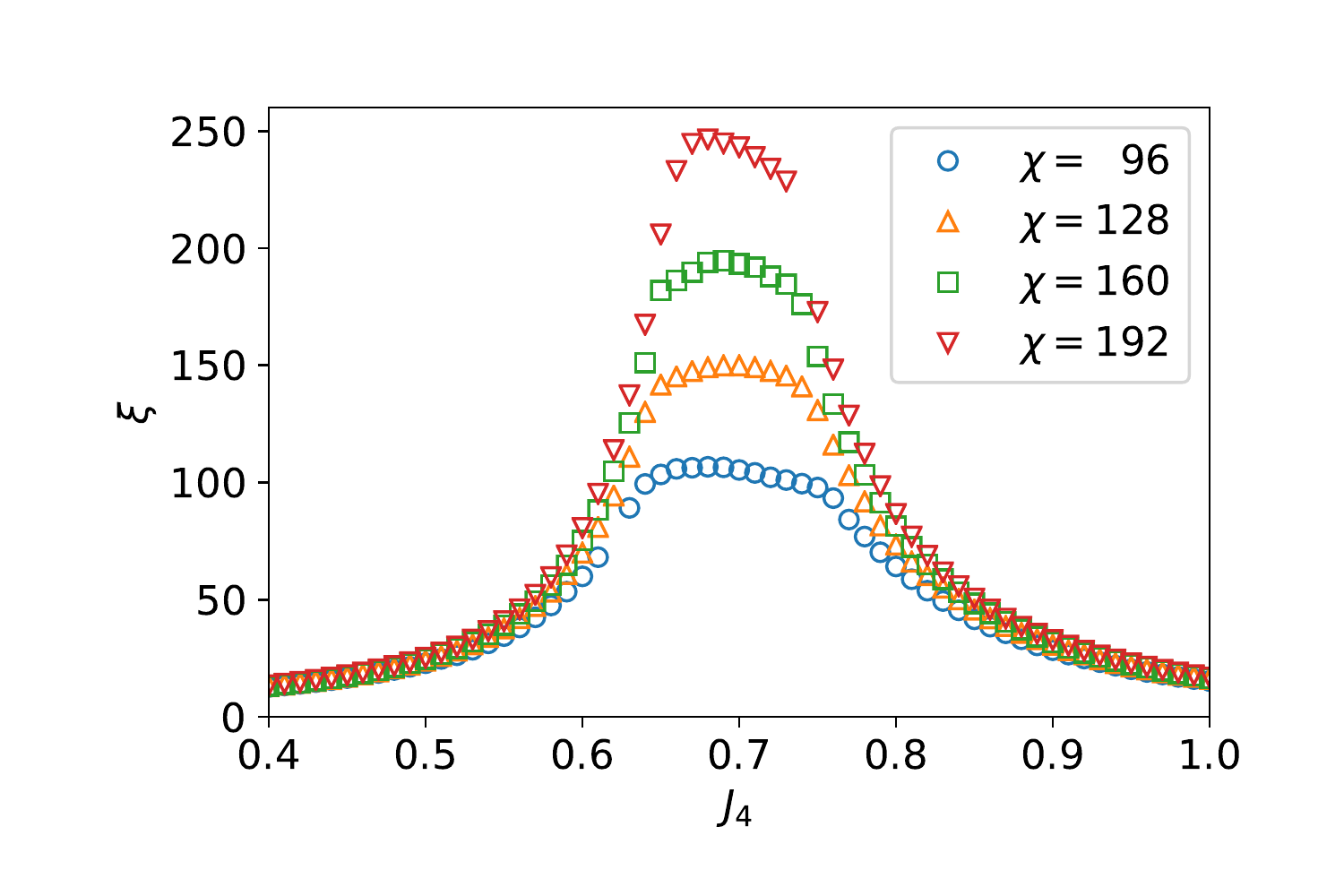}
\caption{
Correlation length $\xi$ as a function of $J_4$ around the RS-Haldane* transition for $\Delta=0.7$ (see the red solid line in Fig.\ \ref{fig:PhaseDiagram1}). 
The correlation length shows a peak that grows consistently with an increase in $\chi$. 
However, the non-sharpness of the peaks does not allow a precise determination of the transition point. 
The transition point is estimated to be $J_{4,c}^\text{RS-H$^*$} = 0.690(5)$ in the analysis of the string correlations in Fig.\ \ref{fig:StringFunction}. 
%
%Although the correlation length does not show a sharp peak, it grows consistently with an increase in $\chi$.
%In the case of $\chi=192$, the correlation length exceeds $200$ lattice spacings.
%These results are indicative of a continuous transition.
%The critical point is estimated at $J_{4,c} = 0.685 \sim 0.695$ 
%using the  string correlation functions.
}
\label{fig:CorrRSHS}
\end{figure}
%%%%%%%%%%%%%%%%%%%%%%%%%%%%%%%%%%%%%%%%%%%%%%%%%%%%%%%%%%%%%%%%%%%%%%%%%%%%%%%%%%%%%%%%%%%%

%%%%%%%%%%%%%%%%%%%%%%%%%%%%%%%%%%%%%%%%%%%%%%%%%%%%%%%%%%%%%%%%%%%%%%%%%%%%%%%%%%%%%%%%%%%%
\begin{figure}
\includegraphics[width=90mm]{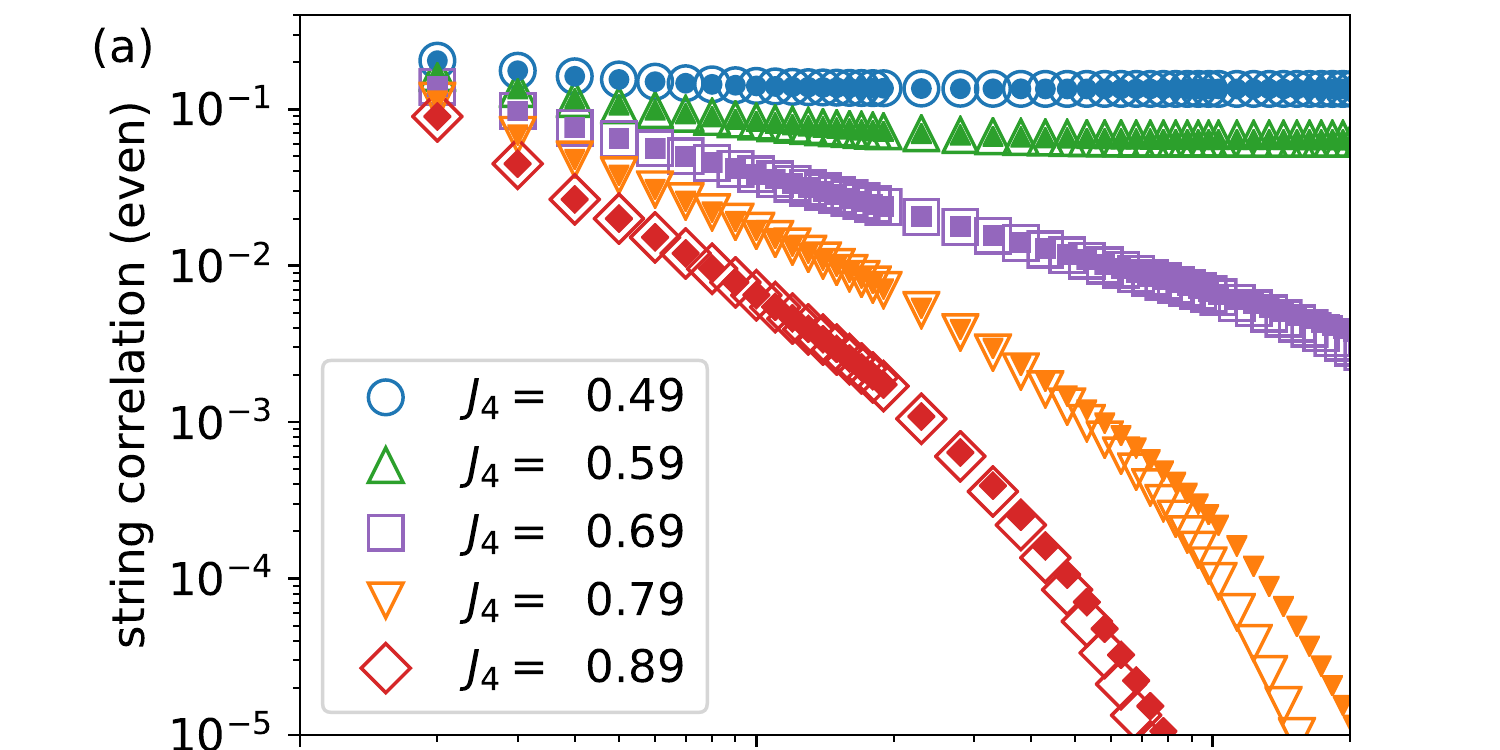}
\includegraphics[width=90mm]{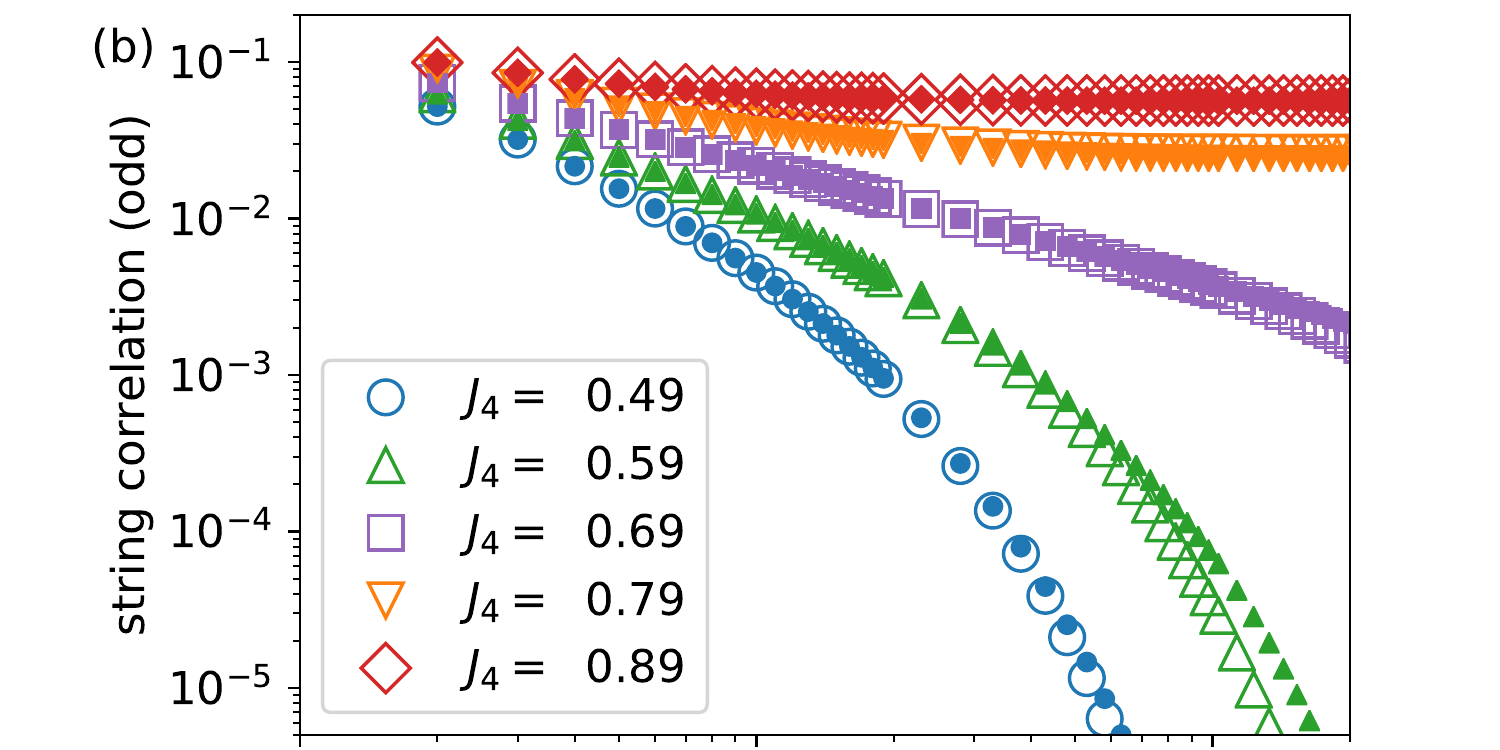}
\includegraphics[width=90mm]{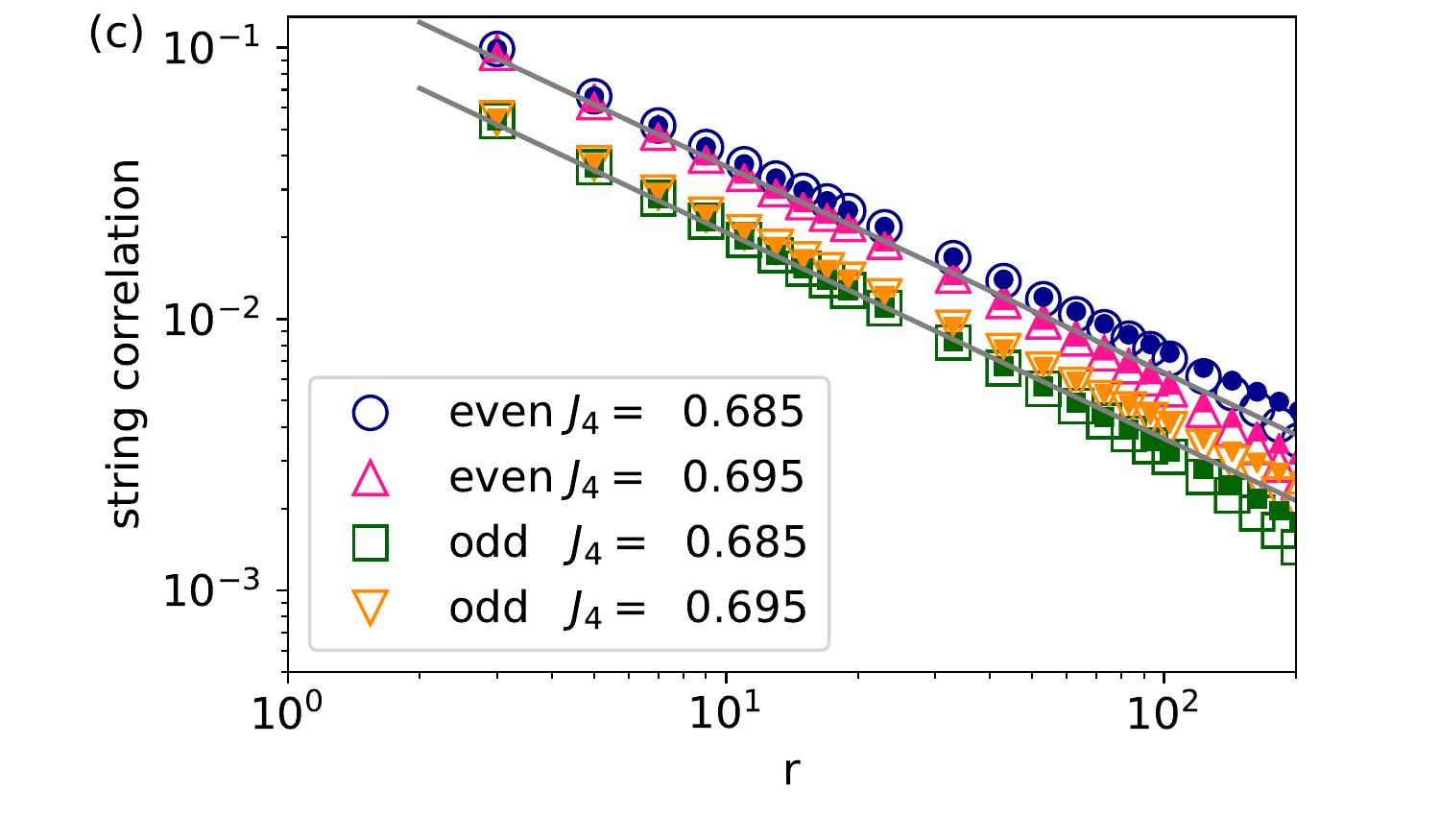}
\caption{
Two types of string correlation functions \eqref{eq:stringOP} around the RS-Haldane* transition for $\Delta=0.7$ in Fig.\ \ref{fig:PhaseDiagram1}. 
Logarithmic scales are used for both axes. 
Large open and small filled symbols show the data for the bond dimensions $\chi=96$ and $192$, respectively. 
We focus on the range of $r$ where the dependence on $\chi$ is not significant. 
(a) $O^z_{\text{even}}(r)$ is non-decaying at long distances in the RS phase with $J_4\lesssim 0.69$. 
(b) $O^z_{\text{odd}}(r)$ is non-decaying in the Haldane* phase with $J_4\gtrsim 0.69$. 
(c) The two correlation functions show a power-law decay with the same exponent $K_+$ at the transition 
if the transition belongs to the Gaussian universality class. 
As analyzed in this panel, this should occur between $J_4=0.685$ and $0.695$, 
and the exponent is estimated to be $K_+=0.76(3)$. 
Two parallel gray solid lines are guides to the eye and their slope is $-0.76$.
%
%String correlation functions around the RS-Haldane* transition for $\Delta=0.7$.
%Large open symbols represent $\chi=96$ and small filled symbols represent $\chi=192$.
%(a) The string correlation $\expval{\mathcal{O}_{\text{even}}(r)}$.
%(b) The string correlation $\expval{\mathcal{O}_{\text{odd}}(r)}$.
%The string correlation functions at $J_4 = 0.69$ exhibit power-law behavior.
%(c) The string correlation $\expval{\mathcal{O}_{\text{even}}(r)}$ and $\expval{\mathcal{O}_{\text{odd}}(r)}$.
%Two parallel gray solid lines are guide to the eye and the slope is $-0.76$.
%At the critical point, the parameter $K_+$ is equal to
%the gradient of both $\expval{\mathcal{O}_{\text{even}}(r)}$ and $\expval{\mathcal{O}_{\text{odd}}(r)}$
%Therefore, we obtain the critical point $J_{4,c} = 0.685 \sim 0.695$ and the TLL parameter $K_+ = 0.76(3)$.
}
\label{fig:StringFunction}
\end{figure}
%%%%%%%%%%%%%%%%%%%%%%%%%%%%%%%%%%%%%%%%%%%%%%%%%%%%%%%%%%%%%%%%%%%%%%%%%%%%%%%%%%%%%%%%%%%%

%%%%%%%%%%%%%%%%%%%%%%%%%%%%%%%%%%%%%%%%%%%%%%%%%%%%%%%%%%%%%%%%%%%%%%%%%%%%%%%%%%%%%%%%%%%%
\begin{figure}
\includegraphics[width=90mm]{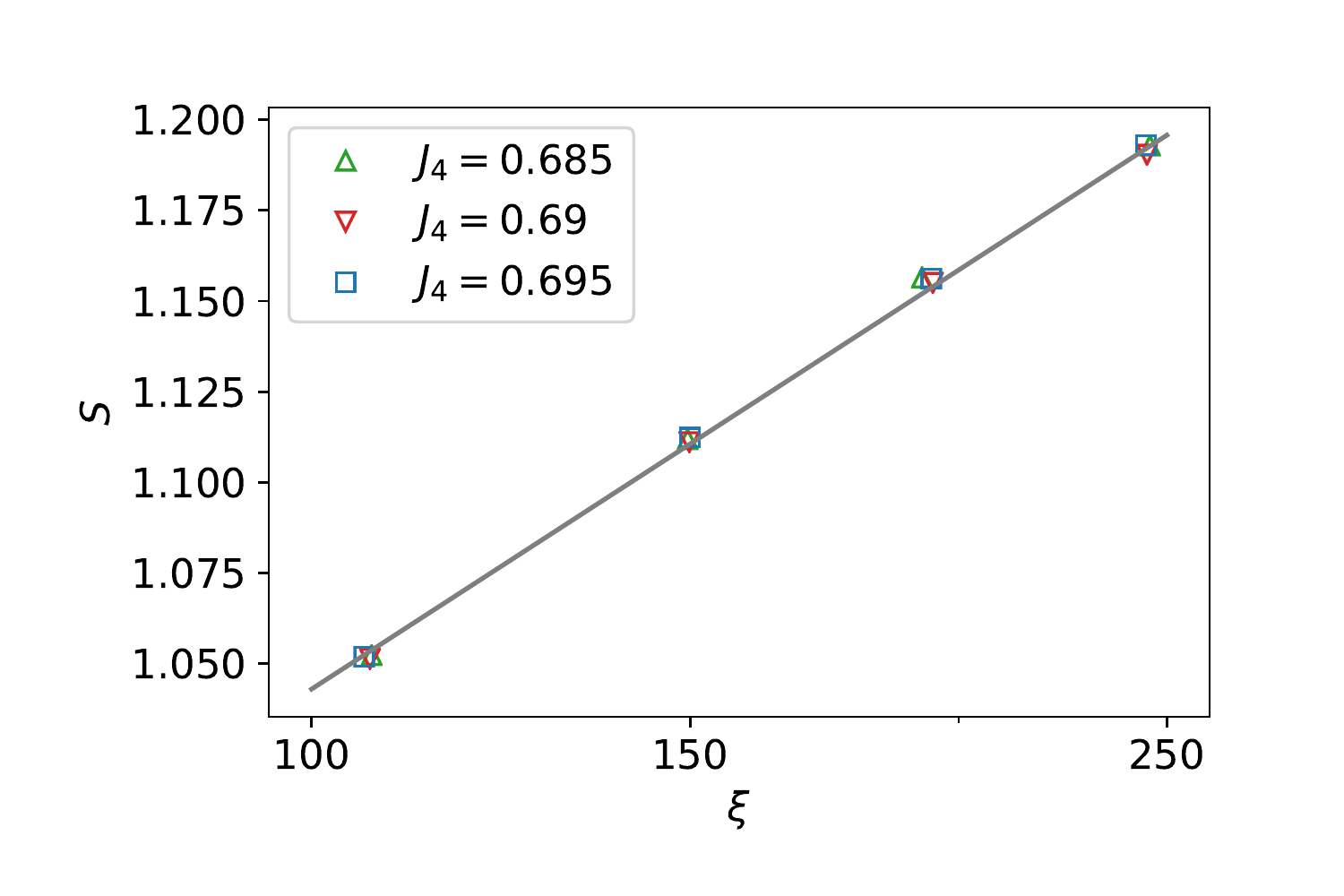} 
\caption{
Entanglement entropy $S(\chi)$ versus the correlation length $\xi(\chi)$ for the bond dimensions $\chi = 96,128,160,192$.
These are calculated at $J_4 = 0.685$, $0.690$, and $0.695$, around which the RS-Haldane* transition is expected to occur (see Fig.\ \ref{fig:StringFunction}).
A logarithmic scale is used for the horizontal axis.
The gray straight line is a guide to the eye and its slope is $1/6$, %which is consistent with 
which corresponds to 
the central charge $c=1$ [see Eq.\ \eqref{eq:cardy}].
}
\label{fig:CCJ40690}
\end{figure}
%%%%%%%%%%%%%%%%%%%%%%%%%%%%%%%%%%%%%%%%%%%%%%%%%%%%%%%%%%%%%%%%%%%%%%%%%%%%%%%%%%%%%%%%%%%%

% [ Correlation length ]--------------------------------------
Here we focus on the RS-Haldane* transition along the red solid line with $\Delta=0.7$ in Fig.\ \ref{fig:PhaseDiagram1}. 
In our VUMPS calculations, the ground state is obtained in the form of a period-2 MPS with the finite bond dimension $\chi$. 
We can extract the correlation length $\xi$ from it in the following way. 
Let $A(k)^{s_j} \in \mathbb{R}^{\chi \times \chi}$ be the matrix for the state $s_j$ at the $j$-th effective site. 
Using the matrices at the two neighboring sites $j=1$ and $2$, 
we construct the combined matrix $\mathbb{A}^{s} := A(1)^{s_1}A(2)^{s_2} \in \mathbb{R}^{\chi \times \chi}$ for the combined state $s=(s_1,s_2)$. 
The correlation length is then calculated as $\xi(\chi) = - 2 / \ln|\epsilon_2(\chi)|$, 
where $\epsilon_2(\chi)$ is the second largest 
absolute eigenvalue of the transfer matrix $T:= \sum_s \mathbb{A}^{s\dagger}\otimes\mathbb{A}^{s}$.
The obtained correlation length $\xi(\chi)$ is plotted for different bond dimensions $\chi$ in Fig.~\ref{fig:CorrRSHS}. 
It shows a peak that grows consistently with an increase in $\chi$, 
and the peak value exceeds $200$ lattice spacings for $\chi=192$. 
However, the peaks in Fig.~\ref{fig:CorrRSHS} are not as sharp as those found for the N\'eel-SD transition in Ref.~\cite{ogino2020continuous}. 
Therefore, while Fig.~\ref{fig:CorrRSHS} suggests the occurrence of a phase transition, 
it does not allow a precise determination of the RS-Haldane* transition point. 
We note that the non-sharpness of peaks has also been found for the RS-SD transition in the isotropic case $\Delta=1$ \cite{ogino2020continuous}, 
and is likely to be due to nontrivial effects of finite $\chi$ on transitions between valence bond phases. 
The transition point can instead be determined with reasonable accuracy through the analysis of the string correlations, which is done next. 

%We plot the correlation length $\xi(\chi)$ at $\Delta=0.7$ in Fig.~\ref{fig:CorrRSHS}.
%It shows that the correlation length consistently grows with an increase in $\chi$
%and the peak of the correlation length exceeds $200$ lattice spacings in the case of $\chi=192$.
%As shown in Ref.~\cite{ogino2020continuous}, the N\'eel-SD transition has a sharp peak
%but the Haldane*-SD transition does not have any sharp peak.
%It seems that this phenomenon is caused by the finite bond-dimension effect,
%which is similar to the RS-SD transition in Ref.~\cite{ogino2020continuous}.

% [ String correlations ]--------------------------------------
Figure \ref{fig:StringFunction} shows the numerical results on the two types of string correlation functions in Eq.\ \eqref{eq:stringOP}. 
We find a qualitative difference between the ranges of $J_4\lesssim 0.69$ and $J_4\gtrsim 0.69$. 
For $J_4\lesssim 0.69$, $O^z_{\text{even}}(r)$ is non-decaying at long distances 
while $O^z_{\text{odd}}(r)$ shows a rapid decay. 
For $J_4\gtrsim 0.69$, $O^z_{\text{odd}}(r)$ is non-decaying while $O^z_{\text{even}}(r)$ shows a rapid decay. 
These results are consistent with the occurrence of the RS-Haldane* transition. 
However, we note that (non-)decaying behavior of the two correlations only reflects the locking position of the field $\phi_+$ [see Eq.\ \eqref{eq:stringOP_bos}], 
and that more detailed characterization of the phases requires the use of topological indices, which is done in Sec.\ \ref{sec:SPT}. 

As discussed in Sec.\ \ref{sec:bos_critical}, the two string correlation functions show a power-law decay 
with the same exponent $K_+$ at the transition if it belongs to the Gaussian universality class. 
Therefore, the transition point can be identified by linear and parallel behavior of the two correlations plotted in logarithmic scales. 
As seen in Fig.\ \ref{fig:StringFunction}(c), this should occur between $ J_4 = 0.685$ and $ J_4 = 0.695$, 
giving the estimate $J_{4,c}^\text{RS-H$^*$}=0.690(5)$ of the transition point. 

%We estimate the critical point and the TLL parameter $K_+$ assuming that 
%both of the string correlation functions show power-law decay and have the same slope at the critical point.
%We plot the string correlation functions $O^z_{\text{odd/even}}(r)$ in Fig.~\ref{fig:StringFunction}.
%In the RS [Haldane*] phase, the string correlation function $O^z_{\text{odd}}(r)$ goes to zero [a non-zero finite value]
%and $O^z_{\text{even}}(r)$ goes to a non-zero finite value [zero].
%Both of the string correlation functions have the same slope $-K_+ = -0.76(3)$
%between $ J_4 = 0.685$ and $ J_4 = 0.695$.

% [ Central charge ]--------------------------------------
Critical points of a large class of 1D quantum systems are described by the conformal field theory (CFT). 
To investigate the underlying CFT, we calculate the entanglement entropy $S$ 
for a bipartition of the infinite 1D system into two half-infinite chains. 
According to the CFT, the entanglement entropy $S$ and the correlation length $\xi$ %have the relationship
obey the scaling 
\begin{eqnarray}
S = \frac{c}{6} \ln \xi + S_0,
\label{eq:cardy}
\end{eqnarray}
where $c$ is the central charge and $S_0$ is a constant \cite{Calabrese_2004,PhysRevLett.102.255701}. 
The numerical results in Fig.\ \ref{fig:CCJ40690} show a good agreement with this scaling with $c=1$. 
Our results in Figs.\ \ref{fig:StringFunction}(c) and \ref{fig:CCJ40690} thus support the scenario that 
the RS-Haldane* transition belongs to the Gaussian universality class with $c=1$. 

%Figure \ref{fig:CCJ40690} shows $S$ and $\xi$ around the RS-Haldane* transition.

%************************************************
\subsection{Haldane*-SD transition} \label{sec:HSSD}
%************************************************

%%%%%%%%%%%%%%%%%%%%%%%%%%%%%%%%%%%%%%%%%%%%%%%%%%%%%%%%%%%%%%%%%%%%%%%%%%%%%%%%%%%%%%%%%%%%
\begin{figure}
\includegraphics[width=90mm]{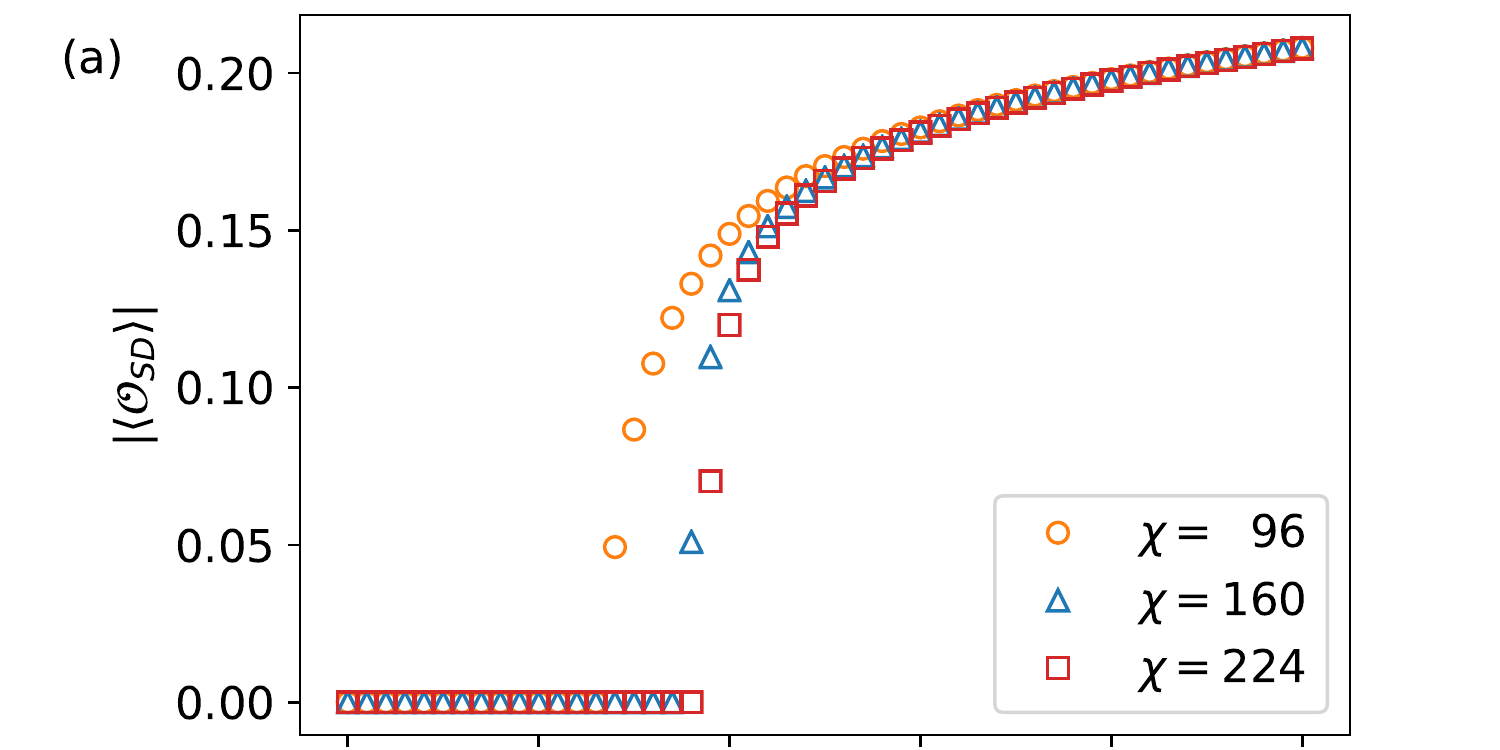} 
\includegraphics[width=90mm]{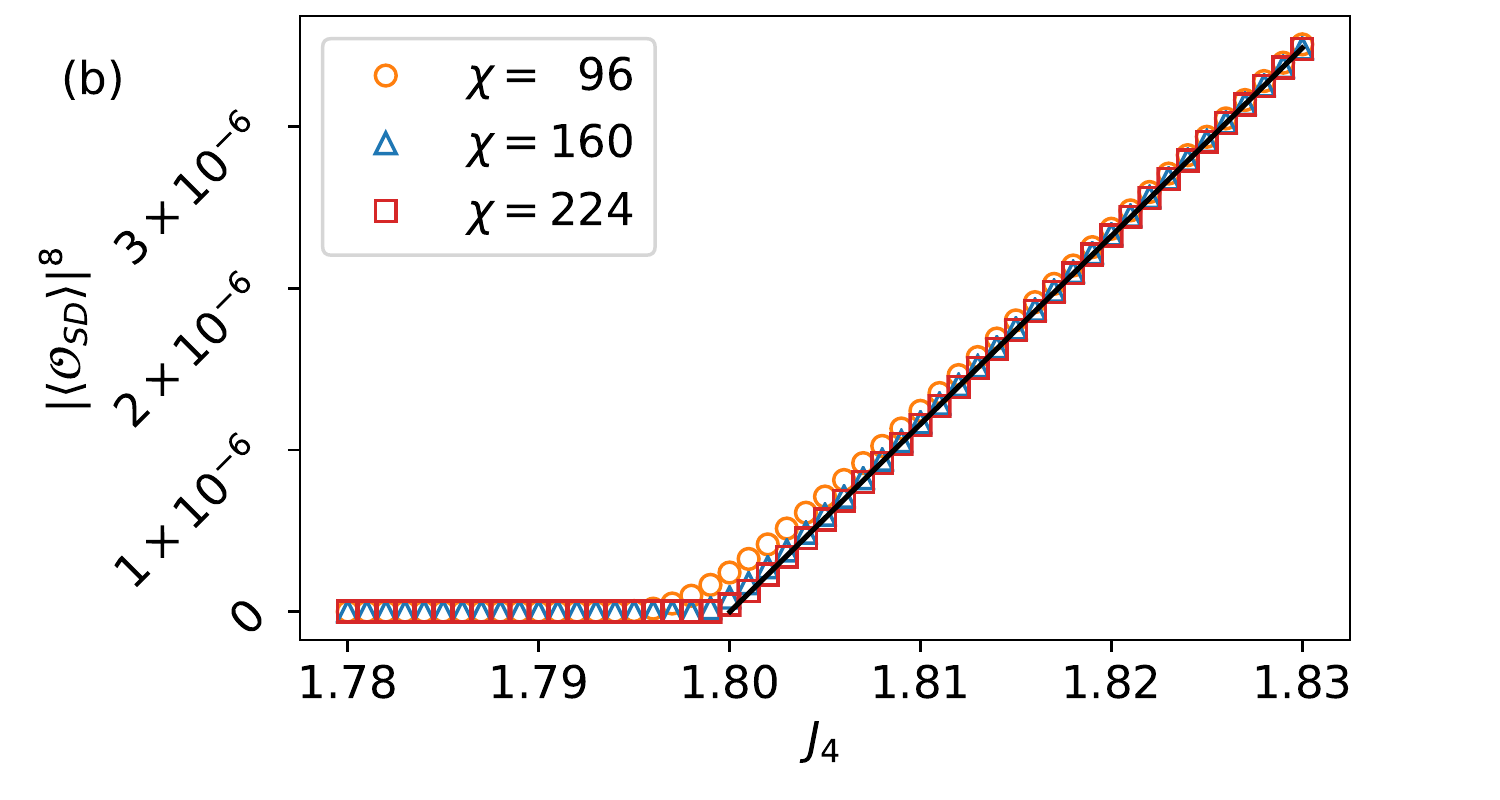}
\caption{
(a) SD order parameter $|\expval{\mathcal{O}_{\text{SD}}}|$ [defined in Eq.\ \eqref{eq:O_SD}] as a function of $J_4$ around the Haldane*-SD transition for $\Delta = 0.9$ 
(see the blue dashed line in Fig.\ \ref{fig:PhaseDiagram1}). 
(b) $|\expval{\mathcal{O}_{\text{SD}}}|^8$ as a function of $J_4$. 
The black solid line shows the linear fitting of the $\chi=224$ data in the range $1.81\le J_4\le 1.83$, where the dependence on $\chi$ is sufficiently converged. 
Its intersect with the horizontal axis gives the estimate $J_{4,c}^\text{H$^*$-SD} \simeq 1.80$ of the transition point. 
%
%Order parameter $|\expval{\mathcal{O}_{\text{SD}}}|$ and $|\expval{\mathcal{O}_{\text{SD}}}|^8$ around the Haldane*-SD transition for $\Delta = 0.9$.
%Using the linear extrapolation of $|\expval{\mathcal{O}_{\text{SD}}}|^8$, we obtain the critical point $J_{4,c} \sim 1.80$.
%The black solid line is a guide to the eye.
}
\label{fig:SDOrderParam}
\end{figure}
%%%%%%%%%%%%%%%%%%%%%%%%%%%%%%%%%%%%%%%%%%%%%%%%%%%%%%%%%%%%%%%%%%%%%%%%%%%%%%%%%%%%%%%%%%%%

%%%%%%%%%%%%%%%%%%%%%%%%%%%%%%%%%%%%%%%%%%%%%%%%%%%%%%%%%%%%%%%%%%%%%%%%%%%%%%%%%%%%%%%%%%%%
\begin{figure}
\includegraphics[width=90mm]{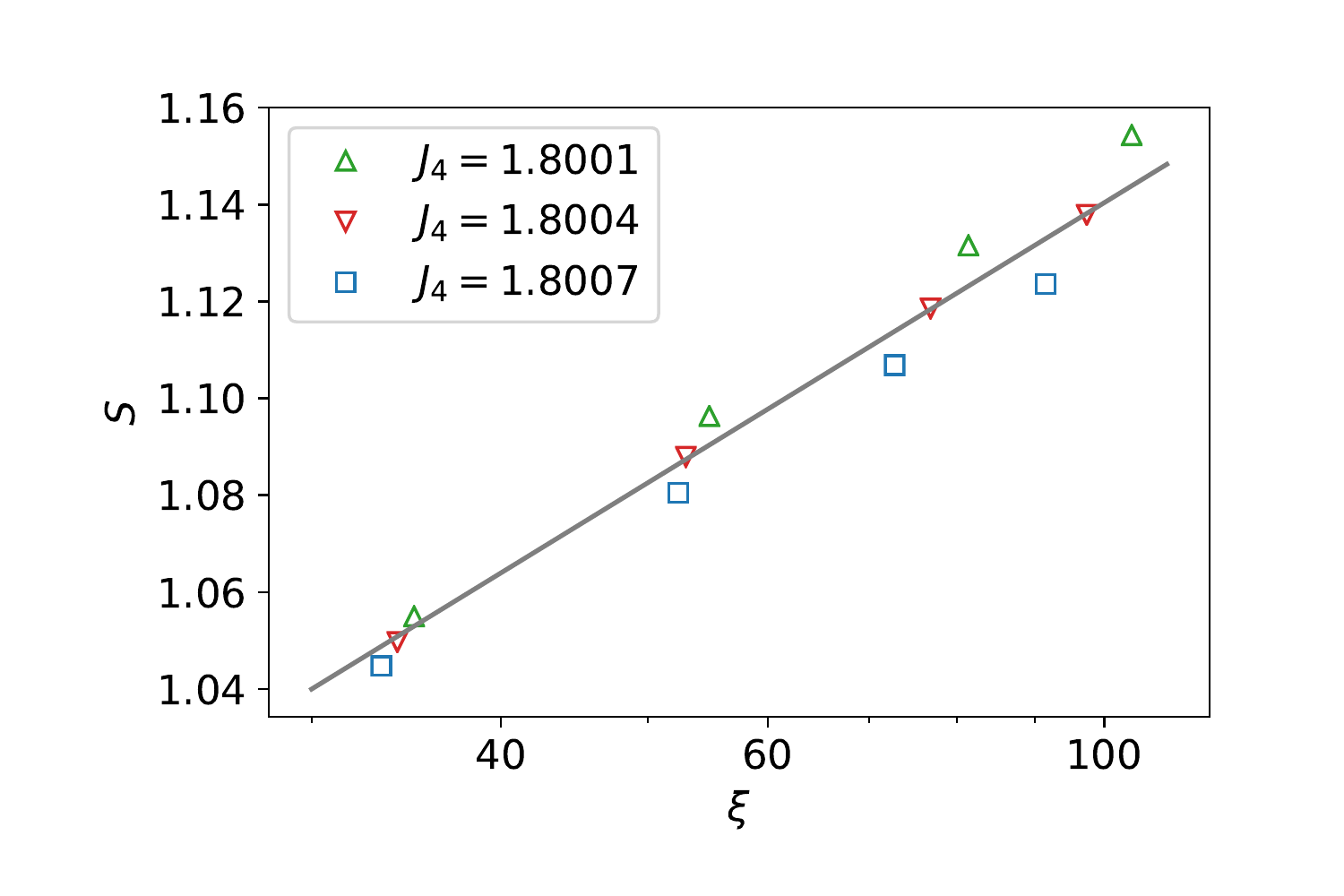} 
\caption{
Entanglement entropy $S(\chi)$ versus the correlation length $\xi(\chi)$ for the bond dimensions $\chi = 96,128,160,192$.
These are calculated at $J_4 = 1.8001$, $1.8004$, and $1.8007$, around which the Haldane*-SD transition is expected to occur (see Fig.\ \ref{fig:SDOrderParam}).
A logarithmic scale is used for the horizontal axis. 
The gray straight line is a guide to the eye and 
its slope is $1/12$, %which is consistent in 
which corresponds to 
the central charge $c=1/2$.
}\label{fig:CCJ418004}
\end{figure}
%%%%%%%%%%%%%%%%%%%%%%%%%%%%%%%%%%%%%%%%%%%%%%%%%%%%%%%%%%%%%%%%%%%%%%%%%%%%%%%%%%%%%%%%%%%%

% [ SD order parameter ]--------------------------------------
Here we focus on the Haldane*-SD transition along the blue dashed line with $\Delta=0.9$ in Fig.\ \ref{fig:PhaseDiagram1}. 
Figure \ref{fig:SDOrderParam}(a) shows the SD order parameter $|\expval{\mathcal{O}_{\text{SD}}}|$, 
which indicates the onset of the SD order at a certain transition point $J_{4,c}^\text{H$^*$-SD}$. 
According to the effective field theory, the Haldane*-SD transition belongs to the Ising universality class with $c=1/2$. 
In this universality class, the critical exponent for the spontaneous order parameter is given by $\beta=1/8$. 
Therefore, it is expected that the order parameter to the eighth power becomes linear around the transition point 
and its intersect with the horizontal axis gives the estimate of the transition point. 
This analysis is performed in Fig.\ \ref{fig:SDOrderParam}(b). 
We indeed find a good agreement with linear behavior in the range where the dependence on $\chi$ is sufficiently converged. 
The transition point is estimated to be $J_{4,c}^\text{H$^*$-SD}\simeq 1.80$. 

%We numerically estimate the critical point by using the order parameter $\expval{\mathcal{O}_{\text{SD}}}$ 
%assuming the transition belongs to the $(1+1)$ dimensional Ising universality class
%with the critical exponent $\beta = 1/8$ and the central charge $c=1/2$.
%The order parameter along the fixed $\Delta$ obeys the scaling 
%\begin{align}
%(-1)^j \expval{\mathcal{O}_{\text{SD}}} = 
%\begin{cases}
%\pm A_{\text{SD}}  ~ |J_4 - J_{4,c}|^{\beta_{\text{SD}} } &(\text{SD phase})\\
%0 &(\text{otherwise}),
%\end{cases}
%\end{align}
%where $J_{4.c}$ is the critical point and $A_{\text{SD}}$ is a constant.
%Therefore, the order parameter to the eighth power is proportional to $\Delta J_4$ in the SD phase:
%\begin{align}
%\expval{\mathcal{O}_{\text{SD}}}^8 = 
%\begin{cases}
%A_{\text{SD}}^8  ~ |J_4 - J_{4,c}| & (\text{SD phase}), \\
%0 &(\text{otherwise}).
%\end{cases}
%\label{eq:SD8}
%\end{align}
%Figure~\ref{fig:SDOrderParam} shows $|\expval{\mathcal{O}_{\text{SD}}}|$ and $|\expval{\mathcal{O}_{\text{SD}}}|^8$
%around the Haldane*-SD transition at $\Delta = 0.9$.
%The linear fitting gives the correctness of our assumption (\ref{eq:SD8})
%and the critical point $J_{4,c} \sim 1.80$.

% [ Entanglement entropy ]--------------------------------------
Figure \ref{fig:CCJ418004} shows the relation between the entanglement entropy $S(\chi)$ and the correlation length $\xi(\chi)$ 
at three points near the estimated critical point. 
We can confirm the consistency with the CFT scaling \eqref{eq:cardy} with the central charge $c=1/2$. 

%The correlation length and the entanglement entropy near the estimated critical point are plotted in Fig.~\ref{fig:CCJ418004}.
%The slope is in agreement with the predicted central charge $c=1/2$.

%%%%%%%%%%%%%%%%%%%%%%%%%%%%%%%%%%%%%%%%%%%%%%%%
\section{Numerical results for $J_\perp = -1$}\label{sec:ResultJpinv}
%%%%%%%%%%%%%%%%%%%%%%%%%%%%%%%%%%%%%%%%%%%%%%%%

%%%%%%%%%%%%%%%%%%%%%%%%%%%%%%%%%%%%%%%%%%%%%%%%%%%%%%%%%%%%%%%%%%%%%%%%%%%%%%%%%%%%%%%%%%%%
\begin{figure}
\includegraphics[width=90mm]{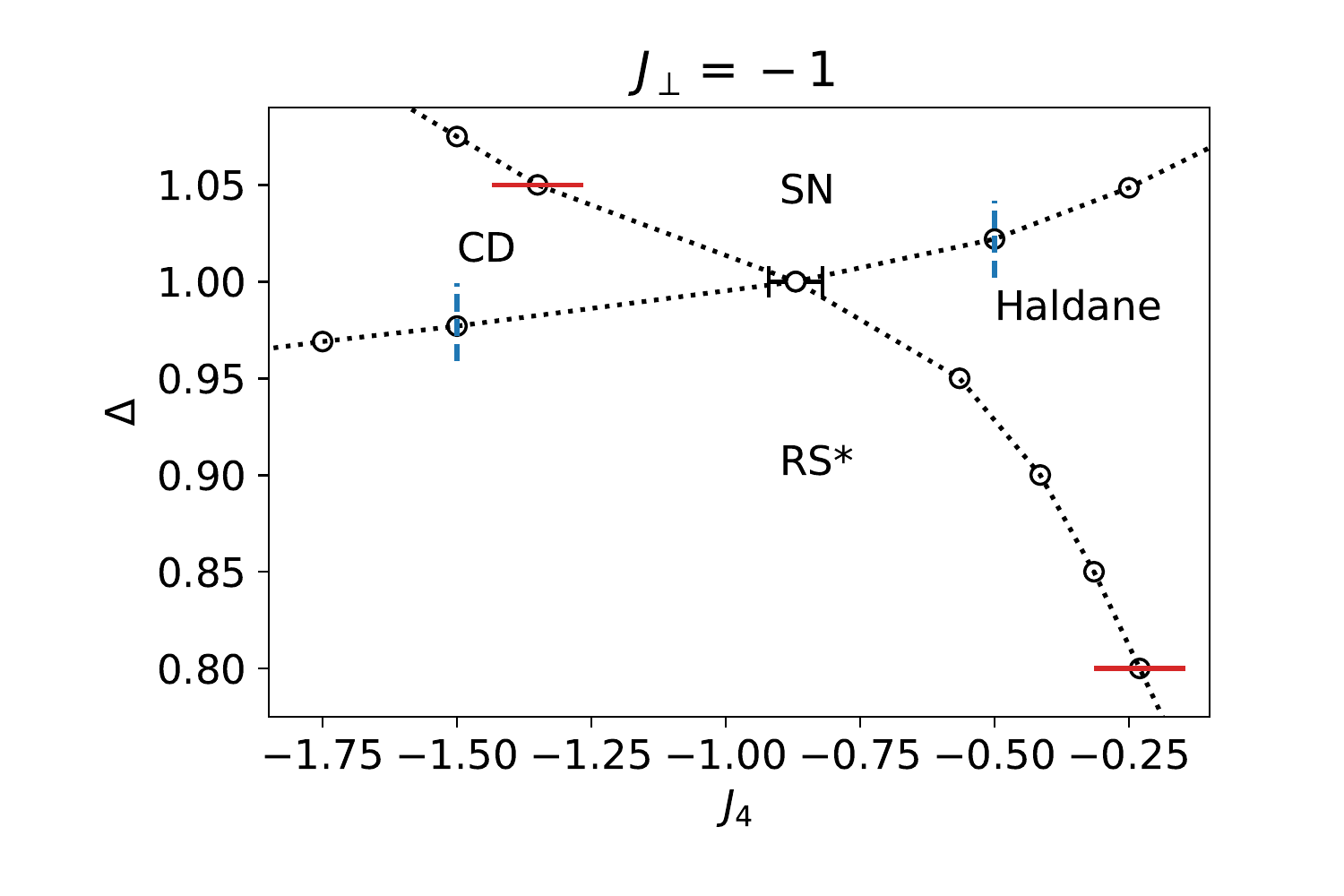}
\caption{
Phase diagram of the model \eqref{eq:modelmain} on the $J_4$-$\Delta$ plane with $J=1$ and $J_\perp=-1$. 
Circles indicate transition points obtained numerically, and dotted lines are our assumption of the phase boundaries. 
We will focus on the transitions along the red solid lines with $\Delta=0.8$ (Sec.\ \ref{eq:H_RSS}) and $1.05$ (Sec.\ \ref{sec:CD_SN})
and those along the blue dashed lines with $J_4=-0.5$ (Sec.\ \ref{sec:H_SN}) and $-1.5$ (Sec.\ \ref{sec:RSS_CD}). 
The transition point $J_{4,c}=-0.87(5)$ in the isotropic case $\Delta=1$ (a circle with an error bar) is estimated in Appendix \ref{App:isotropic}; 
see Ref.~\cite{PhysRevB.88.104403} for an estimate by exact diagonalization. 
%The model with $\Delta = 1$ is calculated in Ref.~\cite{PhysRevB.88.104403}
%and we calculated the region where $\Delta \neq 1$.
%The circles ($\circ$) indicate the transition points obtained 
%through the analysis of the string correlation functions and the order parameters.
%The dotted lines are our assumption of the phase boundaries. 
}
\label{fig:PhaseDiagram2}
\end{figure}
%%%%%%%%%%%%%%%%%%%%%%%%%%%%%%%%%%%%%%%%%%%%%%%%%%%%%%%%%%%%%%%%%%%%%%%%%%%%%%%%%%%%%%%%%%%%

In this section, we fix $ J = 1$ and $J_\perp = -1$. 
The obtained phase diagram in Fig.~\ref{fig:PhaseDiagram2} is qualitatively consistent with the field-theoretical prediction in Fig.\ \ref{fig:PhaseDiagrambos2}. 
Below we explain how the phase boundaries are obtained and how the critical properties are characterized in our numerical analysis. 

%In this section, we fix $ J = 1$ and $J_\perp = -1$, and present the numerically obtained phase diagram in the $J_4$-$\Delta$ plane in Fig.~\ref{fig:PhaseDiagram2}. 
%The Stripe N\'eel phase and the CD phase breaks different $\mathbb{Z}_2$ symmetries and 
%the RS* phase and the Haldane phase do not break any symmetry.

%************************************************
\subsection{RS*-Haldane transition}\label{eq:H_RSS}
%************************************************

%%%%%%%%%%%%%%%%%%%%%%%%%%%%%%%%%%%%%%%%%%%%%%%%%%%%%%%%%%%%%%%%%%%%%%%%%%%%%%%%%%%%%%%%%%%%
\begin{figure}
\includegraphics[width=90mm]{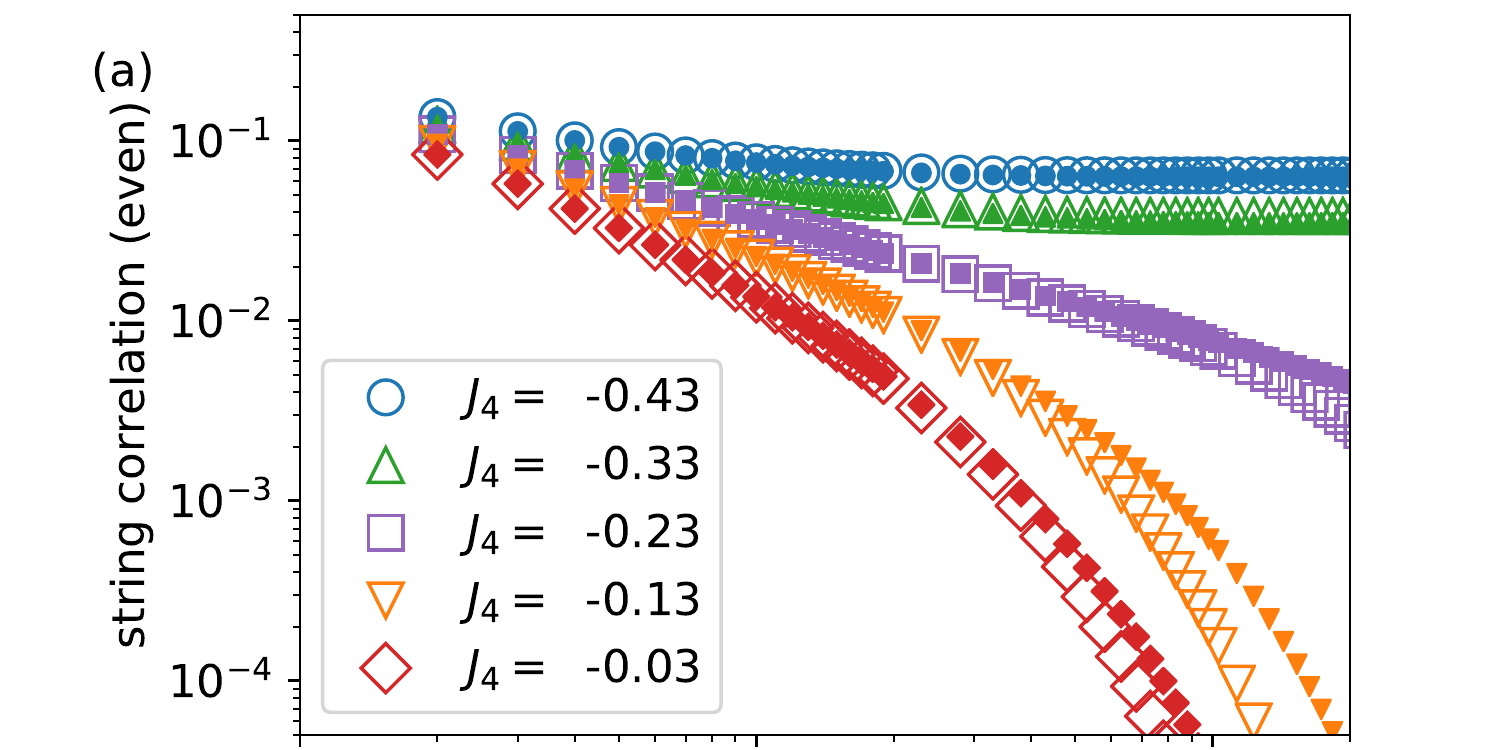}
\includegraphics[width=90mm]{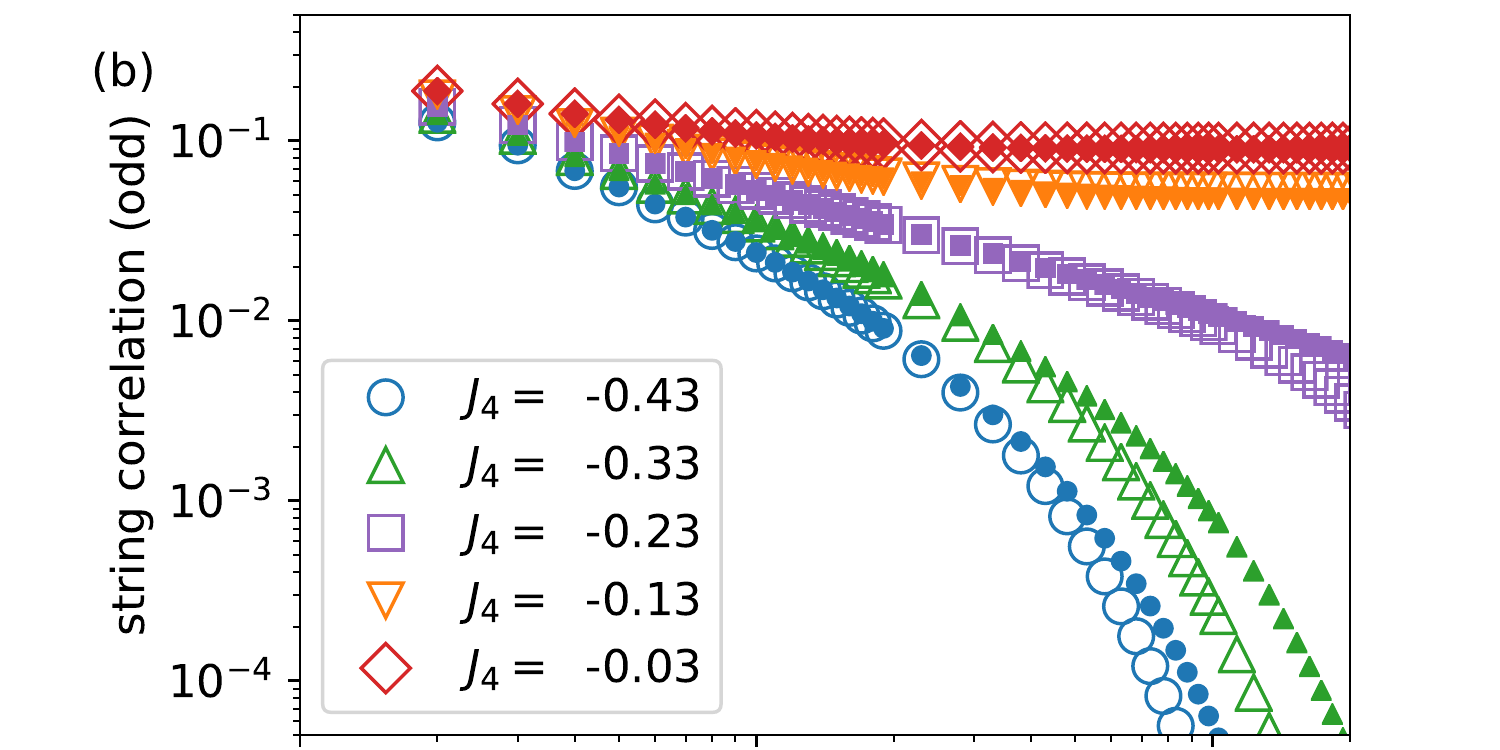}
\includegraphics[width=90mm]{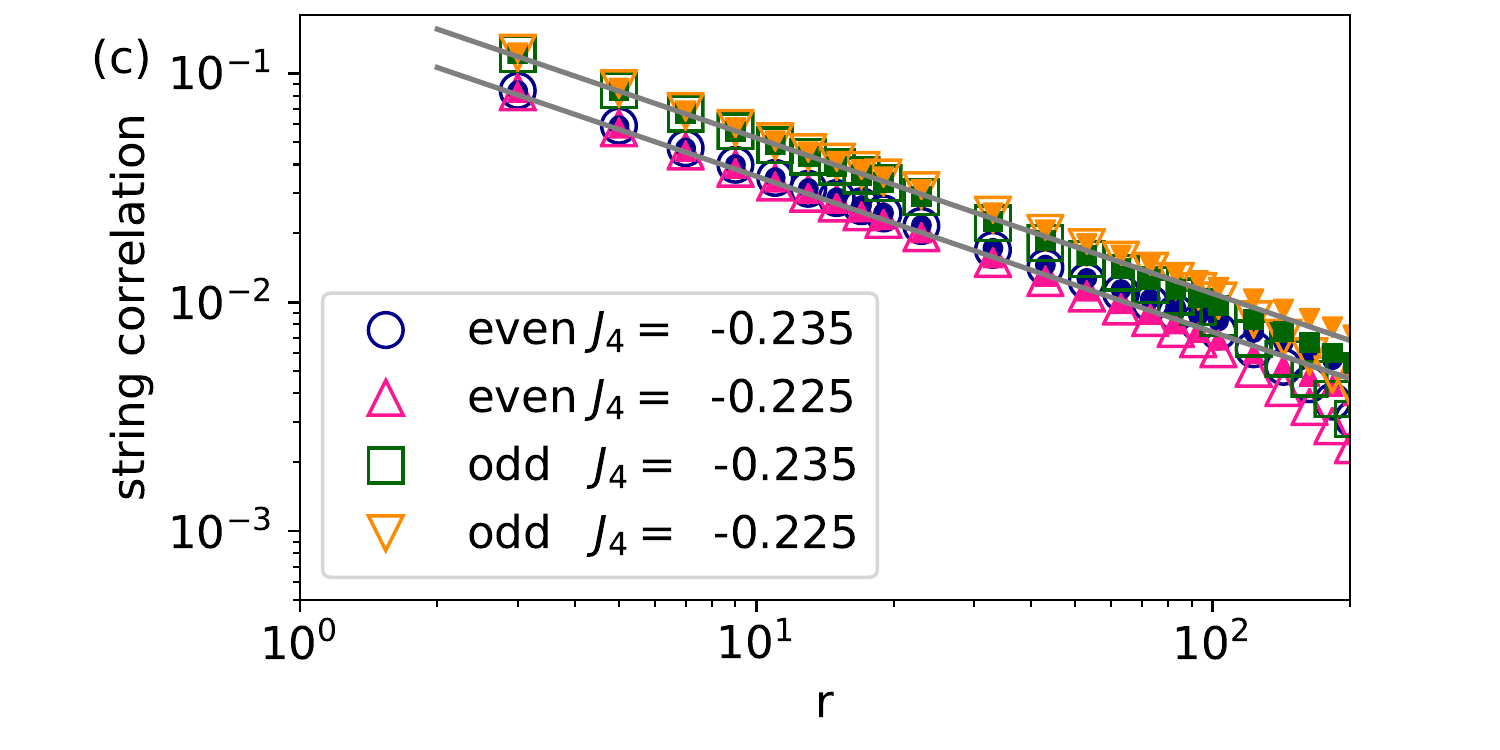}
\caption{
Two types of string correlation functions \eqref{eq:stringOP} around the RS*-Haldane transition for $\Delta=0.8$ in Fig.\ \ref{fig:PhaseDiagram2}. 
%Logarithmic scales are used for both axes. 
Large open and small filled symbols are for $\chi=96$ and $192$, respectively. 
(a) $O^z_{\text{even}}(r)$ is non-decaying at long distances in the RS* phase with $J_4\lesssim -0.23$. 
(b) $O^z_{\text{odd}}(r)$ is non-decaying in the Haldane phase with $J_4\gtrsim -0.23$. 
(c) The two correlation functions show a power-law decay with the same exponent $K_+$ at the transition in the Gaussian universality class. 
As analyzed in this panel, this should occur between $J_4=-0.235$ and $-0.225$ with the exponent $K_+=0.68(2)$. 
Two parallel gray solid lines are guides to the eye and their slope is $-0.68$.
%
%String correlation functions around the RS*-Haldane transition for $\Delta=0.8$.
%Large open symbols represent $\chi=96$ and small filled symbols represent $\chi=192$.
%(a) The string correlation $\expval{\mathcal{O}_{\text{even}}(r)}$.
%(b) The string correlation $\expval{\mathcal{O}_{\text{odd}}(r)}$.
%The string correlation functions at $J_4 = -0.23$ exhibit power-law behavior.
%(c) The string correlation $\expval{\mathcal{O}_{\text{even}}(r)}$ and $\expval{\mathcal{O}_{\text{odd}}(r)}$.
%Two parallel gray solid lines are guide to the eye and the slope is $-0.68$.
%At the critical point, the parameter $K_+$ is equal to
%the gradient of both $\expval{\mathcal{O}_{\text{even}}(r)}$ and $\expval{\mathcal{O}_{\text{odd}}(r)}$
%Therefore, we obtain the critical point $J_{4,c} = -0.235 \sim -0.225$ and the TLL parameter $K_+ = 0.68(2)$.
}
\label{fig:StringFunctionRSSH}
\end{figure}
%%%%%%%%%%%%%%%%%%%%%%%%%%%%%%%%%%%%%%%%%%%%%%%%%%%%%%%%%%%%%%%%%%%%%%%%%%%%%%%%%%%%%%%%%%%%

The RS*-Haldane transition can be analyzed in a similar manner as the RS-Haldane* transition discussed in Sec.\ \ref{sec:RSHS}. 
Figure \ref{fig:StringFunctionRSSH} shows the string correlation functions around the RS*-Haldane transition for $\Delta=0.8$. 
For $J_4\lesssim -0.23$, $O^z_{\text{even}}(r)$ is non-decaying at long distances 
while $O^z_{\text{odd}}(r)$ shows a rapid decay. 
For $J_4\gtrsim -0.23$, $O^z_{\text{odd}}(r)$ is non-decaying while $O^z_{\text{even}}(r)$ shows a rapid decay. 
These results are consistent with the occurrence of the RS*-Haldane transition. 
More detailed characterization of these phases is given in terms of topological indices in Sec.\ \ref{sec:SPT}. 
Assuming the Gaussian universality class, the transition point $J_{4,c}^\text{RS$^*$-H}$ can be identified 
by linear and parallel behavior of the two correlation functions plotted in logarithmic scales. 
This should occur between the two points examined in Fig.\ \ref{fig:StringFunctionRSSH}(c), 
giving the estimate $J_{4,c}^\text{RS$^*$-H}=-0.230(5)$. 
We have also confirmed the consistency with the central charge $c=1$ in a similar manner as in Fig.\ \ref{fig:CCJ40690} (not shown). 

%We estimate the critical point and the TLL parameter $K_+$
%assuming that both of the string correlation functions show power-law decay 
%and have the same slope at the critical point as discussed in Sec.~\ref{sec:ResultJp}.
%We plot the string correlation functions $O^z_{\text{odd/even}}(r)$ at $\Delta = 0.8$ in Fig.~\ref{fig:StringFunctionRSSH}.
%In the RS* [Haldane] phase, the string correlation function $O^z_{\text{odd}}(r)$ goes to zero [a non-zero finite value]
%and $O^z_{\text{even}}(r)$ goes to a non-zero finite value [zero].
%Both of the string correlation functions have the same slope  $-K_+ = -0.68(2)$ between $J_4 = -0.235$ and $J_4 = -0.225$.
%By fitting Eq.~(\ref{eq:cardy}) to the data of the entanglement entropy versus the correlation length, we obtain the central charge $c\simeq1$.
%We calculate the entanglement spectrum and only the Haldane* phase has the double degeneracy (Fig.~\ref{fig:ES2}).

%************************************************
\subsection{CD-SN transition} \label{sec:CD_SN}
%************************************************

%%%%%%%%%%%%%%%%%%%%%%%%%%%%%%%%%%%%%%%%%%%%%%%%%%%%%%%%%%%%%%%%%%%%%%%%%%%%%%%%%%%%%%%%%%%%
\begin{figure}
\includegraphics[width=90mm]{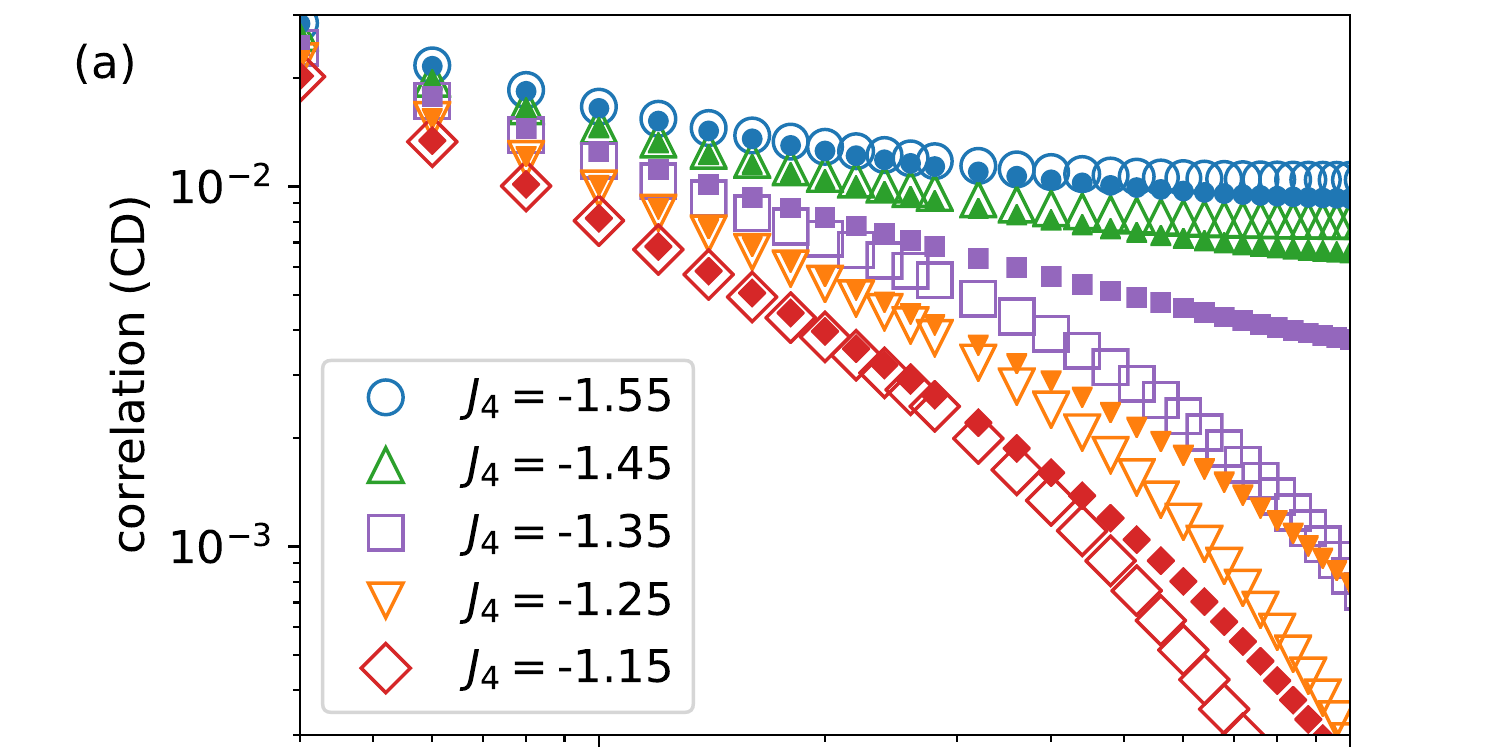}
\includegraphics[width=90mm]{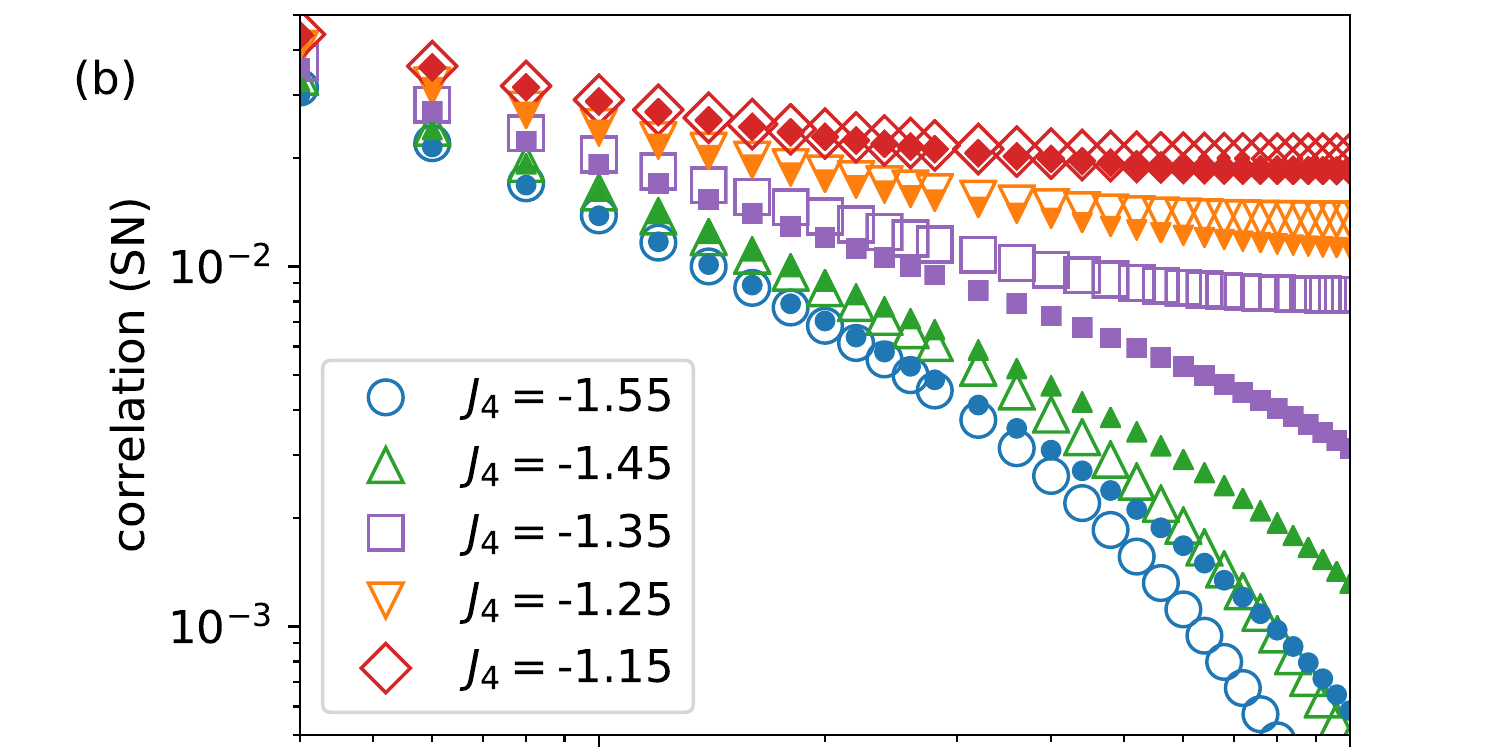}
\includegraphics[width=90mm]{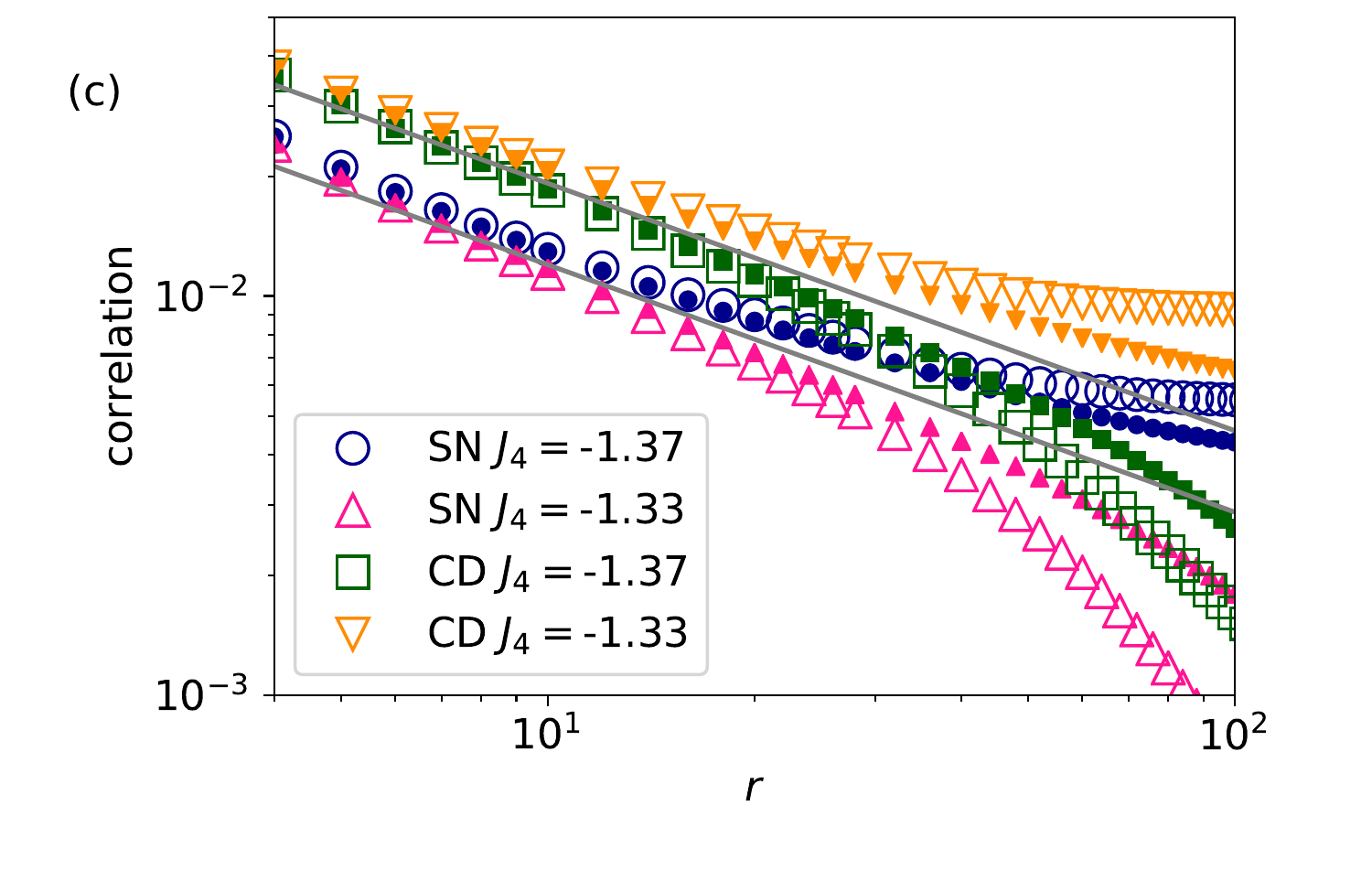}
\caption{
Correlation functions \eqref{eq:corr_SN_CD} around the CD-SN transition for $\Delta=1.05$ in Fig.\ \ref{fig:PhaseDiagram2}. 
%Logarithmic scales are used for both axes. 
Large open and small filled symbols are for $\chi=96$ and $192$, respectively. 
(a) The CD correlation function $C_{\text{CD}}(r)$ plotted in logarithmic scales is convex for $J_4\lesssim -1.35$, signifying the CD long-range order.  
(b) The SN correlation function $C_{\text{SN}}(r)$ plotted in logarithmic scales is convex for $J_4\gtrsim -1.35$, signifying the SN long-range order. 
(c) The two correlation functions show a power-law decay with the same exponent $K_+$ at the transition in the Gaussian universality class. 
As analyzed in this panel, this should occur between $J_4=-1.37$ and $-1.33$ with the exponent $K_+=0.62(8)$. 
Two parallel gray solid lines are guides to the eye and their slope is $-0.62$. 
%
%Large open symbols represent $\chi=96$ and small filled symbols represent $\chi=192$.
%(a) The CD correlation functions $C_{\text{CD}}(r)$.
%(b) The SN correlation functions $C_{\text{SN}}(r)$.
%The correlation functions at $J_4 = -1.35$ exhibit power-law behavior.
%(c) The orrelation functions $C_{\text{CD}}(r)$ and $C_{\text{SN}}(r)$.
%Two parallel black solid lines are guide to the eye and the slope is $-0.62$.
%At the critical point, the parameter $K_+$ is equal to
%the gradient of both $C_{\text{CD}}(r)$ and $C_{\text{SN}}(r)$.
%Therefore, we obtain the critical point $J_{4,c} = -1.37 \sim -1.33$ and the TLL parameter $K_+ = 0.62(8)$.
}
\label{fig:CorrFunctionCDSN}
\end{figure}
%%%%%%%%%%%%%%%%%%%%%%%%%%%%%%%%%%%%%%%%%%%%%%%%%%%%%%%%%%%%%%%%%%%%%%%%%%%%%%%%%%%%%%%%%%%%

Figure \ref{fig:CorrFunctionCDSN} shows the CD and SN correlation functions \eqref{eq:corr_SN_CD} around the CD-SN transition for $\Delta=1.05$ in Fig.\ \ref{fig:PhaseDiagram2}. 
The data show the tendency of the CD and SN long-range orders for $J_4\lesssim -1.35$ and $J_4\gtrsim -1.35$, respectively. 
As discussed in Sec.\ \ref{sec:bos_critical}, the two correlation functions show a power-law decay with the same exponent $K_+$ at the transition point $J_{4,c}^\text{CD-SN}$ 
if the transition belongs to the Gaussian universality class. 
This should occur between the two points examined in Fig.\ \ref{fig:CorrFunctionCDSN}(c), giving the estimate $J_{4,c}^\text{CD-SN}=-1.35(2)$. 

%We numerically estimate the critical point and the TLL parameter $K_+$
%assuming that both of the correlation functions show power-law decay and have the same slope at the critical point.
%We plot the correlation functions $O^z_{\text{CD/SN}}(r)$ at $\Delta = 1.05$ in Fig.~\ref{fig:CorrFunctionCDSN}.
%In the CD [SN] phase, the correlation function $O^z_{\text{SN}}(r)$ goes to zero [a non-zero finite value]
%and $O^z_{\text{CD}}(r)$ goes to a non-zero finite value [zero].
%Both of the correlation functions have the same slope $-K_+ = -0.62(8)$ between $J_4 = -1.37$ and $J_4 = -1.33$.

%************************************************
\subsection{Haldane-SN transition}\label{sec:H_SN}
%************************************************

%%%%%%%%%%%%%%%%%%%%%%%%%%%%%%%%%%%%%%%%%%%%%%%%%%%%%%%%%%%%%%%%%%%%%%%%%%%%%%%%%%%%%%%%%%%%
\begin{figure}
\includegraphics[width=90mm]{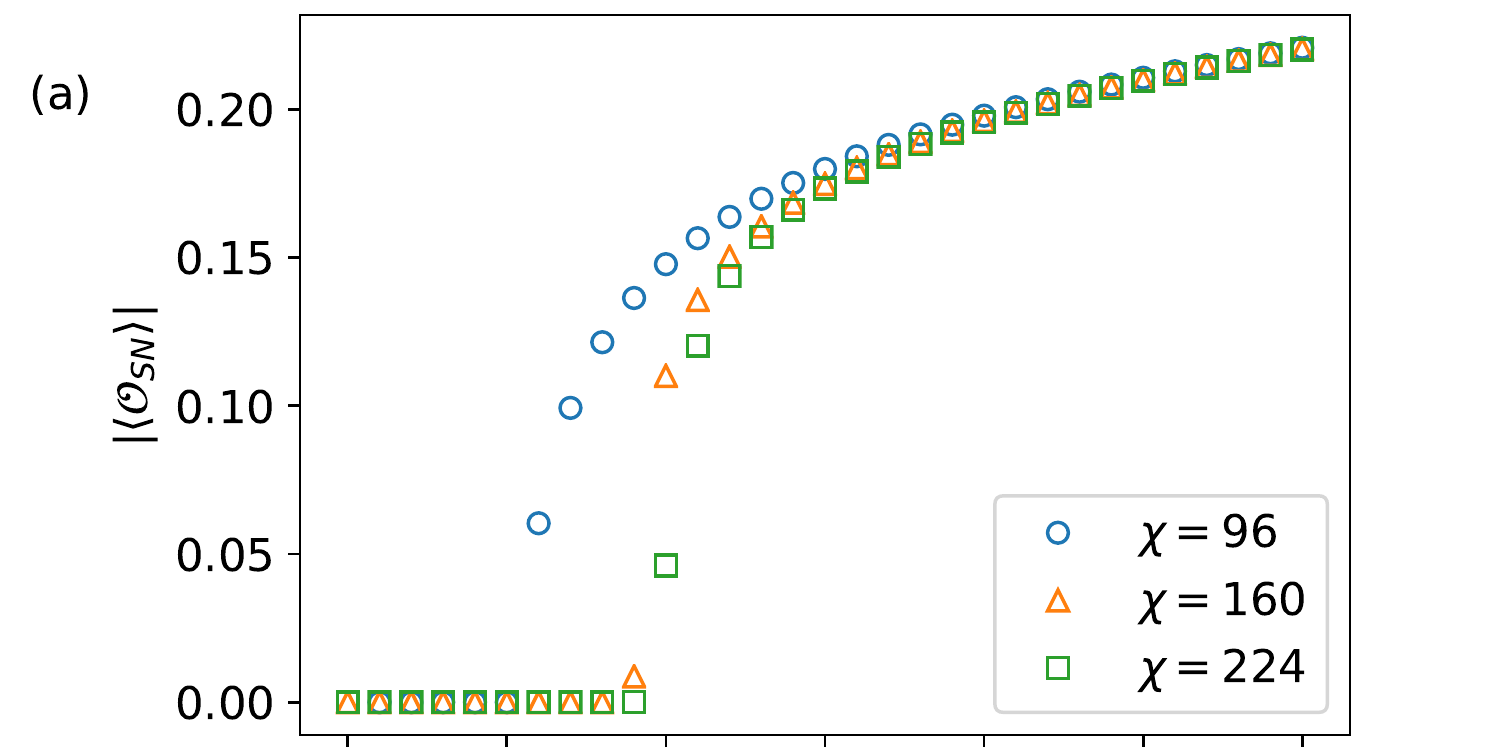} 
\includegraphics[width=90mm]{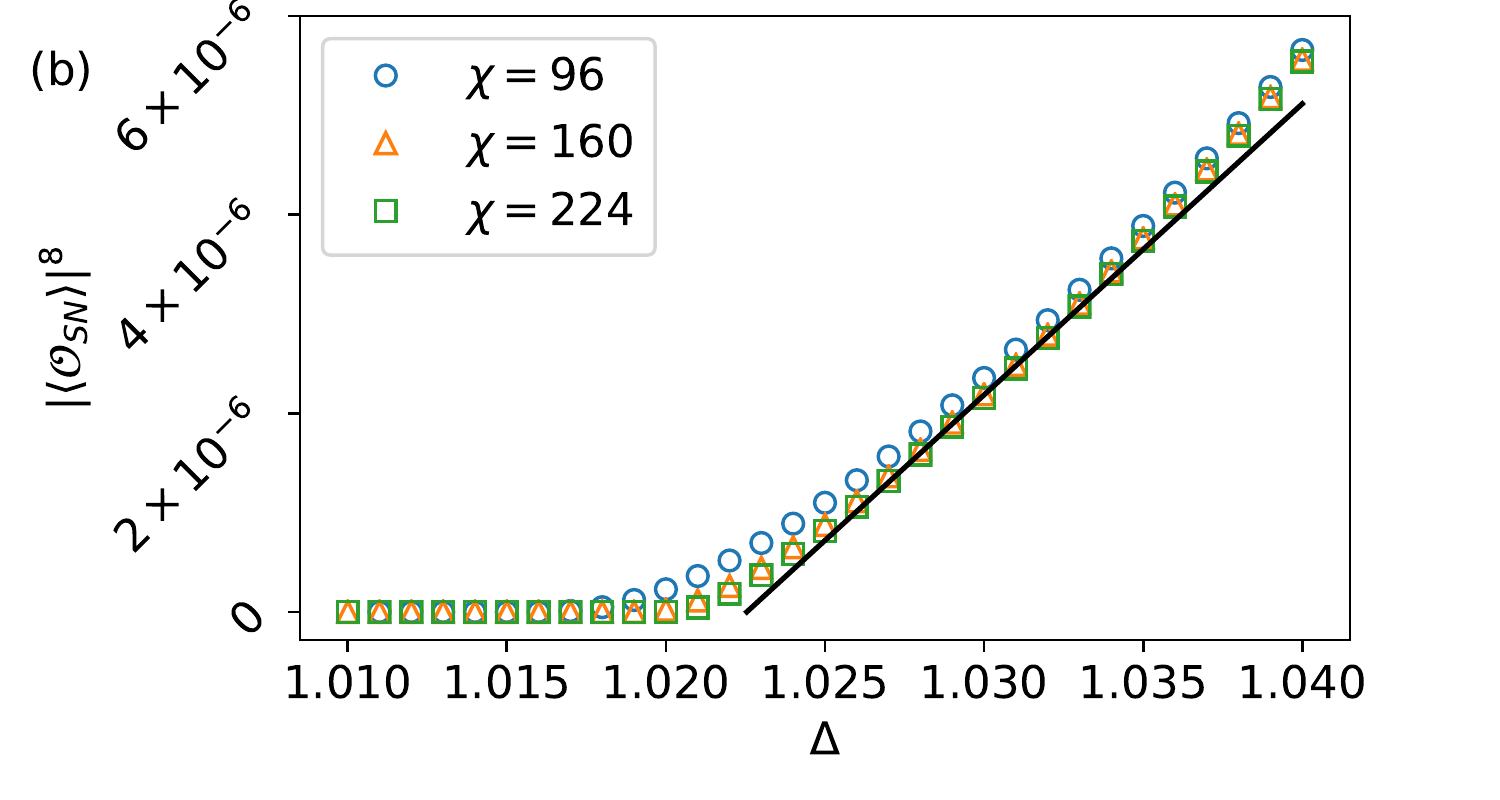}
\caption{
(a) SN order parameter $|\expval{\mathcal{O}_{\text{SN}}}|$ [defined in Eq.\ \eqref{eq:O_SN}] as a function of $\Delta$ 
around the SN-Haldane transition for $J_4=-0.5$ in Fig.\ \ref{fig:PhaseDiagram2}. 
(b) $|\expval{\mathcal{O}_{\text{SN}}}|^8$ as a function of $\Delta$. 
The black solid line shows the linear fitting of the $\chi=224$ data in the range $1.025<\Delta<1.035$. 
Its intersect with the horizontal axis gives the estimate $\Delta_{c}^\text{H-SN} \simeq 1.022$ of the transition point. 
%
%Order parameter $|\expval{\mathcal{O}_{\text{SN}}}|$ and $|\expval{\mathcal{O}_{\text{SN}}}|^8$ around the SN-Haldane transition for $J_4=-0.5$.
%Using the linear extrapolation of $|\expval{\mathcal{O}_{\text{SN}}}|^8$, we obtain the critical point $\Delta_{c} \sim 1.022$.
%The black solid line is a guide to the eye.
}
\label{fig:SNOrderParam}
\end{figure}
%%%%%%%%%%%%%%%%%%%%%%%%%%%%%%%%%%%%%%%%%%%%%%%%%%%%%%%%%%%%%%%%%%%%%%%%%%%%%%%%%%%%%%%%%%%%

The Haldane-SN transition can be analyzed in a similar manner as the Haldane*-SD transition discussed in Sec.\ \ref{sec:HSSD}. 
Here, we fix $J_4=-0.5$ and vary $\Delta$. 
Figure \ref{fig:SNOrderParam} shows our result on the SN order parameter $|\expval{\mathcal{O}_{\text{SN}}}|$. 
In the Ising universality class, $|\expval{\mathcal{O}_{\text{SN}}}|^8$ is expected to be linear around the transition point $\Delta_{c}^\text{H-SN}$ . 
By performing the linear fitting of the data of $|\expval{\mathcal{O}_{\text{SN}}}|^8$ versus $\Delta$ and finding the intersect with the horizontal axis, 
we obtain the estimate $\Delta_{c}^\text{H-SN} \simeq 1.022$. 
We also confirmed the consistency with the central charge $c=1/2$ in a similar manner as in Fig.\ \ref{fig:CCJ418004} (not shown). 

%We numerically estimate the critical point by using the order parameter $\expval{\mathcal{O}_{\text{SN}}}$.
%The SN-Haldane transition belongs to the $(1+1)$ dimensional Ising universality class 
%with $\beta = 1/8$ and $c=1/2$ as shown in Sec.~\ref{sec:EFT}.
%We numerically estimate the critical point by using the order parameter $\expval{\mathcal{O}_{\text{SN}}}$ 
%assuming the transition belongs to the $(1+1)$ dimensional Ising universality class.
%In the same way as is shown in Eq.~(\ref{eq:SD8}),
%the order parameter to the eighth power $|\expval{\mathcal{O}_{\text{SN}}}|^8$ along the fixed $J_4$ 
%is proportional to $|\Delta - \Delta_c|$, where $\Delta_c$ is the critical point, in the SN phase.
%Therefore, the order parameter to the eighth power along the fixed $J_4$ satisfies
%\begin{eqnarray}
%\expval{\mathcal{O}_{\text{SN}}}^8
%= 
%\begin{cases}
%A_{\text{SN}}^8 ~ ( \Delta - \Delta_c ) &(\text{SN phase})\\
%0 &(\text{otherwise}),
%\end{cases}
%\end{eqnarray}
%where $A_{\text{SN}}$ is a constant.
%The order parameter and the estimation of the critical point are shown in Fig.~\ref{fig:SNOrderParam}.
%Along the phase boundary, we obtain the central charge $c\simeq 0.5$.
%These results are consistent with the Ising universality.

%************************************************
\subsection{RS*-CD transition}\label{sec:RSS_CD}
%************************************************

%%%%%%%%%%%%%%%%%%%%%%%%%%%%%%%%%%%%%%%%%%%%%%%%%%%%%%%%%%%%%%%%%%%%%%%%%%%%%%%%%%%%%%%%%%%%
\begin{figure}
\includegraphics[width=90mm]{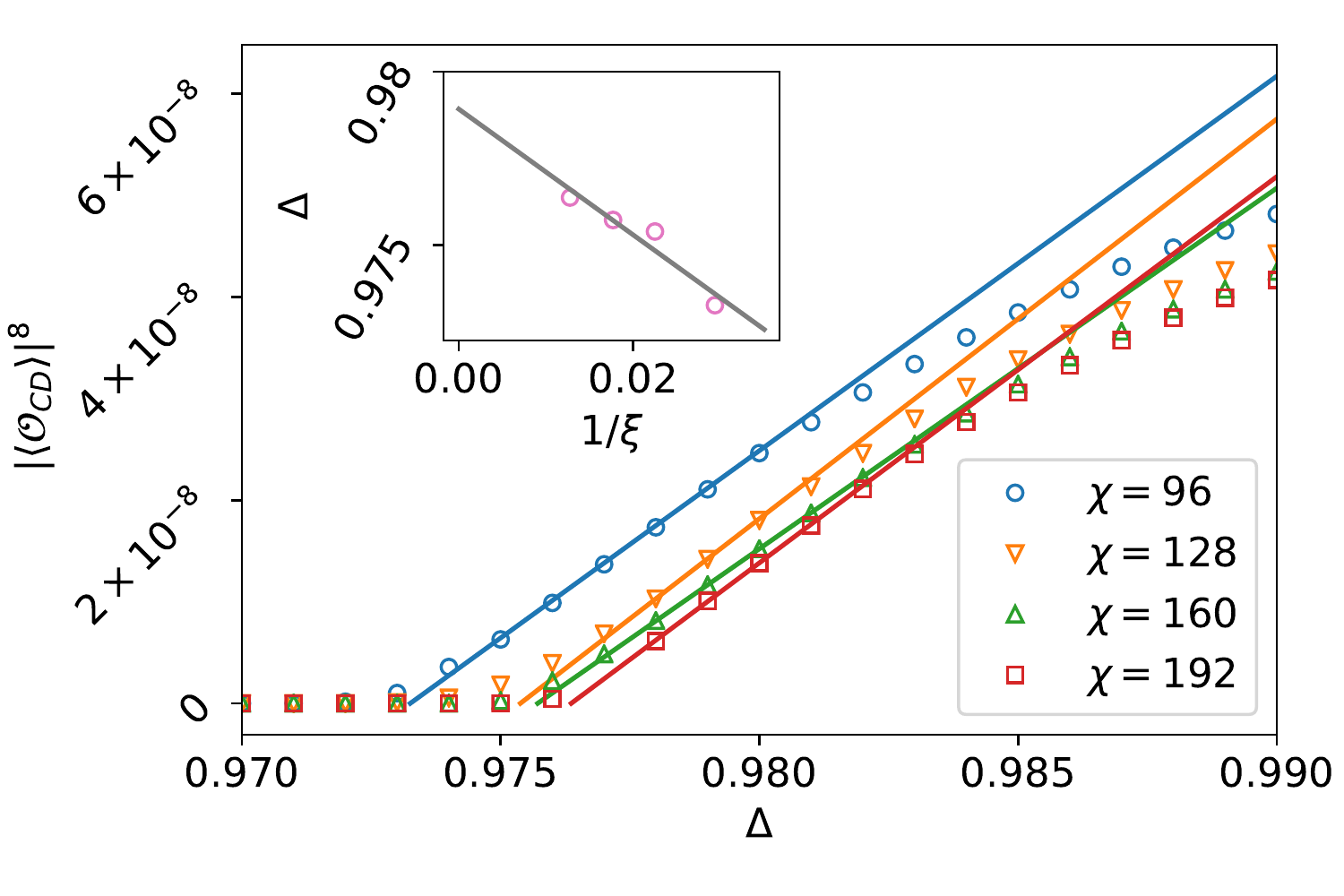} 
\caption{
CD order parameter \eqref{eq:O_CD} to the eighth power, $|\expval{\mathcal{O}_{\text{CD}}}|^8$, as a function of $\Delta$ 
around the RS*-CD transition for $J_4=-1.5$ in Fig.\ \ref{fig:PhaseDiagram2}. 
%Using the linear extrapolation of $|\expval{\mathcal{O}_{\text{CD}}}|^8$ at each bond dimension $\chi$,
By performing the linear extrapolation individually for each bond dimension $\chi$, 
we estimate the pseudo-critical point $\Delta_c(\chi)$.
As shown in the inset, we estimate the critical point $\Delta_c(\chi\rightarrow\infty) \simeq 0.979$ 
by using the bond-dimensional scaling in Eq.~(\ref{eq:scaling2}) with $\nu=1$. 
}\label{fig:CDOrderParam}
\end{figure}
%%%%%%%%%%%%%%%%%%%%%%%%%%%%%%%%%%%%%%%%%%%%%%%%%%%%%%%%%%%%%%%%%%%%%%%%%%%%%%%%%%%%%%%%%%%%

The RS*-CD transition can also be analyzed similarly. 
Here we fix $J_4=-1.5$, and plot the CD order parameter to the eighth power, $|\expval{\mathcal{O}_{\text{CD}}}|^8$, as a function of $\Delta$ in Fig.\ \ref{fig:CDOrderParam}. 
The numerical data show a rather significant dependence on the bond dimension $\chi$. 
This behavior is specific to the RS*-CD transition, and is likely to be caused by its proximity to the CD-SN transition. 
We therefore perform the linear fitting individually for each bond dimension $\chi$, 
find its intersect with the horizontal axis, and obtain the $\chi$-dependent pseudo-critical point $\Delta_c^\text{RS$^*$-CD}(\chi)$. 

%%%%%%%%%%%%%%%%%%%%%%%%%%%%%%%%%%%%%%%%%%%%%%%%%%%%%%%%%%%%%%%%%%%%%%%%%%%%%%%%%%%%%%%%%%%%
\begin{figure}
\includegraphics[width=90mm]{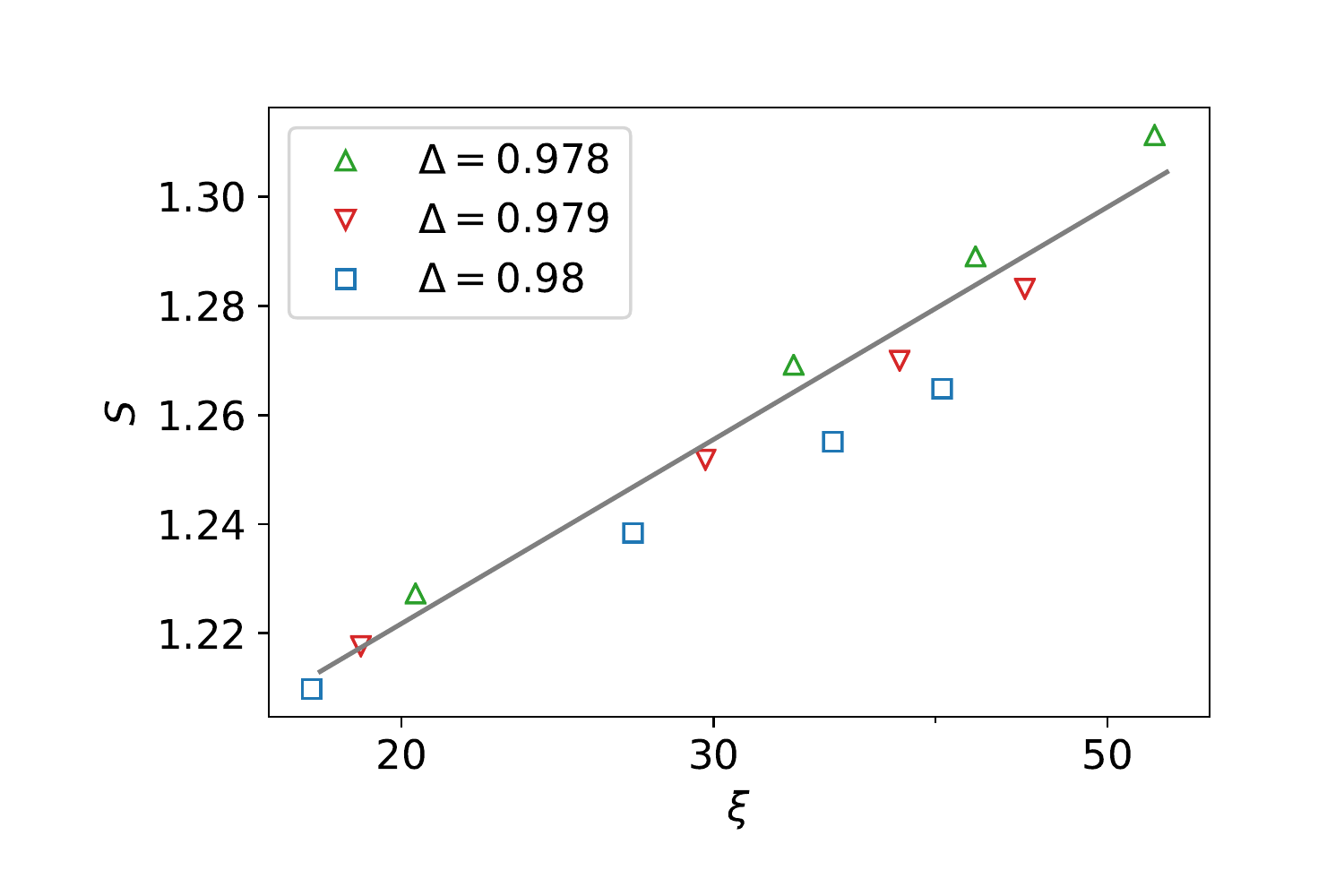} 
\caption{
Entanglement entropy $S(\chi)$ versus the correlation length $\xi(\chi)$ for the bond dimensions $\chi = 96,128,160,192$.
These are calculated at $\Delta = 0.978$, $0.979$, and $0.980$, around which the RS*-CD transition is expected to occur.
A logarithmic scale is used for the horizontal axis.
The gray straight line is a guide to the eye and 
its slope is $1/12$, which %is consistent in 
corresponds to 
the central charge $c=1/2$.
}\label{fig:CCRSSCD}
\end{figure}
%%%%%%%%%%%%%%%%%%%%%%%%%%%%%%%%%%%%%%%%%%%%%%%%%%%%%%%%%%%%%%%%%%%%%%%%%%%%%%%%%%%%%%%%%%%%

We extrapolate the pseudo-critical point to infinite $\chi$ by using the finite-bond-dimensional scaling.
Finite-size scaling predicts that the true critical point $\Delta_{c,\text{true}}$ and the pseudo-critical point $\Delta_c(L)$ satisfy
\begin{eqnarray} 
\Delta_{c,\text{true}} - \Delta_c(L) \sim L^{-\frac{1}{\nu}}.
\label{eq:scaling}
\end{eqnarray}
We assume that Eq.~(\ref{eq:scaling}) holds by replacing the system size $L$ with the correlation length $\xi$:
\begin{eqnarray}
\Delta_{c,\text{true}} - \Delta_c(\chi) \sim \xi(\chi)^{-\frac{1}{\nu}}.
\label{eq:scaling2}
\end{eqnarray}
We also assume the Ising universality class with $\nu=1$. 
Using this scaling, we obtain the critical point $\Delta_c^\text{RS$^*$-CD} \simeq 0.979$ as shown in the inset in Fig.~\ref{fig:CDOrderParam}. 
Around the estimated critical point, we have confirmed the consistency with $c=1/2$ 
by examining the entanglement entropy versus the correlation length as shown in Fig.\ \ref{fig:CCRSSCD}. 

\section{Topological distinction among the four featureless phases 
%Classification of the phases
} \label{sec:SPT}
%************************************************

% [ Section introduction ]--------------------------------------
In the preceding sections, we have used two types of string correlations \eqref{eq:stringOP} in analyzing the RS-Haldane* and RS*-Haldane transitions. 
However, these correlations cannot be used to distinguish between the RS and RS* phases or between the Haldane and Haldane* phases. 
In this section, we demonstrate that topological indices \cite{PhysRevB.81.064439, PhysRevB.86.125441, PhysRevB.83.035107, PhysRevB.84.235128} 
associated with the $D_2\times \sigma$ symmetry and the translational symmetry can distinguish all the four featureless phases. 
Below we briefly review the basic formalism for classifying 1D SPT phases and calculating topological indices
(especially, those of Pollmann and Turner \cite{PhysRevB.86.125441}), 
and then apply it to the present ladder system. 
While the full classification of SPT phases in a spin-$\frac12$ ladder with the $D_2\times\sigma$ symmetry 
has been achieved by Liu {\it et al}.\ \cite{PhysRevB.86.195122}, our results illustrate how those phases are 
%actually identified 
identified unambiguously
in numerical calculations. 

%%%%%%%%%%%%%%%%%%%%%%%%%%%%%%%%%%%%%%%%%%%%%%%%%%%%%%%%%%%%%%%%%%%%%%%%%%%%%%%%%%%%%%%%%%%%
\begin{figure}
\includegraphics[width=80mm]{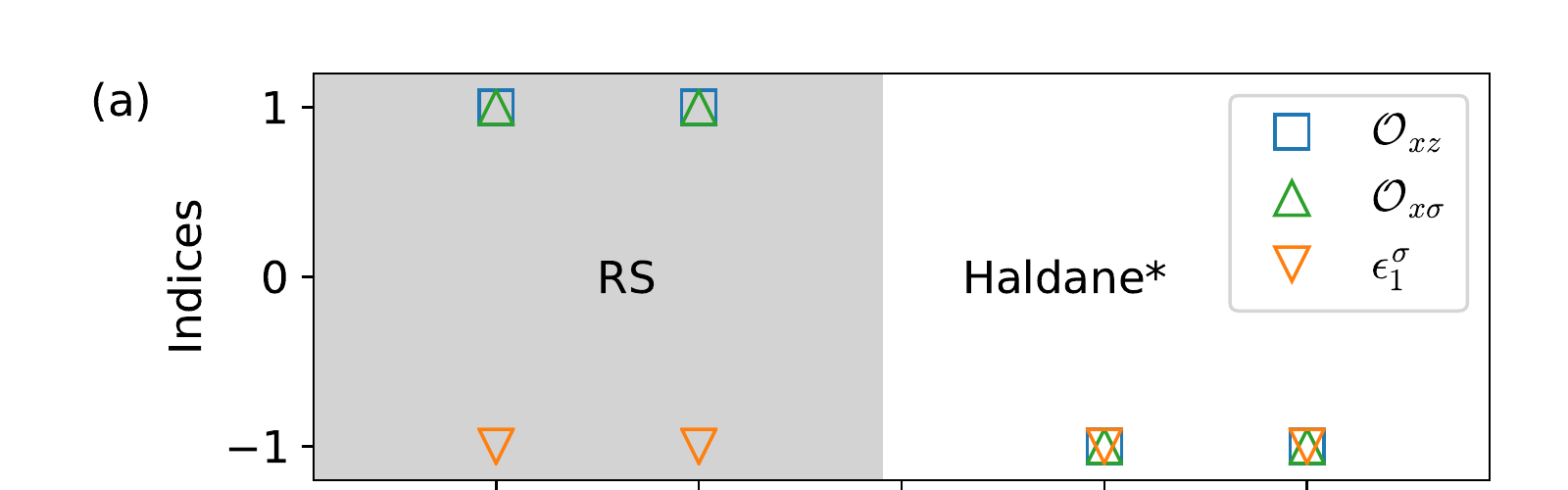} 
\includegraphics[width=80mm]{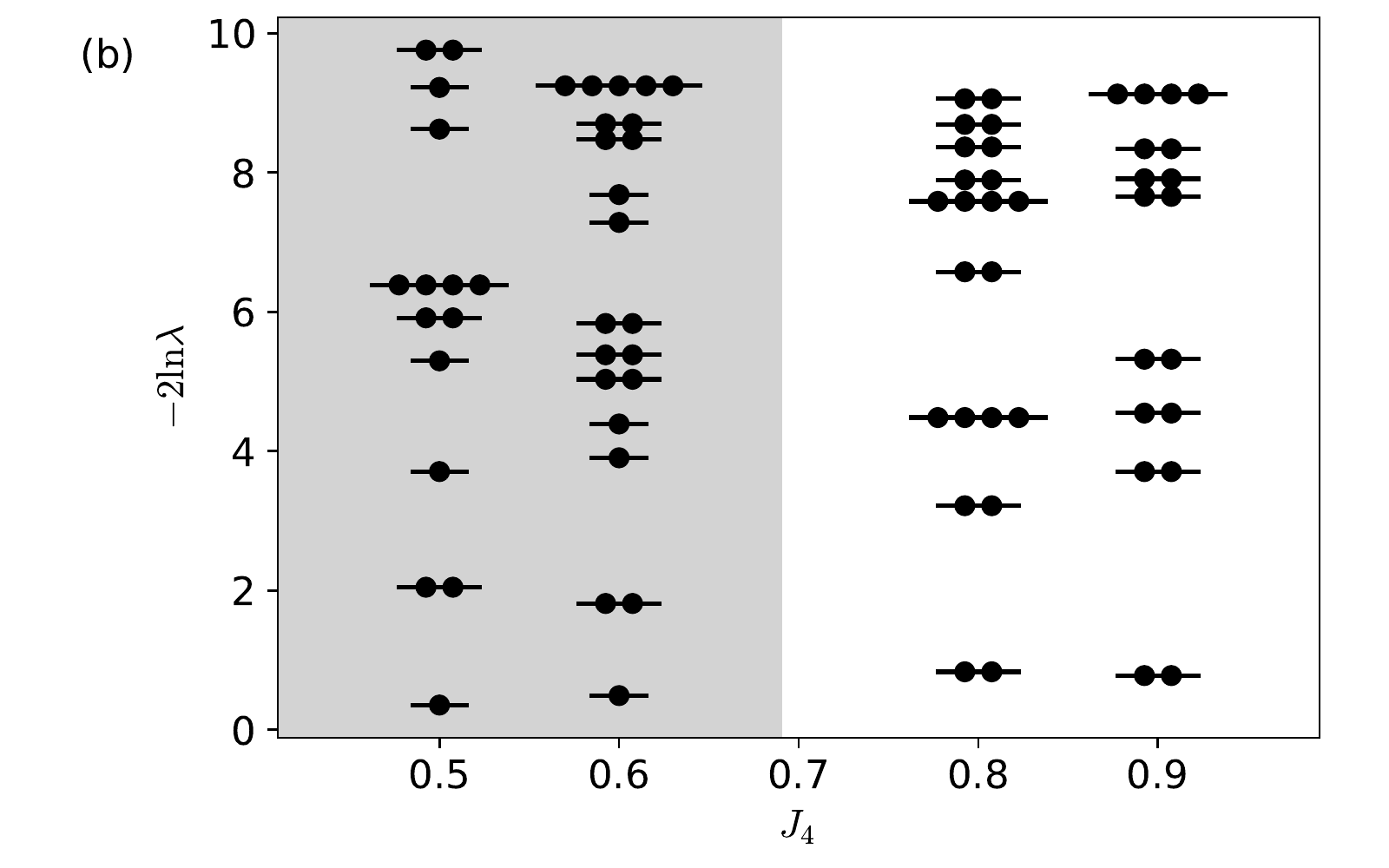} 
\caption{
(a) Topological indices and (b) entanglement spectra $\{-2\ln \lambda_\alpha\}$ around the RS-Haldane* transition at $J_\perp = 1$ and $\Delta = 0.7$; 
see the red solid line in Fig.\ \ref{fig:PhaseDiagram1}. 
%and the RS*-Haldane transition at $J_\perp = -1$ and $\Delta = 0.8$.
%The Haldane, Haldane*, RS, and RS* phase are distinguished 
%by $\epsilon^\sigma_1$, $\mathcal{O}_{\sigma}$, and $\mathcal{O}_{Z_2 \times Z_2}$.
}
\label{fig:SPT1}
\end{figure}
%%%%%%%%%%%%%%%%%%%%%%%%%%%%%%%%%%%%%%%%%%%%%%%%%%%%%%%%%%%%%%%%%%%%%%%%%%%%%%%%%%%%%%%%%%%%
%%%%%%%%%%%%%%%%%%%%%%%%%%%%%%%%%%%%%%%%%%%%%%%%%%%%%%%%%%%%%%%%%%%%%%%%%%%%%%%%%%%%%%%%%%%%
\begin{figure}
\includegraphics[width=80mm]{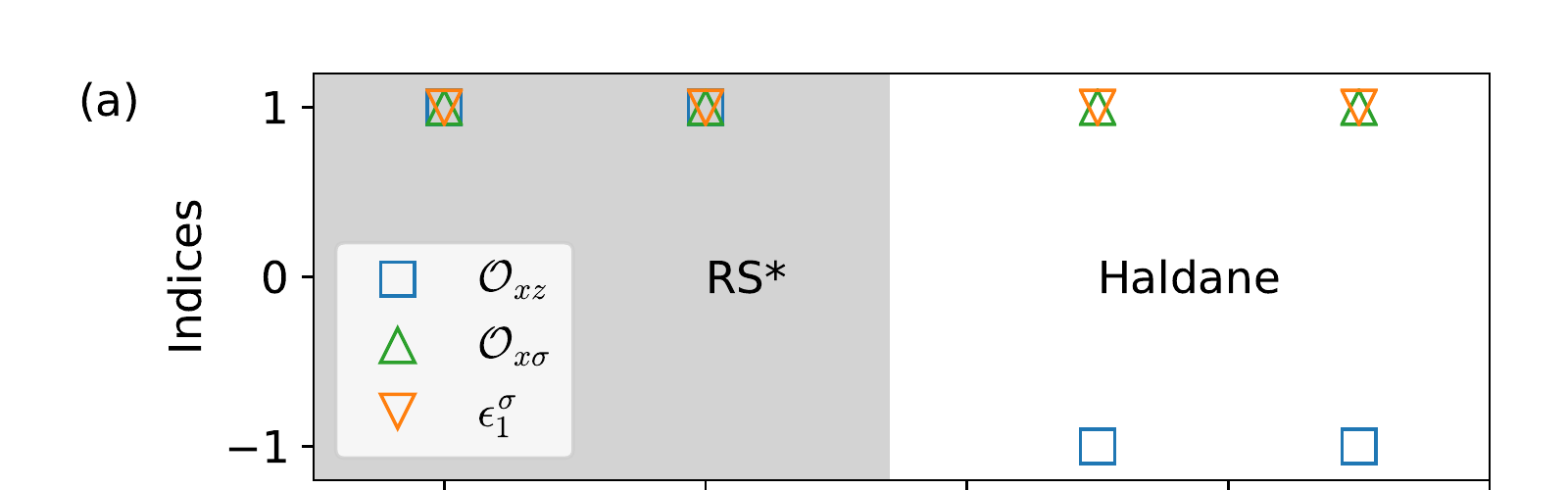} 
\includegraphics[width=80mm]{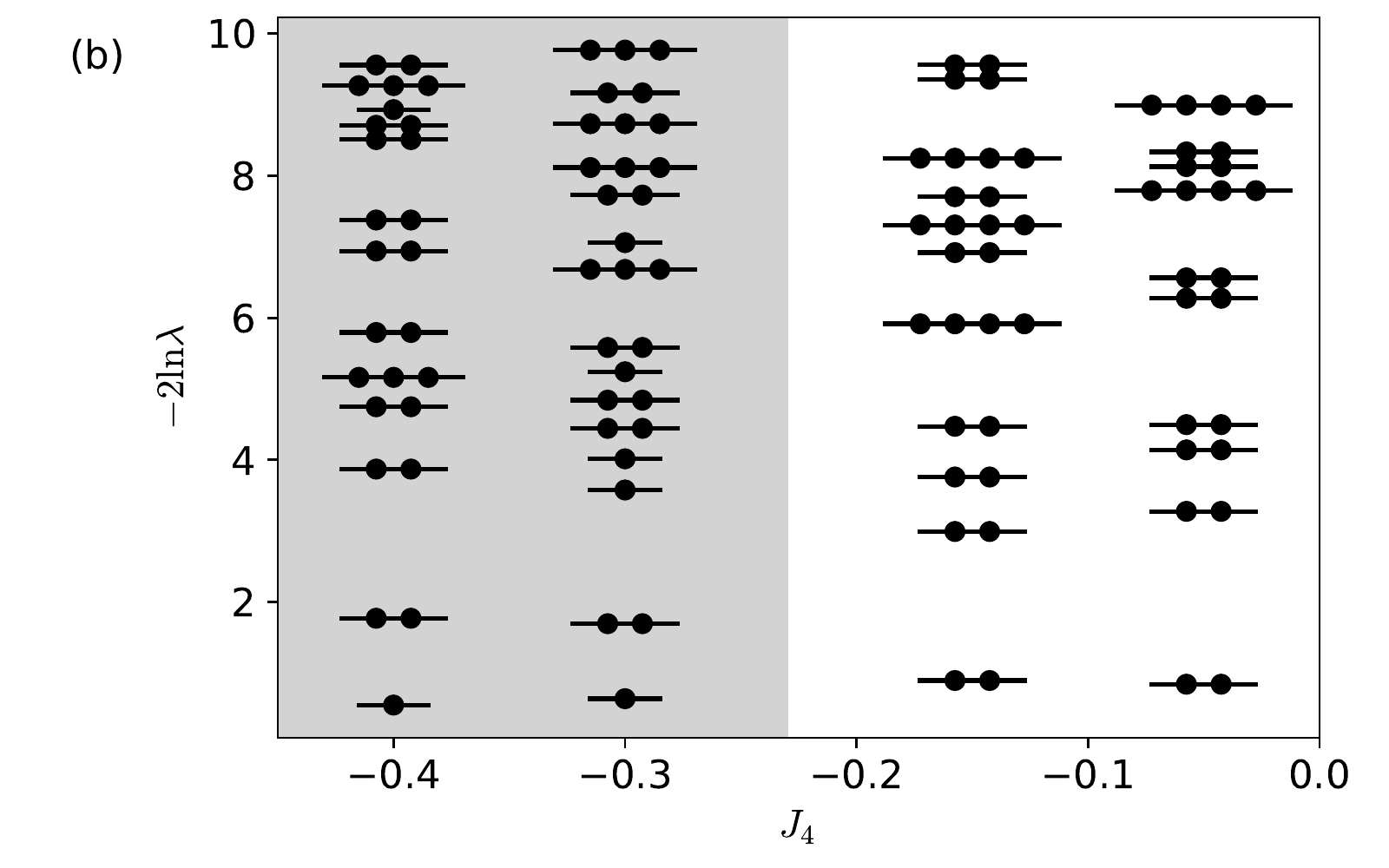} 
\caption{
(a) Topological indices and (b) entanglement spectra $\{-2\ln \lambda_\alpha\}$ around the Haldane*-RS transition at $J_\perp = -1$ and $\Delta = 0.8$; 
see the red solid line in Fig.\ \ref{fig:PhaseDiagram2}. 
%and the RS*-Haldane transition at $J_\perp = -1$ and $\Delta = 0.8$.
%The Haldane, Haldane*, RS, and RS* phase are distinguished 
%by $\epsilon^\sigma_1$, $\mathcal{O}_{\sigma}$, and $\mathcal{O}_{Z_2 \times Z_2}$.
}
\label{fig:SPT2}
\end{figure}
%%%%%%%%%%%%%%%%%%%%%%%%%%%%%%%%%%%%%%%%%%%%%%%%%%%%%%%%%%%%%%%%%%%%%%%%%%%%%%%%%%%%%%%%%%%%
%%%%%%%%%%%%%%%%%%%%%%%%%%%%%%%%%%%%%%%%%%%%%%%%%%%%%%%%%%%%%%%%%%%%%%%%%%%%%%%%%%%%%%%%%%%%
%\begin{figure}
%\includegraphics[width=90mm]{SPTindices_RS_HaldaneS.pdf} 
%\includegraphics[width=90mm]{SPTindices_RSS_Haldane.pdf} 
%\caption{
%The topological indices around the RS-Haldane* transition at $J_\perp = 1$ and $\Delta = 0.7$,
%and the RS*-Haldane transition at $J_\perp = -1$ and $\Delta = 0.8$.
%The Haldane, Haldane*, RS, and RS* phase are distinguished 
%by $\epsilon^\sigma_1$, $\mathcal{O}_{\sigma}$, and $\mathcal{O}_{Z_2 \times Z_2}$.
%}
%\label{fig:SPT}
%\end{figure}
%%%%%%%%%%%%%%%%%%%%%%%%%%%%%%%%%%%%%%%%%%%%%%%%%%%%%%%%%%%%%%%%%%%%%%%%%%%%%%%%%%%%%%%%%%%%

% [ The matrix U ]--------------------------------------
Classification of 1D SPT phases can conveniently be discussed using the MPS representation 
\cite{PhysRevB.81.064439,PhysRevB.85.075125,PhysRevB.83.035107,PhysRevB.84.235128,PhysRevB.84.165139}. 
Assuming the translational invariance, we represent the (normalized) ground state of the infinite system in the form of a canonical MPS as 
\begin{align}
\ket{\Psi} = \sum_{\dots,l,m,n,\dots} 
[ \dots \Lambda \Gamma_{l} \Lambda \Gamma_{m} \Lambda \Gamma_{n} \dots] \ket{\dots l,m,n, \dots}. 
\end{align}
Here, $\Gamma_m$ is a $\chi$-by-$\chi$ matrix with $m$ being the spin state at a site,
and $\Lambda=\mathrm{diag}(\lambda_1,\dots,\lambda_\chi)$ is a diagonal matrix comprised of the Schmidt values associated with a bipartition of the system into half-infinite chains. 
In our application to the spin-$\frac12$ ladder, $m$ runs over the four spin states on a rung. 
Suppose that $\ket{\Psi}$ is invariant under an on-site unitary transformation, 
which is represented in the spin basis as a unitary matrix $\Sigma_{mm'}$ acting on every site. 
Namely, we have $|\langle\Psi | \tilde{\Psi}\rangle|=1$, where $|\tilde{\Psi}\rangle$ is the state after the transformation. 
Then the $\Gamma_m$ matrices can be shown to satisfy \cite{PhysRevLett.100.167202} 
\begin{align}
\sum_{m'} \Sigma_{mm'} \Gamma_{m'} = e^{i\theta_\Sigma} U_\Sigma^\dagger \Gamma_m U_\Sigma,
\label{eq:defU}
\end{align}
where $e^{i\theta_\Sigma}$ is a phase factor and $U_\Sigma$ is a $\chi$-by-$\chi$ unitary matrix which commutes with $\Lambda$. 
The phase factors $\{e^{i\theta_\Sigma}\}$ form a 1D representation of the symmetry group. 
The matrices $\{U_\Sigma\}$ form a $\chi$-dimensional projective representation of the symmetry group. 
Namely, it may differ from the 
%conventional 
linear representation by phase factors: 
\begin{equation}
 U_{\Sigma_1} U_{\Sigma_2} = e^{i\rho(\Sigma_1,\Sigma_2)} U_{\Sigma_1\Sigma_2}. 
\end{equation}
The phases $\rho(\Sigma_1,\Sigma_2)$, called the factor set of the representation, can be used to classify different phases. 
Specifically, if these phases cannot be gauged away by redefining the phases of $U_\Sigma$, 
the state $\{\ket{\Psi}\}$ belongs to a nontrivial SPT phase. 
When the translational symmetry is further imposed, $e^{i\theta_\Sigma}$ can also be used to classify phases \cite{PhysRevB.84.235128}. 
% When $\Sigma$ is a $Z_2$ symmetry, one can easily show $e^{i\theta_\Sigma}=\pm 1$. 

% [ How to calculate the matrix U and index alpha ]--------------------------------------
To obtain the matrix $U_\Sigma$ and the phase factor $e^{i\theta_\Sigma}$ numerically, we introduce a generalized transfer matrix 
\begin{align}
T^\Sigma_{\alpha\alpha';\beta\beta'} := \sum_m \qty( \sum_{m'} \Sigma_{mm'} \Gamma_{m',\alpha\beta}) \qty(\Gamma_{m,\alpha'\beta'})^* \lambda_\beta \lambda_{\beta'}.
\label{eq:defextendedTM}
\end{align}
Let $X$ be the right eigenvector with the largest absolute eigenvalue $\epsilon^\Sigma_1$: 
\begin{align}
T^\Sigma_{\alpha\beta;\alpha'\beta'} X_{\beta\beta'} = \epsilon^\Sigma_1 X_{\beta\beta'},
\label{eq:extendedTM}
\end{align}
where $X$ is normalized such that $XX^\dagger = I$.
Since we assumed $|\langle\Psi | \tilde{\Psi}\rangle|=1$,  $\epsilon^\Sigma_1$ must have unit modulus. 
Conversely, if $|\epsilon^\Sigma_1|<1$, we have $|\langle\Psi | \tilde{\Psi}\rangle|=0$. 
We now regard the vector $X$ with $\chi^2$ elements as a $\chi$-by-$\chi$ matrix. 
One can show that if $|\epsilon^\Sigma_1| = 1$, $X$ and $\epsilon^\Sigma_1$ are equal to $U^\dagger$ and $e^{i\theta_\Sigma}$, 
respectively, specifically, $U_{\beta\beta'}=X_{\beta'\beta}^*$ \cite{PhysRevB.86.125441,PhysRevLett.100.167202}.

%If $|\epsilon^\Sigma_1| < 1 $, the inner product of the original quantum state and the transformed quantum state is zero.
%Thus, the original quantum state is not invariant under the transformation.
%In this section, we focus on $|\epsilon^\Sigma_1| = 1$, that is to say, the quantum states which are invariant under the transformation.
%In the presence of the transformation $\Sigma$ and the time-reversal symmetry,
%the 1D representation of $\Sigma$ satisfies $\epsilon^\Sigma_1 = \pm 1$ \cite{PhysRevB.84.235128}.

% [ Case of a ladder ]--------------------------------------
We now focus on the case of a spin-$\frac12$ ladder with the $D_2\times\sigma$ symmetry studied by Liu {\it et al.} \cite{PhysRevB.86.195122}. 
The symmetry transformations acting on each rung are given by the spin rotations 
$\Sigma^x := \exp[i\pi (S^x_1 + S^x_2)]$ and $\Sigma^z := \exp[i\pi (S^z_1 + S^z_2)]$ 
and the inter-chain exchange $\Sigma^\sigma:=\sum_{\alpha,\beta=\uparrow,\downarrow}\ket{\alpha\beta}\bra{\beta\alpha}$. 
For these transformations, we can introduce the unitary matrices $U_x:=U_{\Sigma^x}$, $U_z:=U_{\Sigma^z}$, and $U_\sigma:=U_{\Sigma^\sigma}$ 
and the corresponding indices $\epsilon_1^x$, $\epsilon_1^z$, and $\epsilon_1^\sigma$ as explained above. 
While $\Sigma^x$, $\Sigma^z$, and $\Sigma^\sigma$ commute with one another, 
the commutation relations among $U_x$, $U_z$, and $U_\sigma$ may involve nontrivial phase factors. 
Such phase factors can be used as fingerprints of different phases. 
They can conveniently be detected by introducing traced commutators \cite{PhysRevB.86.125441} 
\begin{equation} \label{eq:comD2}
% \mathcal{O}_{Z_2\times Z_2} := 
\mathcal{O}_{xz} := 
\begin{cases}
0 & \text{if}~~ |\epsilon_1^x| < 1 ~\text{or}~ |\epsilon_1^z| < 1; \\
\frac{1}{\chi} \tr( U_x U_z U_x^\dagger U_z^\dagger ) & \text{if}~~ |\epsilon_1^x| = |\epsilon_1^z| = 1,
\end{cases}  
\end{equation}
and
\begin{equation} \label{eq:comsigma}
% \mathcal{O}_{\sigma} := 
\mathcal{O}_{x\sigma} := 
\begin{cases}
0 & \text{if} ~~ |\epsilon^x| < 1 ~\text{or}~ |\epsilon_1^\sigma| < 1; \\
\frac{1}{\chi} \tr( U_x U_\sigma U_x^\dagger U_\sigma^\dagger ) & \text{if}~~|\epsilon_1^x| = |\epsilon_1^\sigma| = 1.
\end{cases}
\end{equation}
As explained above, we have $|\epsilon_1^\Sigma|<1$ if the state is not invariant under the transformation $\Sigma$; 
%in this case, traced commutators involving $\Sigma$ becomes zero. 
in this case, $\mathcal{O}_{\Sigma\Sigma'}$ is zero by definition.

%\begin{eqnarray}
%\Sigma^\sigma := \mqty[ 1 & 0 & 0 & 0 \\ 0 & 0 & 1 & 0 \\ 0 & 1 & 0 & 0 \\ 0 & 0 & 0 & 1 ],
%\end{eqnarray}
%which is represented in the spin basis $\{\ket{\uparrow\uparrow}, \ket{\uparrow\downarrow}, \ket{\downarrow\uparrow}, \ket{\downarrow\downarrow}\}$. 

%\begin{align}
%\sum_{j'} \Sigma^x_{jj'} \Gamma_{j'} = \alpha_x U_x^\dagger \Gamma_j U_x, \label{eq:SPTx}\\
%\sum_{j'} \Sigma^z_{jj'} \Gamma_{j'} = \alpha_z U_z^\dagger \Gamma_j U_z, \label{eq:SPTz}\\
%\sum_{j'} \Sigma^\sigma_{jj'} \Gamma_{j'} = \alpha_\sigma U_\sigma^\dagger \Gamma_j U_\sigma. \label{eq:SPTs}
%\end{align}

% [ Commutation relations among U's ]--------------------------------------
The commutation relations among $U_x$, $U_z$, and $U_\sigma$ can be read off from Table I in Ref.\ \cite{PhysRevB.86.195122}, 
in which possible projective representations of $D_2\times\sigma$ and the corresponding SPT phases are summarized. 
If we do not impose the translational symmetry, the RS and RS* phases (termed rung-$\ket{0,0}$ and rung-$\ket{1,z}$ in Ref.\ \cite{PhysRevB.86.195122}) 
fall into the trivial phase as they both include rung-factorized product states. 
In this trivial phase, $U_x$, $U_z$, and $U_\sigma$ commute with one another, 
indicating 
% $\mathcal{O}_{Z_2\times Z_2}=\mathcal{O}_\sigma=1$. 
$\mathcal{O}_{xz}=\mathcal{O}_{x\sigma}=1$. 
In contrast, the Haldane and Haldane* phases (termed $t_0$ and $t_z$ in Ref.\ \cite{PhysRevB.86.195122}) 
show the nontrivial relation $U_x U_z = - U_z U_x$, indicating 
% $\mathcal{O}_{Z_2\times Z_2}=-1$. 
$\mathcal{O}_{xz}=-1$. 
Furthermore, the Haldane* phase shows $U_x U_\sigma = - U_\sigma U_x$, indicating 
% $\mathcal{O}_\sigma=-1$. 
$\mathcal{O}_{x\sigma}=-1$. 
If we further impose the translational symmetry, the RS and RS* phases can be distinguished by the index $\epsilon^\sigma_1$. 
Indeed, by multiplying $\Sigma_\sigma$ to a singlet 
$\ket{0,0}:=\left( \ket{\uparrow\downarrow}-\ket{\downarrow\uparrow} \right)/\sqrt{2}$ 
and a twisted singlet 
$\ket{1,z}:=\left( \ket{\uparrow\downarrow}+\ket{\downarrow\uparrow} \right)/\sqrt{2}$ on a rung, 
we find $\epsilon^\sigma_1=-1$ and $\epsilon^\sigma_1=1$ in the RS and RS* phases, respectively. 
Therefore, we can distinguish the four featureless phases by using the indices 
% $\mathcal{O}_{Z_2\times Z_2}$, $\mathcal{O}_{\sigma}$, and $\epsilon^\sigma_1$.
$\mathcal{O}_{xz}$, $\mathcal{O}_{x\sigma}$, and $\epsilon^\sigma_1$.

%We obtain the commutation or anti-commutation relation by using the projective representation 
%In the presence of $D_2 ( = \mathbb{Z}_2\times \mathbb{Z}_2)$ symmetry,
%the unitary matrices $U_x$ and $U_z$ satisfy
%$U_x U_z = U_z U_x$ in the RS and RS* phase, and $U_x U_z = - U_z U_x$ in the Haldane and Haldane* phase;
%these relations correspond to $\mathcal{O}_{Z_2\times Z_2} = 1$ and $\mathcal{O}_{Z_2\times Z_2} = -1$, respectively.
%Similarly, in the presence of $D_2 \times \sigma$ symmetry,
%the unitary matrices $U_x$ and $U_\sigma$ satisfy
%$U_x U_\sigma = U_\sigma U_x$ in the RS, RS*, and Haldane phase, and $U_x U_\sigma = - U_\sigma U_x$ in the Haldane* phase;
%these relations correspond to $\mathcal{O}_{\sigma} = 1$ and $\mathcal{O}_{\sigma} = -1$, respectively.
%The unitary matrices $U_z$ and $U_\sigma$ satisfy $U_zU_\sigma = U_\sigma U_z$ in all the four phases.
%Therefore, we distinguish the RS, RS*, Haldane, and Haldane* phase by using $\epsilon^\sigma_1$, $\mathcal{O}_{Z_2\times Z_2}$, and $\mathcal{O}_{\sigma}$.
%The $\mathbb{Z}_2$ index satisfies $\epsilon^\sigma_1 = 1$ in the Haldane and RS* phase
%and $\epsilon^\sigma_1 = -1$ in the Haldane* and RS phase.

% [ Numerical calculations ]--------------------------------------
We have numerically calculated these topological indices in our model. 
Here, we have implemented the VUMPS algorithm not for multi-site unit cells as used in the previous sections 
but for single-site unit cells. 
With this implementation, transitions between featureless phases can be investigated more accurately; 
furthermore, the translational invariance is explicitly imposed, which is useful in applying the formalism described above. 
The procedure goes as follows. 
First, we calculate the ground state $\ket{\Psi}$ represented in a mixed canonical form.
Second, we recast the state into a canonical form. 
Third, we construct the generalized transfer matrix (\ref{eq:defextendedTM}), 
and obtain the largest eigenvalue $\epsilon^\Sigma_1$ and the corresponding right eigenvector $X$,
where the latter is equal to $U^\dagger$ if $|\epsilon^\Sigma_1| = 1$.
Finally, we calculate the traced commutators in Eqs.~\eqref{eq:comD2} and \eqref{eq:comsigma}.
Additionally, we have calculated the entanglement spectrum $\{-2\ln \lambda_\alpha\}$, 
which exhibits nontrivial degeneracy when some of $U_\Sigma$'s are noncommutative \cite{PhysRevB.81.064439}. 
For example, when $U_x U_z = - U_z U_x$ as in the Haldane and Haldane* phases, 
the entire spectrum shows double degeneracy as $\Lambda$ commutes with $U_x$ and $U_z$. 

% [ Numerical results on the indices and ES ]--------------------------------------
Numerical results obtained around the RS-Haldane* and Haldane*-RS transitions are shown in Figs.~\ref{fig:SPT1} and \ref{fig:SPT2}, respectively. 
The numerical values of the indices 
% $\mathcal{O}_{Z_2\times Z_2}$, $\mathcal{O}_{\sigma}$, 
$\mathcal{O}_{xz}$, $\mathcal{O}_{x\sigma}$, 
and $\epsilon^\sigma_1$ shown in the upper panels 
are in perfect agreement with the expected values in the four phases. 
The entanglement spectra shown in the lower panels show the expected double degeneracy in the Haldane and Haldane* phases. 
Physically, the double degeneracy arises if an odd number of (twisted) valence bonds are cut when the ladder is bipartitioned 
(see Appendix \ref{App:ES} for entanglement spectra in the ordered phases). 
In this sense, the degeneracy in the entanglement spectrum has similar information as the string correlation $O_\mathrm{odd}^z$; 
it cannot distinguish between the Haldane and Haldane* phases. 
Yet, more detailed information provided by the topological indices can distinguish all the four featureless phases. 

\section{Summary and outlook}
\label{sec:conclusion}
%%%%%%%%%%%%%%%%%%%%%%%%%%%%%%%%%%%%%%%%%%%%%%%%

In this paper, we have studied the spin-$\frac12$ XXZ model with a four-spin interaction on a two-leg ladder in Eq. \eqref{eq:modelmain}. By means of effective field theory and VUMPS calculations, 
we have obtained rich ground-state phase diagrams that consist of eight distinct gapped phases, 
as shown in Figs.\ \ref{fig:PhaseDiagrambos1}, \ref{fig:PhaseDiagrambos2}, \ref{fig:PhaseDiagram1}, and \ref{fig:PhaseDiagram2}. 
Notably, there are four featureless phases, i.e., the RS, RS*, Haldane, and Haldane* phases, which have a unique bulk ground state and do not break any symmetry. 
While the RS* and Haldane* phases have highly anisotropic nature, they are found to appear even in the vicinity of the isotropic case $\Delta=1$. 
In the obtained phase diagrams, the four featureless phases compete not only with magnetic phases but also with dimer phases that break the translational symmetry. 
We have argued and demonstrated that 
Gaussian transitions with the central charge $c=1$ occur between the featureless phases and between the ordered phases 
while Ising transitions with $c=1/2$ occur between the featureless and ordered phases. 
The two types of transition lines cross 
% in the SU$(2)$-symmetric case, 
at the SU$(2)$-symmetric point,
where the criticality is described by the SU$(2)_2$ WZW theory with $c=3/2$ 
\cite{PhysRevLett.78.3939,PhysRevB.66.134423,PhysRevB.88.104403,PhysRevB.82.214420,PhysRevLett.122.027201,PhysRevB.80.014426}. 
As argued by Liu {\it et al.}\ \cite{PhysRevB.86.195122}, 
the four featureless phases are distinguished in the presence of 
the spin rotational dihedral symmetry $D_2$, the inter-leg exchange symmetry $\sigma$, and the translational symmetry. 
We have demonstrated that these phases are indeed distinguished by topological indices associated with these symmetries. 

An implication from the present work is that a rich phase structure can emerge by introducing anisotropy 
around the $c=3/2$ critical point described by the SU$(2)_2$ WZW theory. 
Such a structure has also been found in the spin-$1$ chain with bilinear and biquadratic interactions 
around the $c=3/2$ critical point known as the Takhtajan-Babujian model \cite{TAKHTAJAN1982479,BABUJIAN1982479,affleck1986exact}: 
in the presence of uniaxial anisotropy $D$, the Haldane, dimer, N\'eel, and large-$D$ phases compete around this point \cite{PhysRevB.84.054451}. 
While the universality classes of the transitions among them have yet to be investigated, 
the phase structure is similar to Fig.\ \ref{fig:PhaseDiagrambos2}. 
It would be interesting to extend this idea to other systems such as a generalized Hubbard ladder \cite{PhysRevB.66.245106} 
and spin chains with higher symmetry \cite{affleck1986exact}. 
Research along this direction may provide opportunities to explore further examples of SPT phases and to investigate their competition with ordered phases. 

%############################
\begin{figure}
\includegraphics[width=83mm]{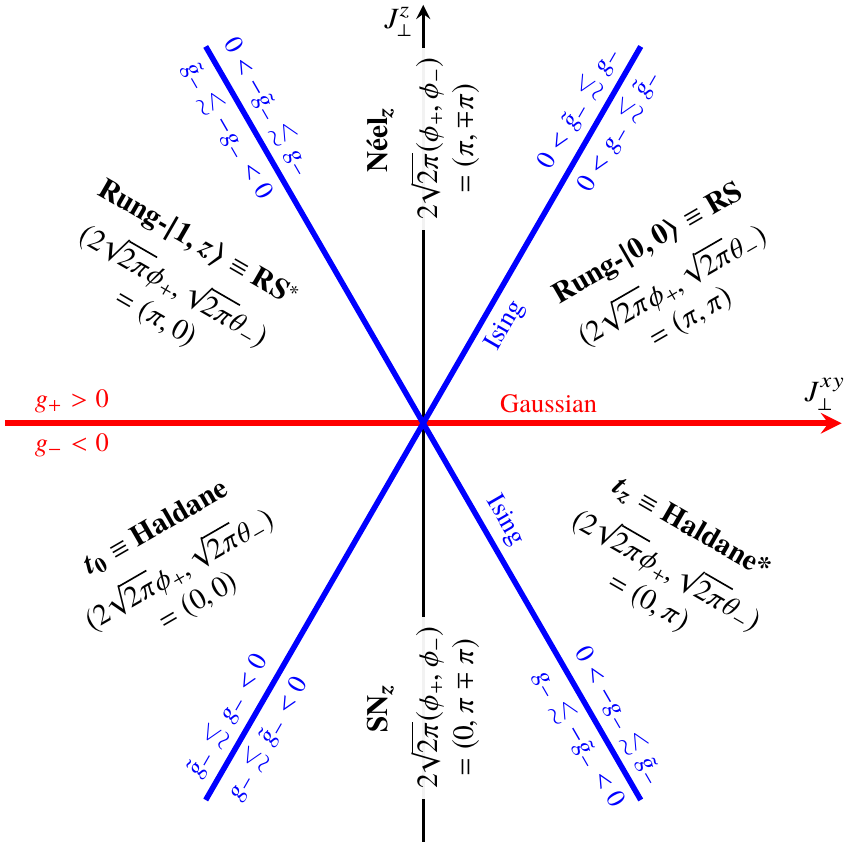} 
\caption{\label{fig:xyz1}
Schematic phase diagram of the model \eqref{eq:model_xyz} with $\Jperp^x=\Jperp^y=:\Jperp^{xy}$ and $|\Jperp^{xy}|,|\Jperp^z|\ll 1$. 
We assume that $\Delta$ is close to unity. 
A Gaussian transition in the symmetric channel is expected on the red solid line with $\Jperp^z=0$. 
Ising transitions in the antisymmetric channel are expected on the blue solid lines with $\Jperp^z \simeq \pm \qty(2b_0^2/a_1^2) \Jperp^{xy}$. 
The N\'eel$_\mu$ and SN$_\mu$ phases ($\mu=x,y,z$) are phases with magnetic orders along the $\mu$ axis; 
for $\mu=z$, they correspond to the N\'eel and SN phases in Figs.\ \ref{fig:PhaseDiagrambos1} and \ref{fig:PhaseDiagrambos2}. 
The phase structure is left-right symmetric as the sign of $\Jperp^{xy}$ can be flipped under the unitary transformation $U_1^z(\pi):=\exp\qty(i\pi\sum_j S_{1,j}^z)$. 
}
\end{figure}
%############################

%############################
\begin{figure}
\includegraphics[width=83mm]{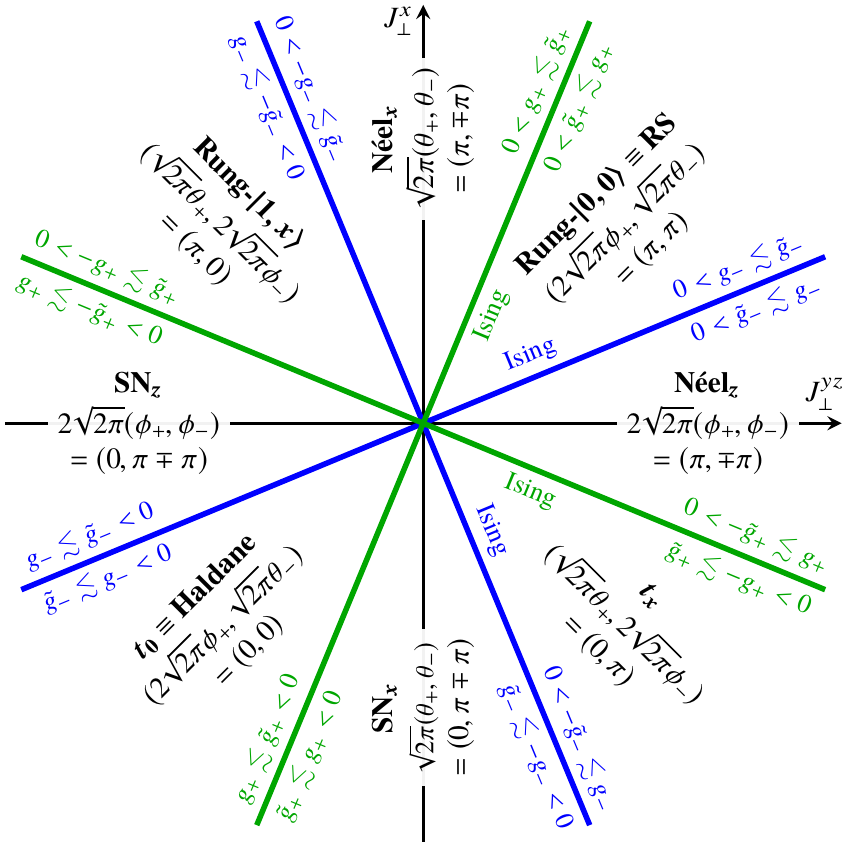} 
\caption{\label{fig:xyz2}
Schematic phase diagram of the model \eqref{eq:model_xyz} with $\Jperp^y=\Jperp^z=:\Jperp^{yz}$ and $|\Jperp^x|,|\Jperp^{yz}|\ll 1$.
We assume that $\Delta$ is close to but less than unity. 
With this assumption, we have $a_1>b_0$; for example, $a_1^2/b_0^2\simeq 1.4$ for $\Delta=0.9$ \cite{LUKYANOV1997571,PhysRevB.58.R583}. 
Ising transitions in the symmetric channel are expected on the green solid lines with $\Jperp^x\simeq \qty(1\pm a_1^2/b_0^2) \Jperp^{yz}$. 
Ising transitions in the antisymmetric channel are expected on the blue solid lines with $\Jperp^x \simeq \qty(-1\pm a_1^2/b_0^2)\Jperp^{yz}$. 
The phase structure is left-right symmetric as the sign of $\Jperp^{yz}$ can be flipped under $U_1^x(\pi)$. 
Under this transformation, the RS and Haldane phases are mapped onto the rung-$\ket{1,x}$ and $t_x$ phases, respectively. 
Our identification of the latter two phases are based on this observation. 
}
\end{figure}
%############################

%%%%%%%%%%%%%%%%%%%%%%%%%%%%%%%%%%%%%%%%%%%%%%%%
% Acknowledgements
%%%%%%%%%%%%%%%%%%%%%%%%%%%%%%%%%%%%%%%%%%%%%%%%
\begin{acknowledgments}
The authors would like to thank 
Y. Fuji, R. K. Kaul, and M. Sato
for useful comments. 
This research was supported by JSPS KAKENHI Grants No.\ JP18K03446, No.\ JP19H01809 and No.\ JP20K03780. 
%\sout{and by MEXT as ``Priority Issue on Post-K computer'' 
%(Creation of New Functional Devices and High-Performance Materials to Support Next-Generation Industries)
% and ``Exploratory Challenge on Post-K computer'' (Challenge of Basic Science---Exploring Extremes through Multi-Physics and Multi-Scale Simulations).}
The numerical computations were performed on computers at the Supercomputer Center, 
the Institute for Solid State Physics (ISSP), the University of Tokyo. 
\end{acknowledgments}

%A.~Furusaki, S.~Iino, R.~K.~Kaul, H.~Kohshiro, K.~Tamai, and L.~Vanderstraeten

\appendix
%%%%%%%%%%%%%%%%%%%%%%%%%%%%%%%%%%%%%%%%%%%%%%%%
\section{Effective field theory for a spin-$\frac12$ XYZ ladder}\label{App:XYZ}
%%%%%%%%%%%%%%%%%%%%%%%%%%%%%%%%%%%%%%%%%%%%%%%%

% [ Purpose and Hamiltonian ]--------------------------------------
In this appendix, we apply the field-theoretical formulation in Sec.\ \ref{sec:EFT} to a highly anisotropic XYZ model on a ladder described by the Hamiltonian 
\begin{align}\label{eq:model_xyz}
H  =J \sum_{\alpha = 1,2} \sum_{j}  (\bm{S}_{\alpha,j}\cdot\bm{S}_{\alpha,j+1})_\Delta 
	 + \sum_{\mu=x,y,z} \Jperp^\mu \sum_j S_{1,j}^\mu S_{2,j}^\mu .
\end{align}
This model is equivalent to the model studied by Liu {\it et al.}\ \cite{PhysRevB.86.195122}, 
and our analysis gives a qualitative description of their numerical results, as we explain in the following. 

% [ Effective Hamiltonian ]--------------------------------------
We set $J=1$ and assume $|\Jperp^\mu| \ll 1~(\mu=x,y,z)$. 
Treating $\Jperp^\mu$'s perturbatively, we obtain the low-energy effective Hamiltonian similar to Eq.\ \eqref{eq:Heff}
but with the additional term $\tilde{g}_+\cos\qty(\sqrt{2\pi}\theta_+)$. 
The coupling constants in this effective Hamiltonian are given by 
\begin{equation}\label{eq:gpm_xyz}
 g_\pm=\frac{a_1^2}{2a} \Jperp^z,~~\tilde{g}_\pm=\frac{b_0^2}{2a} \qty(\Jperp^x\mp\Jperp^y). 
\end{equation}
Assuming that $\Delta$ is close to unity, we can determine the ground-state phase diagram in a similar manner as in Sec.\ \ref{sec:bos_phases}. 

% [ XXZ case ]--------------------------------------
We first consider the case of $\Jperp^x=\Jperp^y=:\Jperp^{xy}$. 
In this case, the model \eqref{eq:model_xyz} has the $U(1)$ spin-rotational symmetry. 
Reflecting this symmetry, the $\tilde{g}_+$ term vanishes and a magnetic order is allowed only along the $z$ axis. 
The obtained phase diagram in Fig.\ \ref{fig:xyz1} exhibits competition among four featureless phases and two phases with magnetic orders along the $z$ axis. 
This diagram agrees qualitatively with Fig. 5 in Ref.\ \cite{PhysRevB.86.195122}, which is for $\Delta=1$. 
In the latter figure, however, the RS-Haldane* and RS*-Haldane phase boundaries are largely bent downward with increasing $|\Jperp^{xy}|$; 
this behavior is beyond the scope of the present analysis for weak inter-chain couplings. 
We refer the reader to Ref.\ \cite{PhysRevB.96.155133} for a related bosonization analysis on a 1D anisotropic Kondo lattice. 

% [ XYZ case ]--------------------------------------
We next consider the case of $\Jperp^y=\Jperp^z=:\Jperp^{yz}$ and $\Delta<1$. 
In this case, the model \eqref{eq:model_xyz} does not have the $U(1)$ spin-rotational symmetry. 
The obtained phase diagram in Fig.\ \ref{fig:xyz2} exhibits competition among four featureless phases and four phases with magnetic orders along the $x$ or $z$ axis. 
In the limit $\Delta\to 1$, where $a_1=b_0$, the N\'eel$_z$ and SN$_z$ phases disappear, and Fig.\ \ref{fig:xyz2} should become equivalent to Fig.\ \ref{fig:xyz1} 
although the roles of the $x$ and $z$ directions are interchanged. 
%Our identification of the rung-$\ket{1,x}$ and $t_x$ phases in Fig.\ \ref{fig:xyz2} are based on this observation. 
Figure \ref{fig:xyz2} agrees qualitatively with Fig. 1 in Ref.\ \cite{PhysRevB.86.195122}, which is for $\Delta=0.9$, if we interchange the $x$ and $z$ directions. 
However, the latter figure indicates that the rung-$\ket{0,0}$ and rung-$\ket{1,x}$ phases are much broader 
and the $t_0$ and $t_x$ phases are much narrower than in Fig.\ \ref{fig:xyz2}. 
We note that our conditions $|g_\pm|\simeq |\tilde{g}_\pm|$ for finding Ising transitions in the dual-field double sine-Gordon model are only approximate ones for $K_\pm \ne 1/2$ 
and can become inaccurate even in the regime of weak inter-chain couplings as $\Delta$ deviates from unity. 

Finally, we note that although the coupling constants in Eq.\ \eqref{eq:gpm_xyz} are subject to the constraint $g_+=g_-$, 
this constraint can be released by adding the four-spin interaction $J_4$, 
as seen in Eq.\ \eqref{eq:coeff_cos}. 
By varying $J_\perp^\mu~(\mu=x,y,z)$ and $J_4$ in the regime of weak inter-chain couplings, 
one can fully control the signs and the relative magnitudes of $g_\pm$ and $\tilde{g}_\pm$. 
This huge space allows one to obtain 16 possible phases: the rung-$\ket{0,0}$, rung-$\ket{1,\mu}$, $t_0$, $t_\mu$, N\'eel$_\mu$, SN$_\mu$ ($\mu=x,y,z$), SD, and CD phases. 

%%%%%%%%%%%%%%%%%%%%%%%%%%%%%%%%%%%%%%%%%%%%%%%%
\section{Entanglement spectra in the ordered phases}\label{App:ES}
%%%%%%%%%%%%%%%%%%%%%%%%%%%%%%%%%%%%%%%%%%%%%%%%

%%%%%%%%%%%%%%%%%%%%%%%%%%%%%%%%%%%%%%%%%%%%%%%%%%%%%%%%%%%%%%%%%%%%%%%%%%%%%%%%%%%%%%%%%%%%
\begin{figure}
\includegraphics[width=90mm]{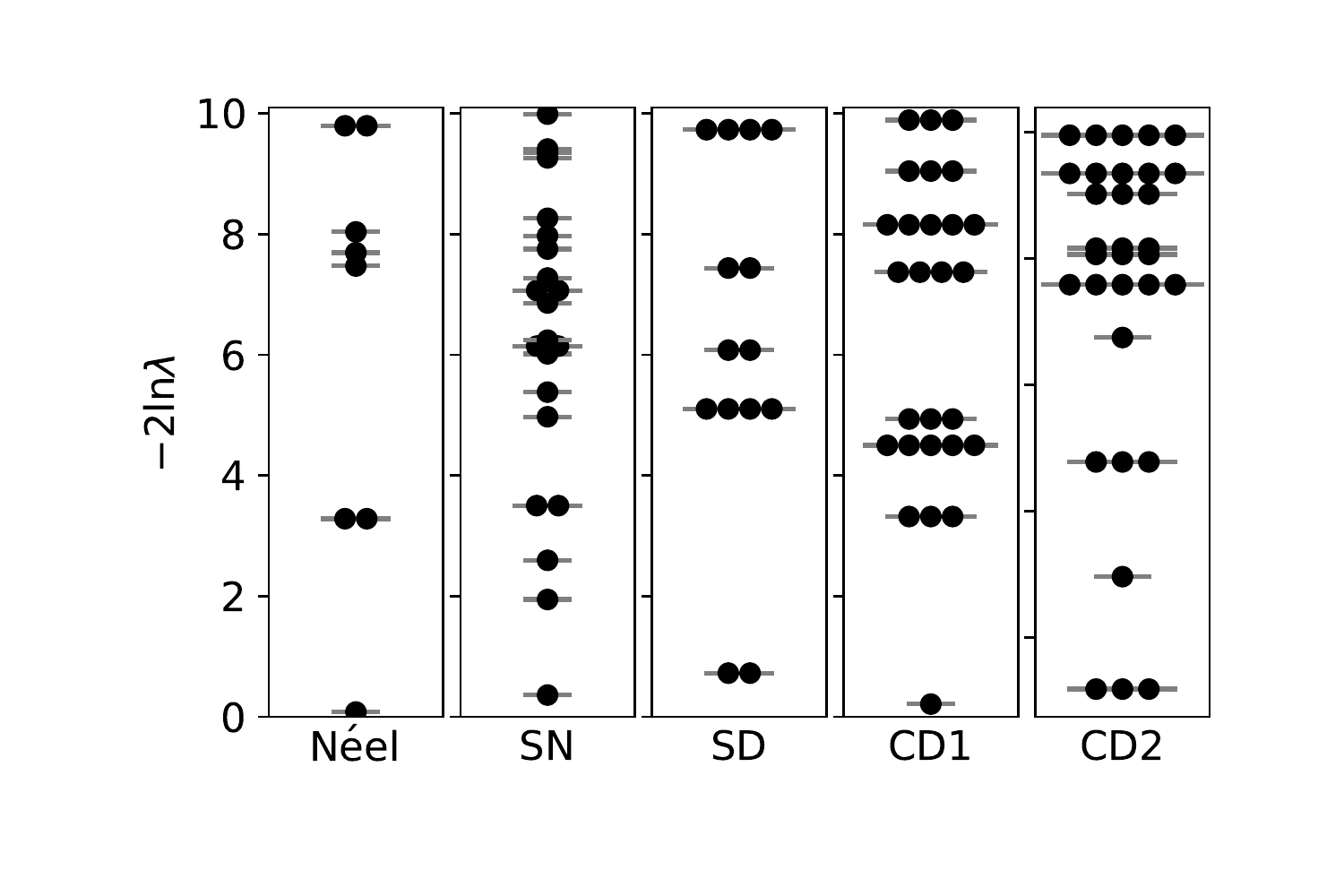} 
\caption{
Entanglement spectra $\{-2\ln \lambda_\alpha\}$ 
at representative points $(J_\perp,J_4,\Delta) = (1,1,1.2)$, $(-1,-1,1.05)$, $(1,1.5,1)$, and $(-1,-1,1)$
in the N\'eel, SN, SD, and CD phases, respectively, 
where $\{\lambda_\alpha\}$ are Schmidt values and the bond dimension is set to $\chi=96$. 
In the CD phase, the entanglement spectra are calculated for two types of cuts as explained in the text. 
%
%The CD phase does not have the one-site translational symmetry 
%so we cut the CD phase at two positions.
%The CD1 and CD2 represent the entanglement spectrum of the CD phase.
%The former cuts no singlet and the latter cuts two singlets.
%The dots show the Schmidt values.
%The N\'eel, SN, SD, and CD phases are calculated with $\chi=96$
%at $(J_4,\Delta) = (1.0,1.2)$, $(-1.0,1.05)$, $(1.5,1.0)$, $(-1.0,1.0)$ respectively.
}
\label{fig:App_ES}
\end{figure}
%%%%%%%%%%%%%%%%%%%%%%%%%%%%%%%%%%%%%%%%%%%%%%%%%%%%%%%%%%%%%%%%%%%%%%%%%%%%%%%%%%%%%%%%%%%%

In Sec.\ \ref{sec:SPT}, we have calculated the entanglement spectra in the four featureless phases, 
as shown in Figs.~\ref{fig:SPT1} and \ref{fig:SPT2}. 
In this appendix, we discuss the entanglement spectra in the four ordered phases. 

All the ordered phases discussed in the paper have the unit cell consisting of two effective sites (i.e., two rungs). 
We have therefore adopted the two-site unit cell implementation of the VUMPS algorithm as explained in Sec.\ \ref{sec:ResultJp}. 
Numerical results on the entanglement spectra are shown in Fig.\ \ref{fig:App_ES}.
It is expected that the entire spectrum shows at least double degeneracy if an odd number of valence bonds are cut when bipartitioning the system \cite{PhysRevB.81.064439}. 
The spectrum in the SD phase indeed shows the expected double degeneracy. 
In the CD phase, the spectra depend on where the system is cut. 
Specifically, we have obtained the CD state with $\expval{\mathcal{O}_{\text{CD}}(0)}>0$ [see Eq.\ \eqref{eq:O_CD} for the definition of the order parameter]. 
The spectrum labeled ``CD1'' is for the case when the cut is placed between the $(-1)$st and $0$th rungs. 
The spectrum labeled ``CD2'' is for the case when the cut is placed between the $0$th and $1$st rungs. 
The entanglement entropy is larger in the latter case. 
In an ideal CD state as shown in the inset in Fig.\ \ref{fig:PhaseDiagrambos2}, 
two valence bonds are cut in the ``CD2'' case, leading to 4-fold degeneracy in the entanglement spectrum. 
However, this degeneracy is split in a generic CD state. 
Therefore, there is no symmetry-protected degeneracy in the entanglement spectrum in the CD phase. 
The N\'eel and SN phases do not exhibit any symmetry-protected degeneracy in the entanglement spectra, either, 
as they include site-factorized product states with no entanglement. 

%The SPT phases can be classified by a double degeneracy of the entanglement spectrum \cite{PhysRevB.81.064439}.
%By using singular value decomposition, a wave function can be written as
%\begin{eqnarray}
%\ket{\Phi} = \sum_\alpha \lambda_\alpha \ket{\alpha L}\ket{\alpha R},
%\end{eqnarray}
%where $\ket{\alpha L}$ and $\ket{\alpha R}$ are orthonormal basis vectors of the left and right parts.
%The entanglement spectrum is $\lambda_\alpha^2$.
%When an odd [even] number of the VBs are cut, 
%the entanglement spectrum degenerate at least doubly [does not degenerate].
%Therefore, the Haldane, Haldane*, and SD phase have the double degeneracy.
%We numerically calculated the entanglement spectrum around the RS-Haldane* transition and the RS*-Haldane transition.
%These four phases have translational invariant so we used one-site unit cell implementation.
%The results are shown in Figs. \ref{fig:SPT1} and \ref{fig:SPT2}.
%On the other hand, the N\'eel, SN, SD, and CD phase
%do not have translational invariance, so we used two-site unit cell implementation.
%The results are shown in \ref{fig:App_ES}.

%%%%%%%%%%%%%%%%%%%%%%%%%%%%%%%%%%%%%%%%%%%%%%%%
\section{Phase transitions in the isotropic case}\label{App:isotropic}
%%%%%%%%%%%%%%%%%%%%%%%%%%%%%%%%%%%%%%%%%%%%%%%%

%%%%%%%%%%%%%%%%%%%%%%%%%%%%%%%%%%%%%%%%%%%%%%%%%%%%%%%%%%%%%%%%%%%%%%%%%%%%%%%%%%%%%%%%%%%%
\begin{figure}
\includegraphics[width=88mm]{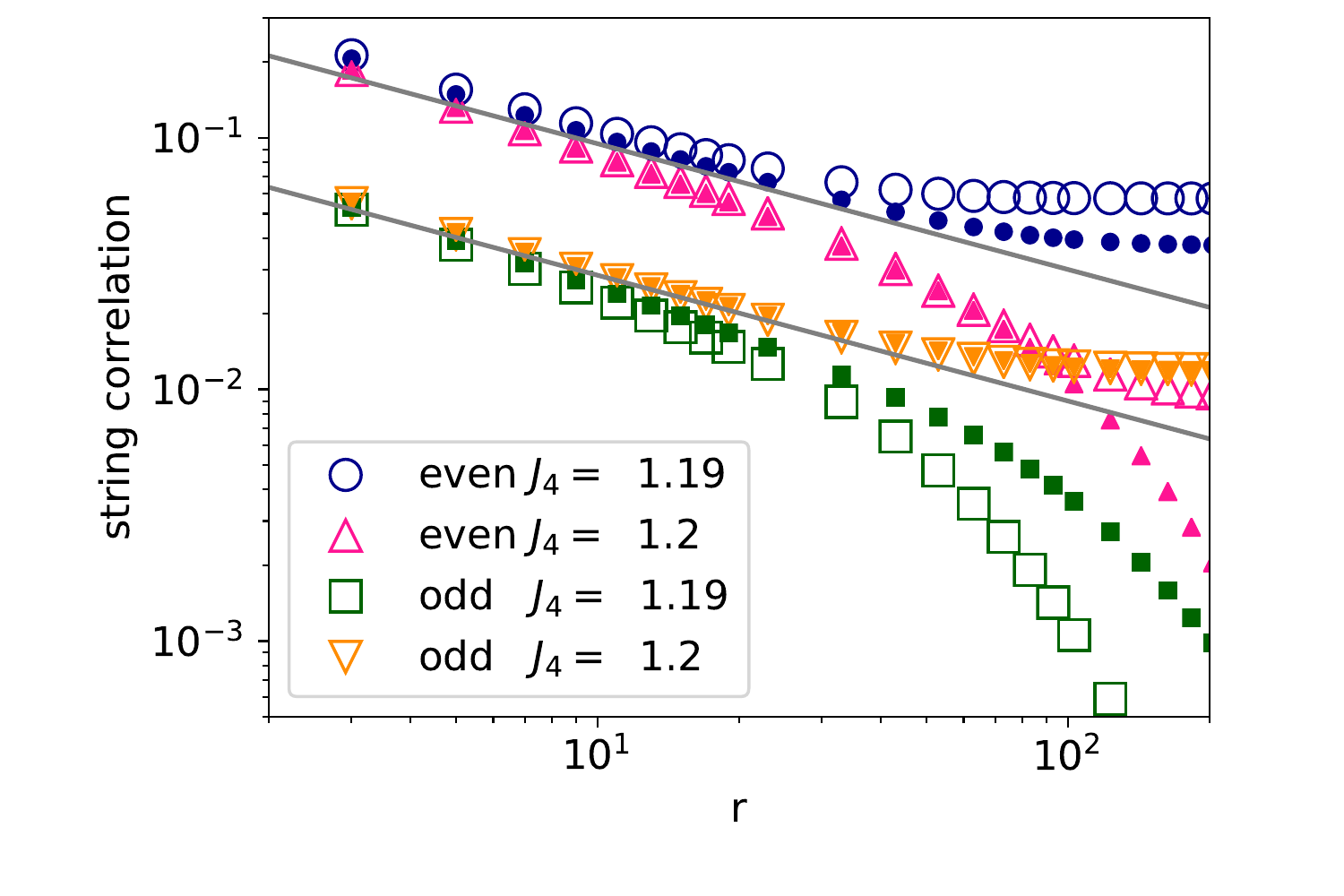} 
\caption{
Two string correlation functions \eqref{eq:stringOP} around the RS-SD transition at $J_\perp=1$ and $\Delta = 1.0$; 
see Fig.\ \ref{fig:PhaseDiagram1}. 
Logarithmic scales are used for both axes. 
Large open and small filled symbols are for $\chi=96$ and $192$, respectively. 
The two correlation functions are expected to show a power-law decay with the same exponent $K_+ = 1/2$ at the transition 
if the criticality is described by the SU$(2)_2$ WZW theory.
Two parallel gray solid lines are guides to the eye and their slope is $-0.5$.
The present figure indicates that the transition should occur between $J_4=1.19$ and $1.20$.
}
\label{fig:RSSD_string}
\end{figure}
%%%%%%%%%%%%%%%%%%%%%%%%%%%%%%%%%%%%%%%%%%%%%%%%%%%%%%%%%%%%%%%%%%%%%%%%%%%%%%%%%%%%%%%%%%%%

%%%%%%%%%%%%%%%%%%%%%%%%%%%%%%%%%%%%%%%%%%%%%%%%%%%%%%%%%%%%%%%%%%%%%%%%%%%%%%%%%%%%%%%%%%%%
\begin{figure}
\includegraphics[width=88mm]{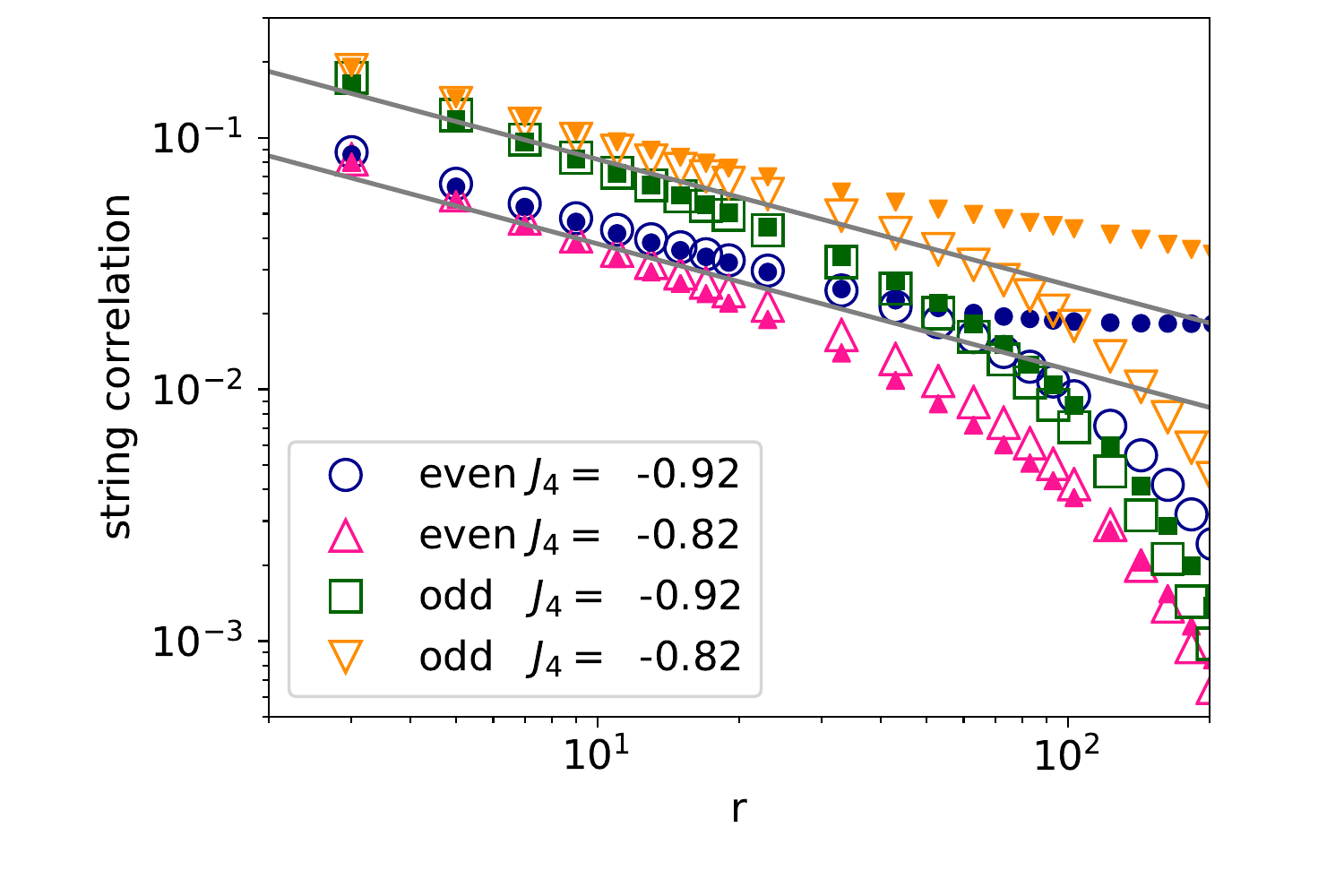} 
\caption{
Two string correlation functions \eqref{eq:stringOP} around the CD-Haldane transition at $J_\perp=-1$ and $\Delta = 1.0$; 
see Fig.\ \ref{fig:PhaseDiagram2}. 
%Logarithmic scales are used for both axes. 
As in the case of Fig.\ \ref{fig:RSSD_string}, large open and small filled symbols are for $\chi=96$ and $192$, respectively. 
%The two correlation functions show a power-law decay with the same exponent $K_+ = 0.5$ at the transition.
Two parallel gray solid lines are guides to the eye and their slope is $-0.5$.
The present figure indicates that the transition should occur between $J_4=-0.92$ and $-0.82$.
}
\label{fig:HaldaneCD_string}
\end{figure}
%%%%%%%%%%%%%%%%%%%%%%%%%%%%%%%%%%%%%%%%%%%%%%%%%%%%%%%%%%%%%%%%%%%%%%%%%%%%%%%%%%%%%%%%%%%%

In this appendix, we estimate the RS-SD and Haldane-CD transition points in the isotropic case $\Delta=1$. 
Field-theoretical analyses for weak inter-chain couplings ($|\Jperp|,|J_4| \ll 1$) 
\cite{PhysRevLett.78.3939,PhysRevB.66.134423,PhysRevB.82.214420,PhysRevLett.122.027201} suggest 
that these transitions are continuous and described by the SU$(2)_2$ WZW theory with the central charge $c=3/2$, 
although a possibility of a first-order transition is not excluded. 
The SU$(2)_2$ WZW theory is equivalent to the combination of 
the Gaussian theory with $K_+=1/2$ in the symmetric channel and the Ising CFT in the antisymmetric channel. 
In estimating the transition points, we use the two types of string correlation functions \eqref{eq:stringOP} 
as they can selectively probe the information in the symmetric channel as seen in Eq.\ \eqref{eq:stringOP_bos}. 
Specifically, these correlation functions are expected to show power-law decay with the same exponent $K_+=1/2$ at the transition 
if the scenario of the SU$(2)_2$ WZW theory is true. 

Figure \ref{fig:RSSD_string} shows the numerical result around the RS-SD transition 
%\tred{ 
at $J_\perp = 1$
%}
. 
The result indicates that a power-law decay of the two correlation functions with the same exponent $K_+=1/2$ is likely to occur between $J_4=1.19$ and $1.20$. 
This result is consistent with the previous estimate $J_{4,c} \simeq 1.19$ of the transition point by exact diagonalization \cite{PhysRevB.80.014426}. 
At this point, we have obtained the central charge $c\simeq 1.55$ by examining the data of the entanglement entropy versus the correlation length 
in our previous work \cite{ogino2020continuous}. 
Our results are thus consistent with the scenario of the SU$(2)_2$ WZW theory. 

Figure \ref{fig:HaldaneCD_string} shows the numerical result around the CD-Haldane transition 
%\tred{ 
at $J_\perp = -1$
%}
.
The result indicates that power-law decay with the same exponent $K_+=1/2$ can occur between $J_4=-0.92$ and $-0.82$. 
We thus obtain the estimate $J_{4,c}=-0.87(5)$ of the transition point, which still has a relatively large error range. 
However, we have found that if we try to reduce the range of $J_4$ further, one of the two correlation functions becomes inconsistent with the $K_+=1/2$ behavior. 
This is likely to be due to insufficient convergence of the numerical data as a function of $\chi$ 
although the breakdown of the scenario of the SU$(2)_2$ WZW theory is not excluded. 
Our estimate of the transition point could be compared with the exact diagonalization result 
of Hijii and Sakai \cite{PhysRevB.88.104403} (in particular, Figs. 5 and 6 therein).
However, because of a relatively large dependence on the system size in Ref.\ \cite{PhysRevB.88.104403}, 
a detailed consistency check is not possible. 

%In this appendix, we estimate the critical points and the TLL parameter $K$ in the isotropic cases
%by using the two string correlation functions \eqref{eq:stringOP}.
%The procedure is essentially the same as the estimation of the anisotropic cases.

%The RS-SD transition occurs at $J_{4,c} \simeq 1.19$ and is described by the $SU(2)_2$ WZW with $K=0.5$ \cite{PhysRevB.80.014426}.
%Figure \ref{fig:RSSD_string} shows that the critical point is between $1.19$ and $1.20$ and we obtain $K=0.58(9)$.

%The Haldane-CD transition occurs at $J_{4,c} \simeq -0.7$ and is described by the $SU(2)_2$ WZW
%with $K=0.5$ \cite{PhysRevB.88.104403}.
%Figure \ref{fig:HaldaneCD_string} shows that the critical point is between $-0.92$ and $-0.82$
%and we obtain the TLL parameter $K=0.58(10)$.
%At first glance, this result seems to contradict the previous study.
%However, we expect that our estimation is consistent with Fig. 5 in Ref. \cite{PhysRevB.88.104403}
%by choosing suitable parameters in Eq. (5) in Ref. \cite{PhysRevB.88.104403}.

%%%%%%%%%%%%%%%%%%%%%%%%%%%%%%%%%%%%%%%%%%%%%
%\section*{References}
%%%%%%%%%%%%%%%%%%%%%%%%%%%%%%%%%%%%%%%%%%%%%

\bibliography{references}
%\begin{thebibliography}{99}
%\end{thebibliography}

%\newpage
%\onecolumngrid
%%%%%%%%%%%%%%%%%%%%%%%%%%%%%%%%%%%%%%%%%%%%%%%%
%\section{Notes for Sec.\ \ref{sec:EFT} (not for publication)}
%%%%%%%%%%%%%%%%%%%%%%%%%%%%%%%%%%%%%%%%%%%%%%%%

\end{document}